# Exploration of Self-Propelling Droplets Using a Curiosity Driven Robotic Assistant

Jonathan Grizou[1], Laurie J. Points[1], Abhishek Sharma[1] and Leroy Cronin[1]*

[1]WestCHEM, School of Chemistry, University of Glasgow, Joseph Black Building, University Avenue, Glasgow G12 8QQ, U.K. *Corresponding author email: lee.cronin@glasgow.ac.uk

**We describe a chemical robotic assistant equipped with a curiosity algorithm (CA) that can efficiently explore the state a complex chemical system can exhibit. The CA-robot is designed to explore formulations in an open-ended way with no explicit optimization target. By applying the CA-robot to the study of self-propelling multicomponent oil-in-water droplets, we are able to observe an order of magnitude more variety of droplet behaviours than possible with a random parameter search and given the same budget. We demonstrate that the CA-robot enabled the discovery of a sudden and highly specific response of droplets to slight temperature changes. Six modes of self-propelled droplets motion were identified and classified using a time-temperature phase diagram and probed using a variety of techniques including NMR. This work illustrates how target free search can significantly increase the rate of unpredictable observations leading to new discoveries with potential applications in formulation chemistry.**

The investigation of multicomponent chemical formulation is a laborious and time-consuming effort. The combinatorial explosion, non-linear properties and rare events mean that even an expert experimentalist requires enormous resources to make significant discoveries. Although lab automation has shown a remarkable increase in experimental throughput[1,2], it does not change the relative rate of discoveries (wrt. the rate at which experiments are done) because the paradigm used to select experiments does not change alongside it. An appealing alternative is to implement the curious and knowledge-based inquiry process inherent in scientific researchers within a reliable and high-throughput robotic system[3–5]. Statistical methods were previously introduced to analyse the vast quantities of data generated by laboratory robots[6,7], and recently machine learning algorithms have started to be integrated into laboratory equipment[8,9]. However, most of these methods focus on the optimization of targeted properties[10,11] or require prior knowledge[12,13].

Herein, we focus on exploration for its own sake. We describe an experimental method (Figure 1) that implements state-of-the-art curiosity algorithms (CA) into a laboratory robot (CA-robot, Figure 2). CAs have been developed to replicate curiosity-driven learning in humans[15,16] and make use of knowledge acquired from developmental psychology, neuroscience, artificial intelligence and robotics.[14] CAs have previously been shown very efficient at exploring systems in simulated problems or constrained robotic scenarios[17–19]. Because CAs are designed to actively and autonomously select experiments that maximize the number of new and reproducible observations, applying CAs to the exploration of chemical systems could dramatically improve the rate of new scientific observations in the labs.

Our CA (Figure 1), called random goal exploration,[17,20] is the simplest of its algorithmic family, is easy to describe, and yet performs comparatively to other implementations[17] - making it an ideal candidate for this interdisciplinary didactic study. To select a new experiment, rather than deciding directly on experimental parameters, the CA generates a self-determined temporary target defined on the observation space. This temporary target represents an observation that the CA-robot will try to generate from the chemical system by defining a new experiment. To do so, the CA-robot refers to the dataset of previous experiments performed and builds a temporary model of the physical system using a regression algorithm. The model is used to infer the experimental parameters that are most likely to generate the self-determined target.

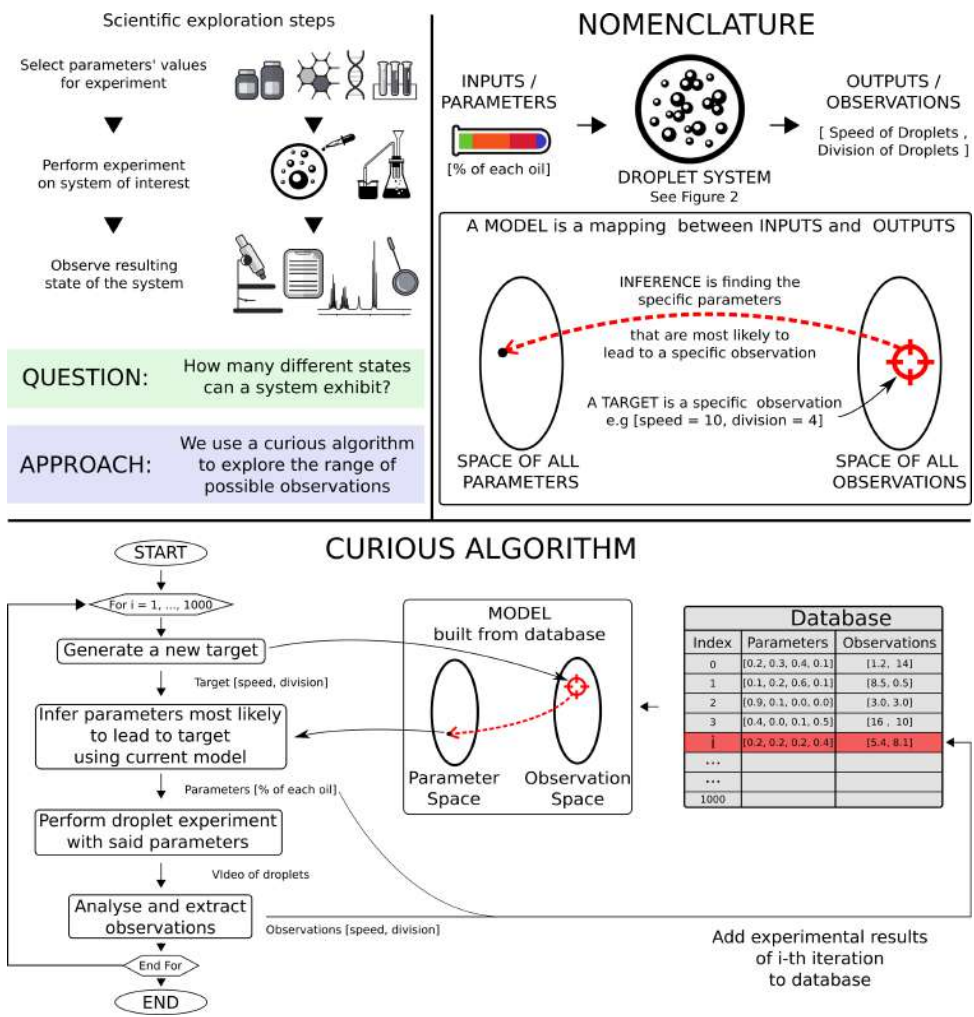

**Figure 1. Description of the curious algorithm (CA) and the exploration methodology.** Top-Left: Explanation of the research question and approach. Top-Right: Nomenclature of the terms used to describe our methodology using the droplet system in study as our example. Bottom: Flow chart of the CA algorithm. At each iteration (1 iteration = 1 experiment), the CA first selects a new temporary target that represent a desired observation. It then collates all the experimental results collected so far and uses them to build a model, which is used to infer the experimental parameters most likely to achieve the temporary target. The said experiment is then tested, and the results are stored in the dataset. The CA repeats this process until the budget allocated to the exploration is used up, 1000 experiments in this work.

The selected experiment is then undertaken leading to a new observation. The experiment results (both parameters and observations) are added to the dataset of previous experiments and will help improve the quality of the model, in turn improving the performance of the CA. The CA-robot repeats this process for a given number of iterations defined by the experimental budget allocated to the project. We highlight that the CA-robot always starts with zero experimental data and build the dataset at the same time as it explores and learns about the system. For the first experiment, the CA cannot use any previous information. To define the N-th experiment, it will be able to reuse the N-1 experiments previously performed. We also clarify that the CA generates a new temporary target for each new experiment. A detailed description of the algorithm is available in Supplementary Information, section 2.2.2.

To understand the benefit of this approach, consider the analogy with learning to play golf for the first time, with no tuition. With each shot, you can vary how you hit the ball and with what club (*your experimental parameters*). Your aim is to learn a wide skill set and discover where you can send the ball (*your observation space*). Every time you play a shot, you learn from how it went, and apply that knowledge to your future shots (*you are building a model from the dataset of past experiences*). The exploration question is: How do you allocate your time? Should you try contracting your muscles randomly and observe where the ball lands (*random parameter search*) or should you try to set yourself a variety of targets to reach and observe how far from these targets the ball lands (*our simple CA, called random goal exploration*)? The problem is the same in experimental sciences, when faced with the task to decipher an unfamiliar system (*hitting a ball with a club*), should we try experiments at random and observe how the system reacts (*contract your muscle randomly*) or should we try to target specific states or properties and observe if we can generate them (*set yourself different targets and learn from the process*). In the first approach (*random*), many experiments will tend to produce no interesting or new effects (e.g. *missing the ball*), in the second approach (*CA*), many targeted states will tend to be out of reach of the system (e.g. *putting the ball on the moon*). However, the strength of the CA approach is that, even if many targets cannot physically be attained, the process of trying to reach them has been shown to generate more varied observations than the random approach without the need of understanding the system in study.[17]

We tested our approach on dynamic oil-in-water droplets – promising protocell models[21,22] displaying an astonishing range of life-like behaviours, including movement, division, fusion and chemotaxis.[22–26] Although these droplets are thought to be driven by Marangoni instabilities originating from surface tension asymmetry,[27] to date, the understanding of even the most simple systems remains limited.[28,29] As such, oil-in-water droplets offer a great example of the challenges in studying complex and poorly understood systems where few components can lead to the emergence of a range of complex properties or behaviours, a topic of great relevance across many industries. Droplet experiments are performed by our CA-robot (Figure 2, Supplementary Movie 1) that can perform droplet experiments, record and analyse the droplets' behaviours, and select the next experiments in full closed-loop autonomy.

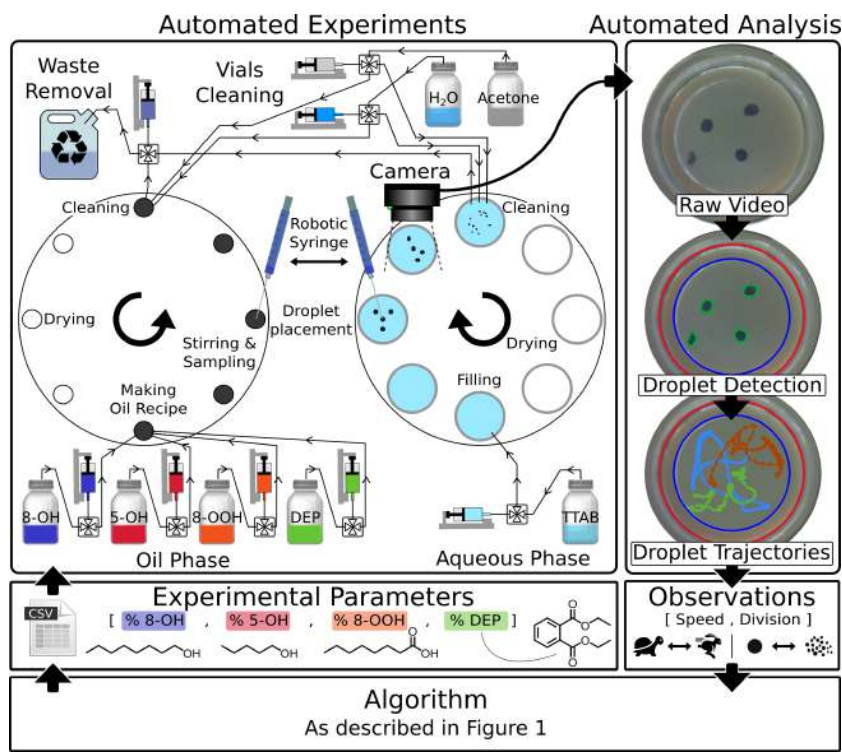

**Figure 2. Diagram of the closed-loop workflow of the robotic platform.** Top-Left: Schematic of new high-throughput droplet generating robot developed for this work. The robot runs the experiments by first mixing the oils accordingly, then prepares the aqueous phase and places droplets in the petri dish using a syringe. The motion of the droplets video is recorded and analysed. Once the experiment is completed, the platform cleans the entire system. Top-Right: Droplet contours and positions are extracted from the video data. Middle-Right: From the trajectories, the average speed and number of droplets generated per experiment was determined. Middle-Left: Experimental parameters are the proportion of each oil composing our droplets, which are then used by the platform to perform the next experiment. Bottom: The curious algorithm learns from the observations and define new experiments to be tested, see Figure 1.

## Discovery of an anomaly

The first objective of this study was to compare the efficiency in generating varied observations from a system between our CA approach and a standard random parameter search (also called screening) used in high-throughput automation. We gave ourselves a finite experimental budget of 1000 experiments and compared the range of behaviours we could observe using the CA or the random algorithm – both algorithms being tested three times. Our parameter space is composed of all possible mixtures of four oils (octanoic acid, diethyl phthalate (DEP), 1-octanol and 1-pentanol) from which our droplets are made. We chose our observation space as the droplet's speed and number of divisions, both selected due to their inherently interesting nature and similarity to the behaviours of simple lifeforms that can move and replicate.

While these specific droplet behavioural metrics were relevant in this context, the methodology and principles applied herein are not specific and could apply to many other metrics or systems. For example, we could consider the shape of droplet as an additional dimension of observation. In pigment mixing experiments, the parameters space could be the composition of a mixture of pigments, and the observation could be the resulting colour after mixing – for example in the Red-Green-Blue space.

To our surprise, during our first set of CA experiments, we noticed a drastic change in the observable outputs for our third repeat compared to the first and second repeats, namely at the third repeat no droplets were observed with speed above 5mm.s-1 (Supplementary Information 1.2). Our expectation was indeed to get roughly the same range of droplet behaviours at each repeat because we considered the same droplet system and the same algorithm. After careful investigation of all possible causes for this anomaly (change in chemicals, experimental conditions, robotic process, tracking algorithm, etc), we identified temperature as the most probable factor behind the observed phenomenon. The temperature in the room might have changed between the second and the third repeat. However, as in all previous reported work on this droplet system[26,30], the temperature was neither recorded, nor controlled, and all experiments were simply performed at room temperature. A new set of questions emerged: (1) Can a change of only a few degrees Celsius really impact our droplet system? If yes, how and to what extend? (2) Was it the CA algorithm that allowed the observation of this anomaly? Or would it have been as likely to make our serendipitous observation with the random algorithm if the temperature had changed too? We answer first the latter questions and then characterize thoroughly the temperature effect on our droplet system.

**Proving that the discovery was enabled by the CA algorithm**

To test whether our discovery was enabled by the CA algorithm, we ran three repeats of both algorithms (CA and random) at 22.6 ±0.5°C and 27.0 ±0.7°C *(mean*±std*)*. At 27°C, and given the same budget of 1000 experiments, the CA-robot generated significantly more varied droplet behaviours than the random parameter search (Figure 3 B and C - notice the higher speed and division of droplets observed using the CA versus the random methodology). We quantified this exploration (Supplementary Information 2.2.3) and found that the CA enables us to observe 73.4 ±15.2% of the total observable space, *ca.* 3.3x more (p=0.039 - Welch's t-test) than a random parameter search (22.5 ±2.1%) within the same experimental budget. Interestingly, after only 128 experiments the CA-robot already generated more varied experiments than random parameter search did in 1000 experiments (Figure 3A), a 7-fold efficiency gain in time and resources given the same hardware setup. Supplementary Movie 2 illustrates the exploration over time using both the CA and random; notice how even after as few as 50 experiments the CA driven exploration is already identifying more extreme cases of droplet behaviour, and this differentiation only increases as more experiments are undertaken. Strikingly the number of active droplet experiments observed (speed > 3mm.s$^{-1}$) is as low as 28.7 ±0.9 for random parameter search but jumps to 395.0 ±16.5 for the CA, a 14-fold improvement (p<0.001), without explicitly asking the robot to generate high speed experiments. This is further visualised in Supplementary Movie 3, which shows videos of the 1$^{st}$, 10$^{th}$ and 50$^{th}$ highest speed recipes from the two approaches.

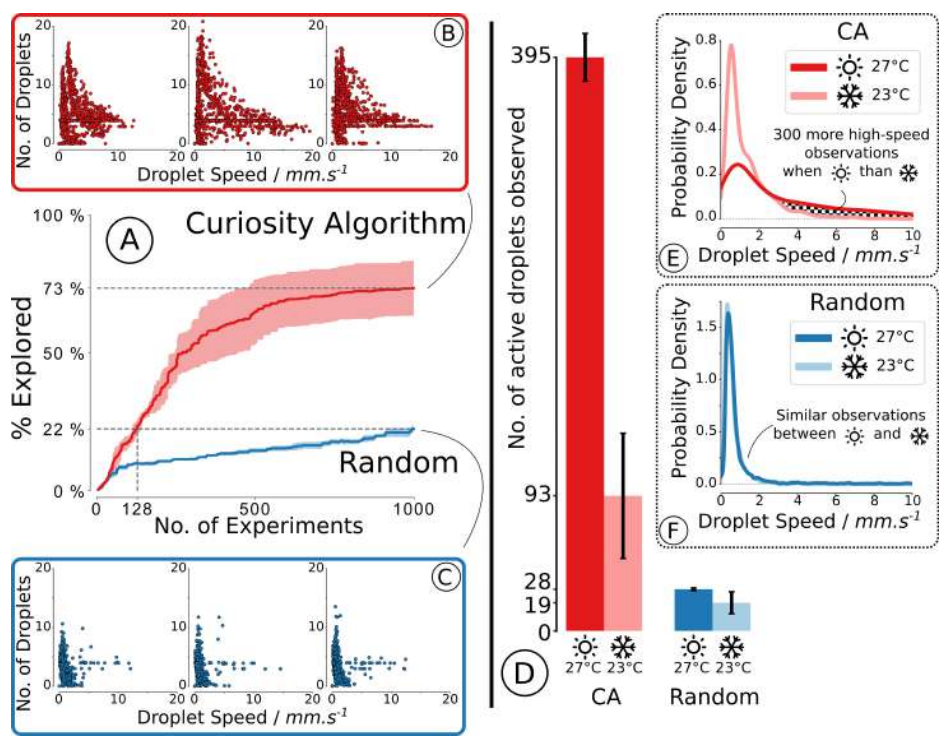

**Figure 3. A summary of the results generated using our CA-robot, illustrating how the CA enables both significantly greater exploration of the behavioural space and the discovery of temperature sensitivity of the droplets.** Left: Comparison of the observed droplet behaviours after 1000 individual experiments for CA and random – average of 3 repeats with shaded area showing 68% confidence interval. (A) Evolution of the percentage of the behaviour space explored between the two methods. CA explored 3.3 times more within the same experimental budget (73% vs 22%) and generated as diverse observations as random after only 128 experiments – a 7-fold reduction in time and financial cost for equivalent results. (B, C) Visualisation of the observations made by each method for each repeat; each scatter dot represents the average speed and number of droplets for a single 90 second droplet experiment. CA (B) leads to much more observations of rare and interesting droplets than random (C). Right: Effect of temperature (22.6 ±0.5°C vs 27.0 ±0.7°C) on the observations made using each algorithm. (D) Number of droplet experiments observed with a speed faster than 3 mms$^{-1}$ for each method and temperature with error bar showing standard deviation. The CA-robot, by performing the same number of experiments, generated 14 times more interesting droplet recipes than random at 27.0°C (395 vs 28, p<0.001), and 5 times more at 22.6°C (93 vs 19, p=0.13). A change of only ca. 4.4°C led to a large and significant difference in the observed droplet behaviours when using the CA (395 vs 93, p=0.005). This difference in effect could not be significantly observed when using random (28 vs 19, p=0.22). This is confirmed by (E) and (F) which show the distribution of observation respectively for CA and random. (E) The distribution of observations has a strong tail indicating a wider exploration from the CA-robot, and there is a significant difference between observations made at 27.0°C and 22.6°C that is not observable with random (F). By focusing on the output space, the CA-robot provides a more accurate picture of the system for the same experimental budget, which allowed the discovery of this delicate temperature effect.

The above result shows convincingly that, at a given temperature of 27°C and with a given budget, the CA enables to observe more varied droplet behaviours that a random parameters search. But could the temperature effect still have been observed by the random parameters search? Figure 3 compares the distribution of the speed of droplet experiments generated by both algorithms at 22.6 ±0.5°C and 27.0

±0.7°C. The *ca.* 4.4°C temperature change has a significant impact on the observations made using the CA (395.0 ±16.5 vs 93 ±43.1 active droplets, p=0.005) whilst a negligible change is observed with random parameter search (28.7 ±0.9 vs 19.3 ±7.6 active droplets, p=0.22). Notice the differences in the distribution of speed observed for each algorithm at both temperature in Figure 3 E and F. This key result allows us to claim that our initial observation of the temperature 'anomaly' was only feasible because of the exploratory benefits that our CA algorithm provides. By extension, we have shown that using a CA over a random parameter search to design exploratory experiments for an unfamiliar system is a better use of a limited experimental budget.

**Characterizing the temperature effect**

To study this newly observed effect in detail, we ran targeted droplet experiments within the range of temperatures accessible in the room (17-30°C). There were significant, unexpected and non-linear variations in the behaviour of the droplets of different compositions due to temperature (Supplementary Information 1.6). Such variations were highly reproducible as, for a given recipe, the observation of droplets' behaviour is enough to infer the room temperature with high accuracy (prediction error of 0.05 ±0.66°C – Supplementary Information 1.7), a testament to both the reproducibility of the droplet behaviours and the existence of a delicate temperature effect. This is rather striking given the complexity of the system, the timescale of an experiment and the relative simplicity of our video-based analysis. One recipe of interest (composed of 1.9% octanoic acid, 47.9% DEP, 13.5% 1-octanol, and 36.7% 1-pentanol) was further analysed. The vast differences of speed observed with this recipe to small temperature changes are illustrated in Supplementary Movie 4. To probe the causes behind these observations we ran longer (15 minute) droplet experiments at a range of temperatures (Supplementary Information 1.9). Surprisingly, as shown in Figure 4A and Supplementary Movie 5, the droplets were seen to exhibit two peaks in their speed-time profile – they accelerate to achieve a first maximum speed, decelerate, and then accelerate again to reach a second maximum speed. The temperature effect on droplet motion can clearly be seen in the variation of their speed profile, with the peak speed timing and magnitude exhibiting clear trends with temperature, with the peaks occurring earlier and with a greater magnitude for hotter experiments.

Utilising droplet displacement data we identified six clear stages of droplet motion: initiation, fluctuation, irregular, deceleration, continuous and saturation, of which characteristic examples may be seen in Figure 4G (P1 to P6). During the initiation stage, the droplet vibrates around a point, showing little locomotion and low speeds. During fluctuation, these vibrations extend and the droplet speed increases before peaking during irregular motion, in which the droplet moves short distances in alternating directions. This is followed by a deceleration stage, during which the droplets slow down and display smoother motion, which then develops into continuous motion, during which concerted movement is seen and resulting in a more circular motion of the droplets around the dish. Eventually the saturation stage is reached, in which the droplets slow down again and come to a halt. The peak speeds are observed for the irregular (purple) and continuous (orange) modes of motion, with the deceleration (green) period existing in between these two. A temperature-time phase diagram was derived showing the times at which each distinct phase of motion occurs at different temperatures (Figure 4C, Supplementary Information 1.10). The temperature-time phase diagram was created by calculating the intercept between cumulative distance travelled plots and linearly-fitted transition times (Figure 4B). The phase-transition times were each defined by characteristic points in the droplet acceleration-time plots. This phase diagram highlights the strong temperature dependence on the duration of each of the phases of motion and can be used to predict the mode of droplet motion observed at any time or temperature within the studied range.

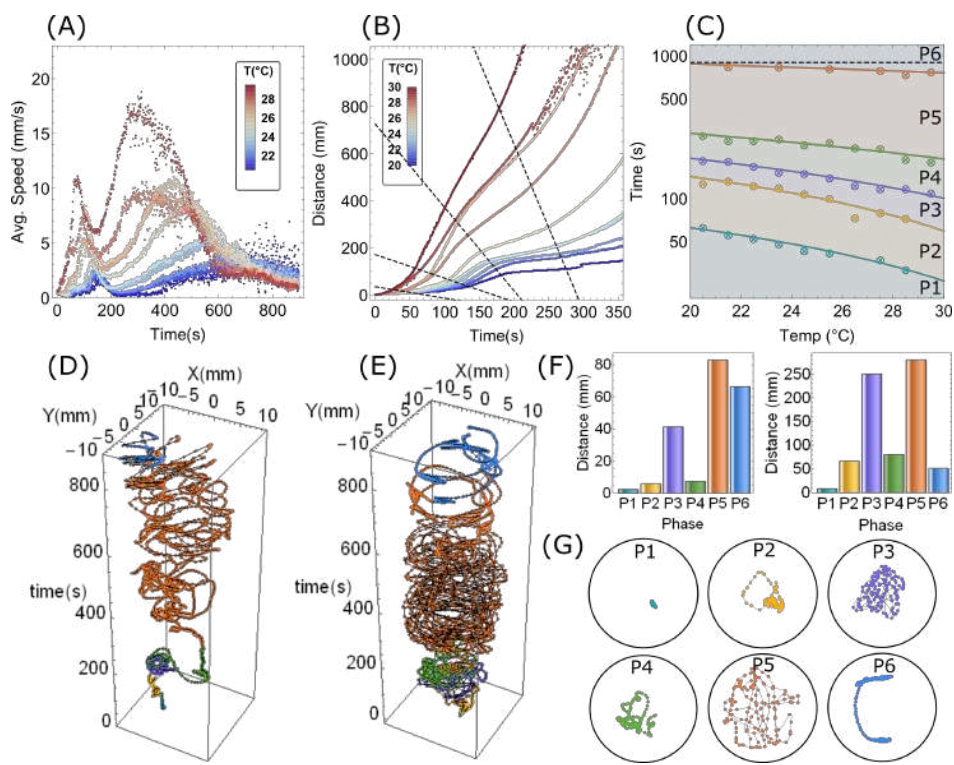

**Figure 4. A summary of the analysis undertaken on a focus recipe, which resulted in the classification of 6 phases of droplet motion and the production of a time-temperature phase diagram.** A) Temperature dependence of droplet speed vs. time. Each colour represents all experiments consisting of four droplets undertaken in a given temperature interval of 1°C. B) Temperature dependence of droplet cumulated distance moved vs. time. The black dashed lines show the phase transitions in droplet motion that are used to estimate the phase diagram and are calculated by linear fitting of maxima and minima in the acceleration profile at each temperature interval. C) Temperature-Time Phase diagram of droplet motion showing different phases, initiation (P1), fluctuation (P2), irregular (P3), deceleration (P4), continuous (P5) and saturation (P6). The marked data points correspond to the intercepts shown in (B). D, E) The trajectory of a single droplet at 21.44°C (D) and 27.39°C (E), with different motion phases highlighted by colour. G) Exemplar 36s segments of each phase of motion, with each point showing the droplet location every 0.25 s at 27.39°C (E). Each example trajectory contains the same number of points to emphasise the differences in distance covered during the different phases, which is quantified in the cumulative distance per phase plots (F) for the droplet trajectories seen in (D - left) and (E -right).

Oil dissolution into the aqueous phase is hypothesised to play a major role in the observed droplet behaviours,[27,29] with oil dissolution impacting the interfacial tension, leading to droplet motion induced by Marangoni instabilities. We utilized a previously reported $^1$H NMR spectroscopic method[24] to quantify the aqueous phase oil concentration during droplet motion at 22.4 ±0.2°C and 27.7 ±0.2°C (Supplementary Information 1.11). A 5°C temperature increase is seen to accelerate the dissolution of all oils (Figure 5).

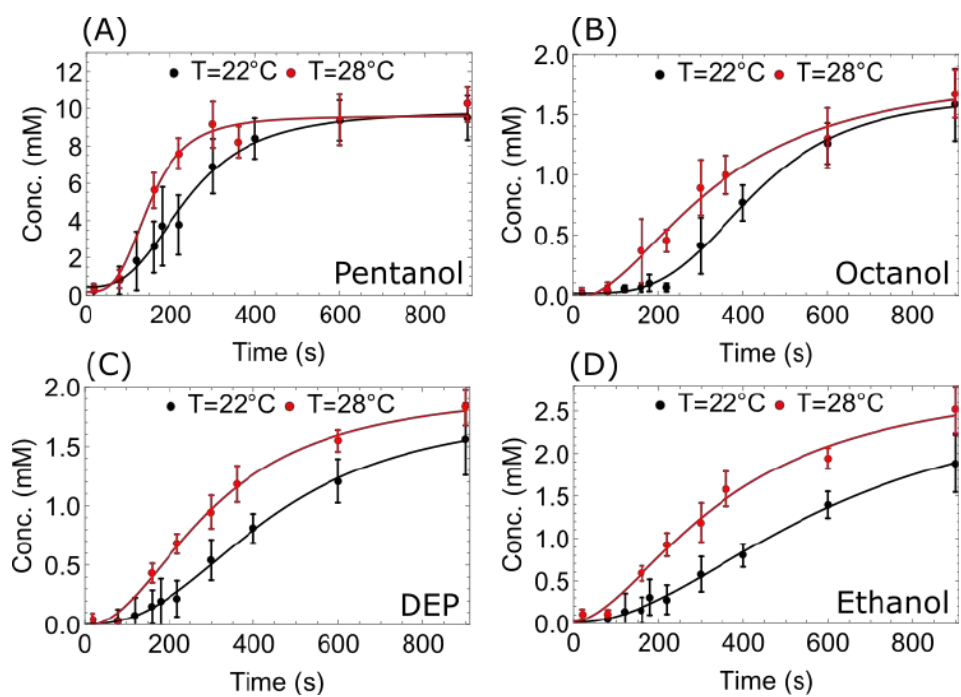

**Figure 5. Oil concentration in the aqueous phase over time at 22°C (black) and 28°C (red), as quantified by $^1$H NMR spectroscopy.** Note how each oil dissolves faster at the higher temperature, whilst DEP and ethanol also dissolve to different final concentrations. Note differences in y-axis scale – pentanol dissolves around 5x more than the other oils. When regulated to a target of 22°C, temperature at the experimental location was 22.4 ±0.2°C. When regulated to a target of 28°C, temperature at the experimental location was 27.7 ±0.2°C.

Pentanol dissolves fastest and to the greatest level, as expected by its relative solubility. Octanoic acid dissolves to a fixed level early in the experiment and then stays constant; this is unsurprising due to its low concentration in the formulation and the fact it will rapidly deprotonate at high pH. As previously reported,[24] we note the presence of ethanol due to the base catalysed hydrolysis of DEP. Interestingly, DEP and ethanol have different final concentrations at the different temperatures, as temperature affects the equilibrium of the hydrolysis reaction, as opposed to only physical processes driving the other oils dissolution. Octanol, DEP and ethanol dissolution are delayed as compared to pentanol dissolution, suggesting that pentanol dissolution is the main contributor to the first peak of droplet motion.

To confirm this hypothesis, we compared the oil dissolution rates with the droplet motion data, as shown in Figure 6A and B and detailed in Supplementary Information 1.12. The rate of pentanol dissolution is seen to be rapidly increasing during the fluctuation and irregular phases, before rapidly decreasing during the deceleration phase. This indicates that pentanol dominates the early stages of droplet motion, and that its dissolution is the primary cause of the fluctuation and irregular forms of motion. As pentanol dissolves so fast in these early stages, it is not surprising that the motion is sporadic, as rapid dissolution in all directions (Figure 6C) prevents the initiation of structure, regular flows and a more continuous form of motion. Because pentanol dissolution has largely ceased by the time of the continuous phase of motion, whilst the other oils are still dissolving to significant levels, it appears that DEP/ethanol and/or octanol are the primary driving force of the continuous period of motion. We hypothesise (Figure 6D) that the more gradual rate of dissolution during the continuous phase of motion allows a positive feedback loop to be setup between oil motion, dissolution and Marangoni flows.[31,32] As the droplet moves in this phase, it advects 'fresh' surfactant solution onto its anterior face (via collision with empty micelles and free surfactant molecules) and leaves a trail of oil filled micelles in its wake (via oil dissolution). Thus, the interfacial tension is higher at the posterior face as there are more oil filled micelles and less free surfactants in this zone. As there is an interfacial tension differential between the anterior and posterior faces of the moving droplet, a Marangoni flow is induced, supporting the forward direction of motion, providing a positive feedback loop for continued forward motion. This hypothesis is also supported by the observation that droplets often avoid following the recent path of other droplets. When the oil dissolution rates begin to saturate, the continuous motion slows and stops.

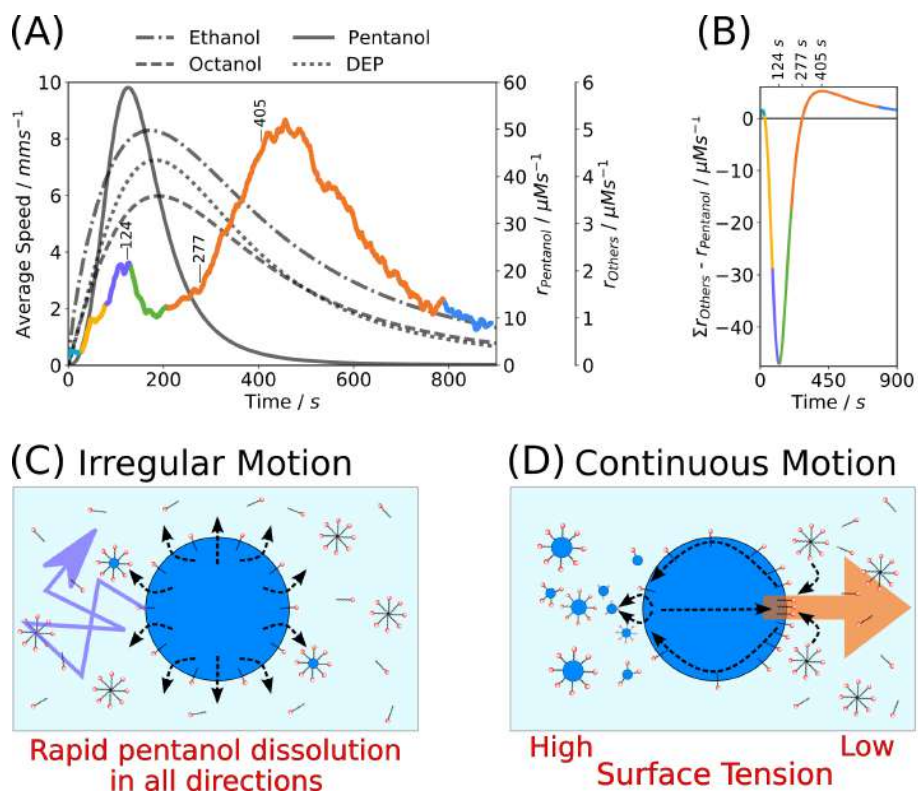

**Figure 6. The correlation between oil dissolution and droplet behaviours and schematics illustrating the proposed mechanisms for the irregular and continuous phases of motion.** A) The average oil droplet speed (coloured plot, left y axis) observed at 28°C (average across 8 experiments processed via a 10 seconds moving average), with the colour corresponding to the phase of motion (cyan-initiation, yellow-fluctuation, purple-irregular, green-deceleration, orange-continuous and blue-saturation). The grey lines illustrate the rates of oil dissolution (right hand y axes) from the fitted $^1$H NMR spectroscopy dissolution data. B) The difference between the sum of the rates of DEP, ethanol and octanol dissolution and the rate of pentanol dissolution against time. Note the peak difference in favour of pentanol at 124 seconds, the point at which the rates are equal at 277 seconds and the peak difference in favour of DEP, ethanol and octanol at 405 seconds. These times are also marked in (A) and correlate closely with the irregular-deceleration transition, rapid acceleration in the continuous phase and the maximum droplet speeds in the continuous phase. C) Schematic illustrating the proposed mechanism for the fluctuation and irregular phases of motion. Rapid pentanol dissolution in all directions (black arrows) into a largely oil free aqueous phase containing many empty micelles and free surfactants leads to no concerted directional motion, but rather erratic motion in various directions (purple arrow). D) Schematic illustrating the proposed mechanism for the continuous phase of motion. At this time, total oil dissolution is slower. The front of the moving droplets contacts 'fresh' aqueous phase whilst the rear of the droplet leaves a trail of 'filled' micelles. Thus, the interfacial tension is lower at the front of the droplet leading to a positive feedback loop of forward motion via Marangoni flows.

We cannot ascertain from the previously discussed data whether it is DEP, ethanol and/or octanol dissolution that is the primary cause of the continuous phase of motion. To discriminate between these, we varied the pH of the surfactant containing aqueous phase, which had a significant impact on the oil droplet behaviour. As the pH and temperature are increased, DEP hydrolysis is significantly accelerated,[33] leading to an earlier and larger second continuous motion peak (Supplementary Figure 36). With increasing pH there is also a $10^6$ fold increase in ionic strength, significantly reducing the aqueous solubility of alcohols,[34] thus lowering the dissolution of pentanol and reducing the irregular motion peak. These results together indicate that DEP hydrolysis is the primary cause of the second movement peak and continuous phase of motion. A range of experiments in which the pentanol-octanol ratio, the alcohol chain length and the number of droplets placed were varied further confirmed the links between pentanol and the first speed peak and DEP and the second speed peak (Supplementary Information 1.13).

As a proof of concept, we investigated the use of droplets as containers with temperature dependent release for active molecules, and showed that the dye methylene blue was released 2.5 times faster at 28.6 ±0.6°C than at 17.6 ±0.2°C (Supplementary Information 1.8, Supplementary Movie 6).

**Conclusion**

By equipping a droplet generating robot with a curiosity algorithm (CA-robot), we were able to uncover the temperature sensitivity of our self-propelled droplet system. We demonstrated that, given the same experimental budget, this temperature effect could not have been observed using a random parameter search. This illustrates that CA-robots can be of significant advantage to assist scientists in revealing properties of unfamiliar systems. Using physical and chemical analysis, we characterized the discovered effect and derived a phase diagram of droplet motion through time and temperature which links to the underlying oil dissolution processes. This chemical analysis revealed the astonishing complexity that underlies the dynamics of our 4-component oil-in-water droplet system. This is the first time a curiosity algorithm has been used for the exploration of a physical system in the lab using a fully automated robotic platform. Future research will focus on constructing the observation dimensions autonomously from the droplet videos in an unsupervised way,[20] as in this work the observation space was designed by the authors which potentially introduces human bias that can limit possible discoveries.

## Methods

**Robotic Platform**

We designed a high-throughput droplet-generating robot (Figure 2) that can execute and record a 90 s droplet experiment every 111 s, including mixing, syringe driven droplet placement, recording, cleaning and drying. Such minimal overhead time was achieved by parallelizing all operations enabling our platform to routinely perform 300 droplet experiments per day in full autonomy. The platform and sequence of operations are fully described in Supplementary Information 2.1.

**Droplet Chemistry**

The oil-in-water system comprises four droplets composed of a mixture of four oils placed onto a surfactant containing aqueous phase in a petri dish[26]. An experiment consists of preparing a formulation of octanoic acid, diethyl phthalate (DEP), 1-octanol and 1-pentanol at a specific ratio determined by the algorithm and dyed with 0.5 mgmL$^{-1}$ of Sudan Black B dye. The oil mixture is sampled by the robot using a 250μL syringe and delivered as 4 x 4μL droplets in a Y pattern from the center of a 32mm petri dish filled with 3.5mL of a 20mM cationic surfactant (myristyltrimethylammonium bromide, TTAB) solution raised to a high pH (ca. 13) using 8gL$^{-1}$ NaOH. The droplet making procedures are fully described in Supplementary Information 2.3.1.

**Image Analysis**

The droplet activity is recorded at 20fps for 90 seconds and analyzed using computer vision. Droplet contours are extracted using a thresholding algorithm and tracked through frames using a proximity rule. The droplets' average speed and the average number of droplets in the dish (droplets can split, fuse or leave the tracking area) are quantified and used as the observation space. The droplet tracking procedures are fully described in Supplementary Information 2.1.5.

**Algorithmic Implementation**

Experimental parameters are generated as a 4-dimensional vector representing the ratio of each oil in the droplet mixture. Observations are represented as a 2-dimensional vector representing the average speed and average number of droplets in an experiment. For the random goal exploration algorithm, the forward model is built uniquely from previous observations using locally weighted linear regression and the inverse model is solved for each target using the CMA-ES algorithm on the learnt forward model. The CA implementation is fully described in Supplementary Information 2.2.

**Data availability**

Due to the large total size of the droplet videos (> 500GB of data), the experimental data used in this work are available upon request to the corresponding author at lee.cronin@glasgow.ac.uk.

**Code availability**

The code used to operate the robotic platform, generate and analyse results are available online in our group GitHub account at https://github.com/croningp and are fully described in the Supplementary Information.

**ASSOCIATED CONTENT**

**Supplementary Information**

The Supplementary Information Appendix contains further results and discussion including more detail on related work, an in-depth comparison of the algorithms and a detailed explanation of the physicochemical analysis undertaken, the modelling of droplet behaviour and the phase diagram preparation. Additional experiments are presented studying the sensitivity of our system to pH, proportion of each oil, chain length of alcohol used and the number of droplets place in the dish, as well as detail given on the dye release and droplets as temperature sensors experiments. The Supplementary Information Appendix also provides detailed information (and the relevant GitHub repositories) about the materials and methods including the full droplet robot design and code, the droplet tracking implementation, a formal description of the curious algorithm and its implementation, and the experimental procedure related to the chemical analysis. Finally, the Supplementary Movies are listed along with their explanatory captions.

## AUTHOR INFORMATION

**Corresponding Author**

Correspondence should be addressed to LC (lee.cronin@glasgow.ac.uk).

**Author Contributions**

L.C. conceived the original idea and J.G. and L.C. together designed the project and the research plan; J.G. designed and built the robotic platform and implemented the computer vision and the algorithms. J.G. and L.J.P. performed droplet experiments. L.J.P. designed and ran the chemical analysis. J.G., L.J.P, A.S. respectively analyzed the algorithm, chemical and droplet motion data. A.S. extracted the phase diagram. L.J.P. designed and tested the chemical payload release system and J.G. quantified the release from the videos. J.G., L.J.P., A.S., and L.C wrote the paper.

**Funding Sources**

The authors gratefully acknowledge financial support from the EPSRC (Grant Nos EP/H024107/1, EP/I033459/1, EP/J00135X/1, EP/J015156/1, EP/K021966/1, EP/K023004/1, EP/K038885/1, EP/L015668/1, EP/L023652/1), the ERC (project 670467 SMART-POM). Some of this research was developed with funding from the Defense Advanced Research Projects Agency (DARPA). The views, opinions and/or findings expressed are those of the author and should not be interpreted as representing the official views or policies of the Department of Defense or the U.S. Government DARPA.

## ACKNOWLEDGMENT

We would like to thank Dave Adams, Pierre-Yves Oudeyer, and all anonymous reviewers for their constructive comments on earlier version of this manuscript. We would like to thank Silke Asche for her

# Exploration of Self-Propelling Droplets Using a Curiosity Driven Robotic Assistant

Jonathan Grizou[1], Laurie J. Points[1], Abhishek Sharma[1] and Leroy Cronin[1]*

[1]WestCHEM, School of Chemistry, University of Glasgow, Joseph Black Building, University Avenue, Glasgow G12 8QQ, U.K. *Corresponding author email: Lee.Cronin@glasgow.ac.uk

# Contents

# 1 Supplementary Results and Discussion

## 1.1 Related Work

A pioneering work on smart laboratory automation was the development of the 'Robot Scientist'[1,2] that can autonomously generate and perform experiments to discriminate between competing molecular biology hypotheses in an active learning[3] scenario. This approach requires background knowledge of the biological problem for generating the hypotheses. Active learning strategies have also recently been used for the exploration of the crystallization conditions of polyoxometalates[4] which required an initial training set of representative experiments to bootstrap the system.

In this work, we describe a novel system that extends the level of autonomy of robotic lab assistants by integrating algorithms for intrinsically motivated learning to a robotic assistant conducting physicochemical experiments in the laboratory with no human intervention. We believe this is the first time intrinsic motivation is added to a laboratory robot to study a physical system.

Intrinsic motivation is the self-desire to acquire new knowledge, skills, or experiences[5]. It pushes animals, including humans, to probe their environment without explicit immediate rewards, a drive characterized by engaging in playful and curious activities. Psychologists believe these activities are essential for sensorimotor and cognitive development throughout our lifespan[5–8]. Intrinsic motivation drives children, and scientists, to open-endedly experiment and play with the world around them.

In the last two decades, fields at the intersection of developmental psychology, neuroscience, artificial intelligence and robotics[9,10], have worked on modelling mechanisms of autonomous learning and intrinsically motivated exploration in humans[11–15] (see also related work in the field of evolutionary computation[16,17]). Interestingly, principles derived from studying intrinsic motivation have been successfully applied to solve difficult problems in the fields of artificial intelligence and robotics, such as sensori-motor coordination in high-dimensional spaces[18], playing ATARI games from raw pixels[19,20] and curriculum learning on a real robot[21]. To our knowledge, however, these algorithms have never been studied outside computer simulations or structured robotic problems.

Recently, in the physical sciences, there has been great interest in liquid droplets under non-equilibrium conditions as simple systems that display an astonishing range of life-like behaviours[22] including movement, division and chemotaxis[23–28]. These droplets are thought to be driven by Marangoni instabilities originating from surface tension asymmetry[29], but to date, the understanding of the driving forces influencing these droplets remains limited and computational models restrained to simple systems (e.g. 2D, single oil, single droplet, simplified interfaces, unbounded fluid)[30,31]. This illustrates the challenges of studying complex systems where few components can lead to the emergence of complexs[32].

The field of evolution-in-materio makes use of artificial evolution to optimize such complex systems directly in the physical world[33,34], with pioneering work on FPGA[35] and X-band antenna design[36]. Evolution-in-materio principles have proven powerful for the optimization of a four-component self-propelled oil-in-water droplet system[26,37,38]. First, the authors designed a droplet-generating robotic platform equipped with image analysis and a genetic algorithm that, starting from a random droplet mixture, was able to autonomously optimize droplet recipes that maximized the movement, division, and vibration of droplets to levels never reported before[26]. In subsequent work it has been shown that the chemical and physical environment of the droplets could be valuable parameters in the optimization of the droplet behaviours[37,38].

## 1.2 Observations leading to the discovery of a temperature effect

Initial experiments focused on evaluating the exploration power of the CA algorithm. These experiments were performed on our droplet system during the summer 2016. The droplets were dyed in red with Sudan Red and the droplet detection and tracking was using machine learning. The space of exploration was droplet speed and droplet dynamic deformation. The temperature was not controlled and not measured. At the time, this was consistent with our previous work and previously reported work on self-propelled droplets were temperature was not controlled neither reported.

Our aim was to test our platform and study whether or not the CA algorithm was able to generate a wider variety of observation than a random parameters search. We performed 3 runs of the CA with a budget of 1000 experiments. Supplementary Figure 1 shows the results of these 3 runs. In both run A and B, the algorithm was able to generate droplets with speed above 5 mm.s-1. However, and to our surprise, on run C no experiment could generate droplets with speed above 5mm.s-1. After careful investigation of the possible causes of this observation, we hypothesised that it might be due to temperature. We tested our hypothesis and indeed uncovered a subtle effect where just a few degrees could dramatically alter the behaviour of droplets.

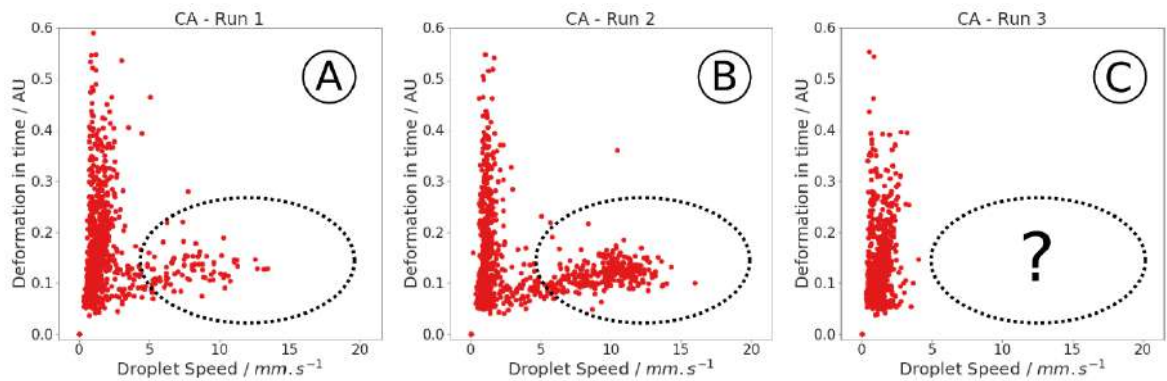

Supplementary Figure 1 – Data from initial experiments leading to the discovery of an anomaly. On run A and B, the CA algorithm generated droplets with high-speed but did not on run C. After investigating all possible causes that might have induced the results of run C, we hypothesised that the temperature might have had an effect on this particular run. We then controlled the temperature precisely and studied its effect on our droplet system. One important contribution of our work is to show that this temperature effect could not have been observed if we were running a random parameter search. Indeed, because the CA is exploring significantly more, it provides us with more representative observations of our system which highlighted the temperature effect (see main manuscript for details). This also implies that to explore an unknown system, using a CA is a better use of a given experimental budget as it is more likely to generate unexpected and novel observations.

However, this temperature effect is not the central claim of this work. We wondered if it was indeed the use of the CA algorithm that made this discovery possible. Was it the CA algorithm that allowed the observation of the initial anomaly? - which subsequently raised our curiosity and our will to investigate. In practice, we had to test whether the probability of making this observation was higher when running the CA than when performing a random search. Concretely, the question becomes: Could we have made this discovery using a random search algorithm if the temperature had changed too?

To test this question, we had to test both algorithms at two different temperatures, which we selected to be close together (within 4°C) and within the range of the lab's temperature in which experiments were typically performed, respectively 22°C and 26°C.

To answer the question, we finally had to measure the difference in observation made by the two algorithms at 22°C and 26°C. If both algorithms would enable us to observe the temperature effect, our discovery could not be claimed to be due to the CA algorithm. The results, as detailed in the main manuscript, turned out to show that a significant difference in observation was made using the CA between 22°C and 26°C, but not with the random search. Hence validating that the discovery was indeed enabled by the CA algorithm. This further implies that to explore an unknown system, using a CA algorithm is a better use of a given experimental budget as it is more likely to generate unexpected and novel observations.

## 1.3 Algorithm Comparison

Resources associated to this section:

1. The code to reproduce each figure can be found at:
   https://github.com/croningp/dropfactory_analysis
2. Larger files, including plots, data and Mathematica files, can be found in the associated release on Github named "SI":
   https://github.com/croningp/dropfactory_analysis/releases/tag/SI
3. Specifically, each type of plots shown in this section have been generated for all experimental runs and for each method, they can be found at:
   https://github.com/croningp/dropfactory_analysis/releases/download/SI/detailed_plots.zip

In this section, we provide a detailed analysis of the comparison between random goal exploration (called 'CA' in this section) and random parameter search (called 'random' in this section) algorithms.

Each run consists of 1000 experiments. We ran three repeats for each method and condition. We had 2 methods (random and CA) and 2 conditions (air conditioning in the room set at 22°C or 26°C –referred to as AC22 and AC26 respectively). In practice, the temperature was recorded at the start of each experiments. For the AC22 condition, the average recorded temperatures were 22.6 ±0.5°C. For the AC26 condition, the average recorded temperatures were 27.0 ±0.7°C.

For each repeat, the random number generator (used both to generate random experimental parameters and in the random goal exploration algorithm) was seeded, enabling the algorithm generating experiments to be restarted at any time in a reproducible way. We used the following random seeds: [110, 111, 112] for each AC26 experiments and [210, 211, 212] for each AC22 experiments. You will see these seeds referred to in the legend or title of the following plots.

In the following, we first compare the differences in terms of experimental observations and explored space between the two algorithms under the AC26 condition. We then compare the distribution of the experiments performed, both in the formulation space and in terms of the physical properties of the droplets. For this section, we focus on the experimental run with seed 111, all the plots for all the runs/seeds are available in the *detailed_plot.zip* file available at:

https://github.com/croningp/dropfactory_analysis/releases/download/SI/detailed_plots.zip.

All runs have the same trends as the specific repeat presented in this section and statistical analysis between repeats are provided in the main manuscript.

Before starting any analysis, we need to validate that the experiments were run at similar temperatures. Temperature was recorded just before droplets were generated thanks to sensor mounted on the robotic platform (more details available in section 2.1.2.16 Environment Monitoring).

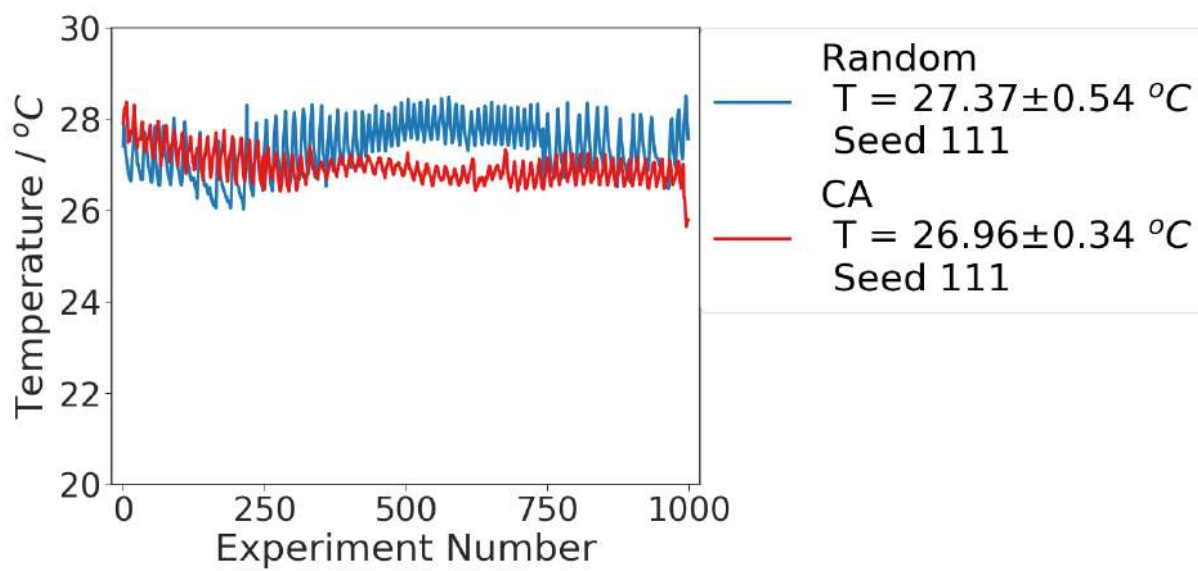

Supplementary Figure 2 – The experimental temperatures recorded for the random and CA algorithms for the run with seed number 111 under the AC26 condition.

As explained before, the whole room was maintained at the desired temperature using air-conditioning, such that all the liquids were kept at the experimental temperature. We can see in Supplementary Figure 2 the cycles of cooling and heating of the air conditioning system during the runs, which take several days. We remind the reader that around 300 experiments were run each day, usually starting around 9am and finishing automatically at 10pm. The temperature between these random experiments and CA runs are similar, with a respective temperature average of 27.37 ± 0.54 °C and 26.96 ± 0.35°C, thus a comparison of the exploration profile is meaningful.

We start by observing the range of droplet behaviours each algorithmic method was able to generate within the 1000 experiments budget, as shown in Supplementary Figure 2.

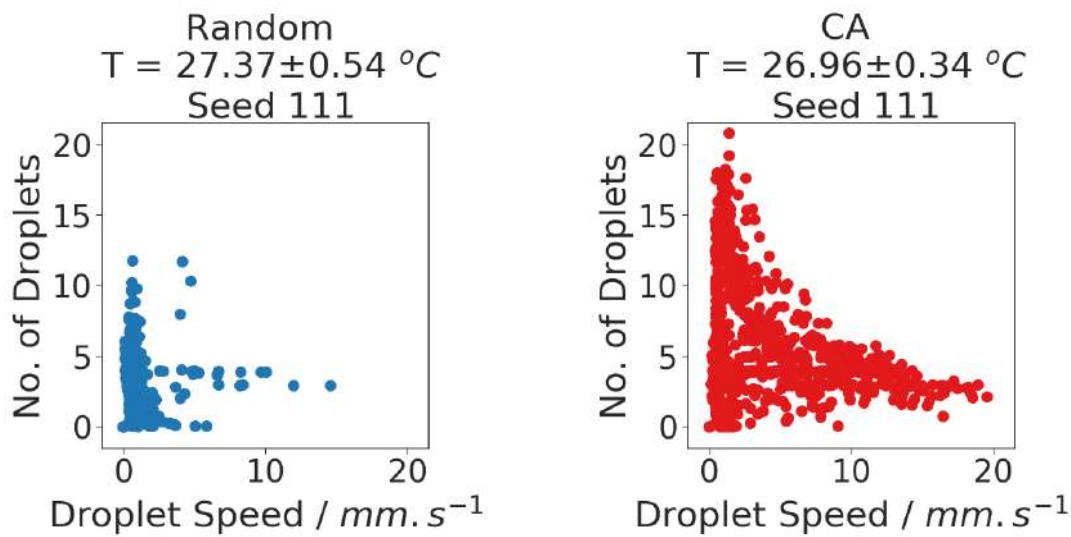

Supplementary Figure 3 - Comparison of the observed droplet behaviours after 1000 individual experiments for random (left) and CA (right). Each scatter dot represents the average speed and number of droplets for a single 90 seconds droplet experiment, with 1000 points plotted on each subplot.

Visually, the CA algorithm was able to generate much more varied observations. Note that the density of points in the random experiment (left) is very high on the bottom-left corner (particularly at 4 droplets and low speed, or 0 droplet and 0 speed) as many experiments produce the same outcome (droplets sinking at the bottom of the dish or dissolving immediatly). This can be better observed on density plots, as shown below on Supplementary Figure 3 and Supplementary Figure 4. The density plots were generated using a Gaussian kernel density estimation (see python library scipy.stats.gaussian_kde) with the bandwidth parameter set to 0.3.

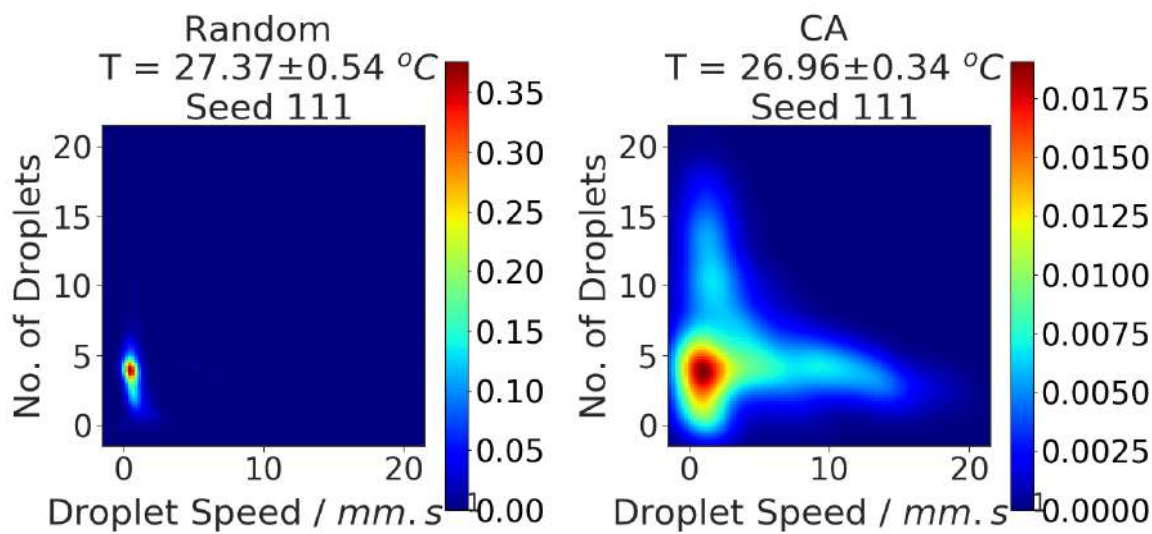

Supplementary Figure 4 – Comparison of the observed droplet behaviours as a density map after 1000 individual experiments for random (left) and CA (right).

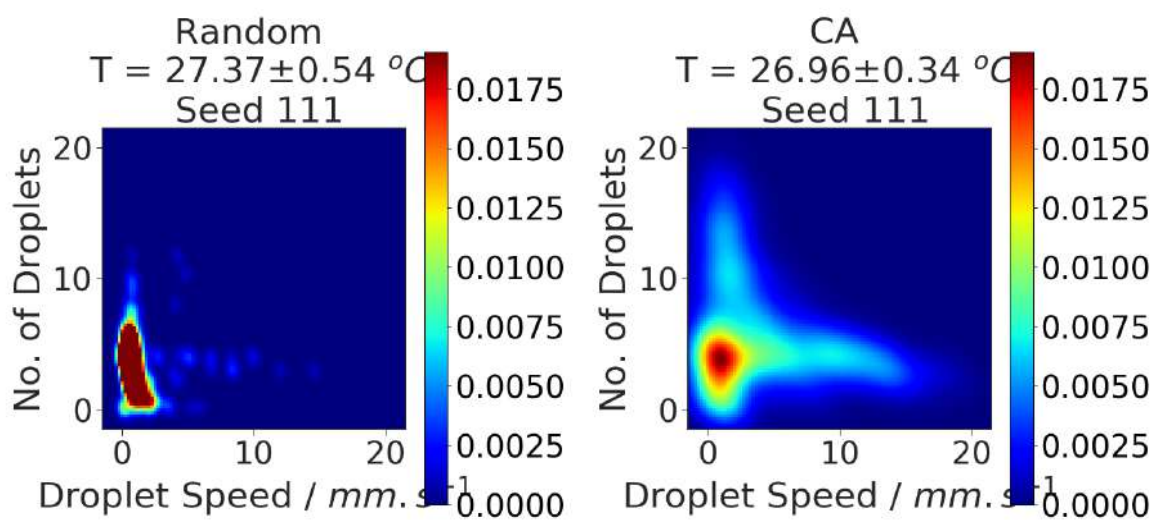

Supplementary Figure 5 - Comparison of the observed droplet behaviours as a density map after 1000 individual experiments for random (left) and CA (right), with equally density-colour scales.

Notice the difference of scale of the colours between the top and the bottom figures. A large majority of the experiments performed following the random experiment method have very low speed and do not divide (see left plot on Supplementary Figure 3, very intense and localised high-density area). The observations coming from the CA algorithm are much more spread. The same distribution analysis can be done on each individual dimension of exploration, respectively speed and division, as shown in Supplementary Figure 5.

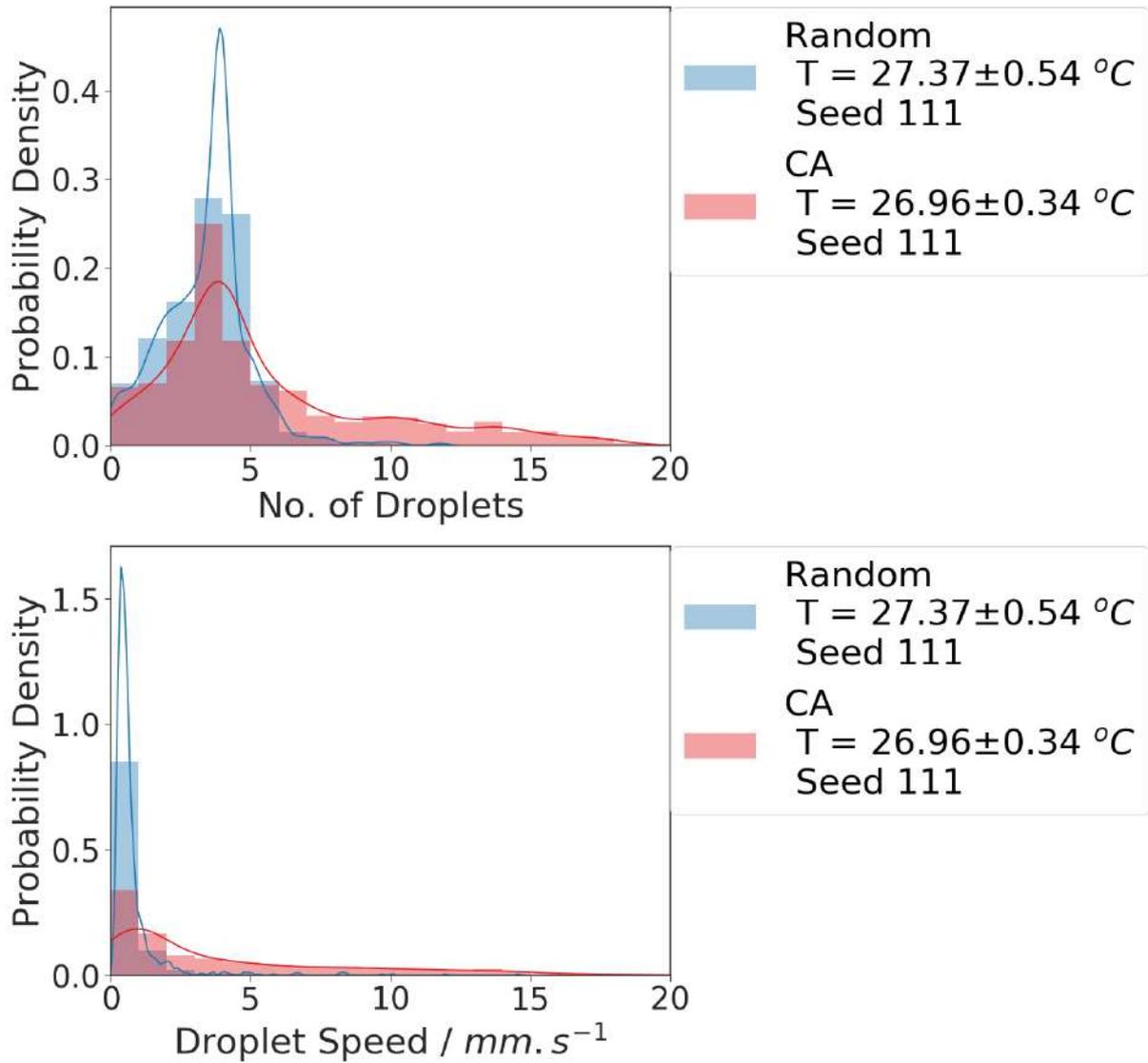

Supplementary Figure 6 – A comparison of the spread of the average number of droplets (upper) and droplet speed (lower) for the random and CA algorithms.

In both cases, we notice that the tail of the observation distribution is much stronger for CA, extending past 15 droplets in division and way past 5 mm.s$^{-1}$ in speed. This difference between the patterns of exploration is confirmed by our exploration metrics as shown below.

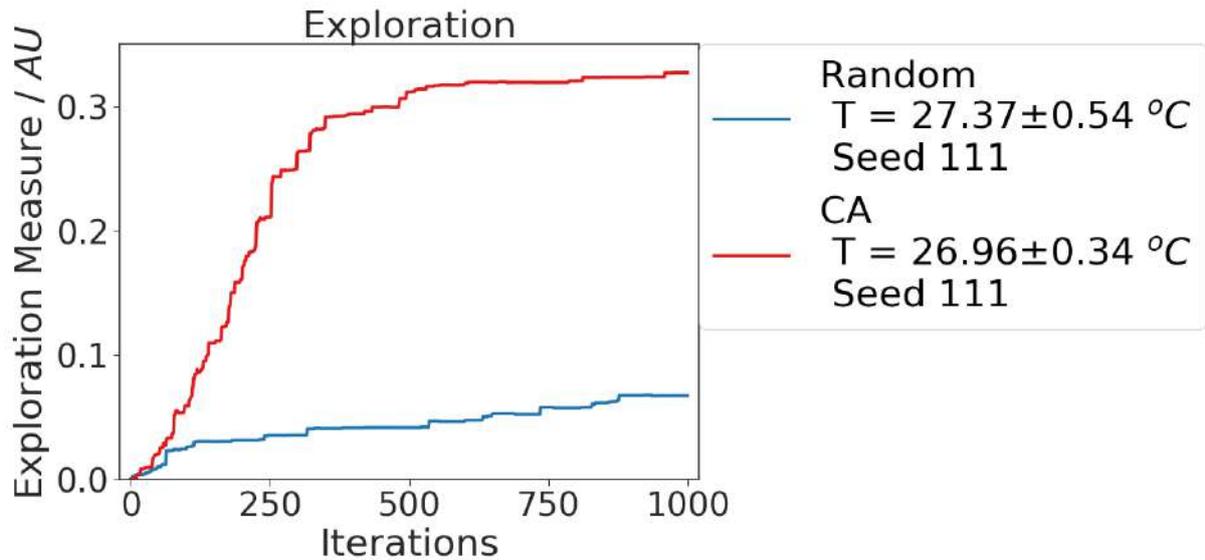

Supplementary Figure 7 – The evolution of the exploration measure throughout experiments for the random and CA algorithms.

After only 150 experiments the CA method has already explored more in terms of droplet behaviours than the random method did in 1000 experiments. We can look in the patterns of exploration into more details by plotting the results of the experiments at various iteration of the algorithms. Supplementary Figure 7 compares the exploration of random and CA algorithms through iterations, every 100 steps. CA already explored most of the reachable space after about 300 iterations. CA exploration dynamic also shows directional preferences, where at t=100 mostly high speed has been discovered and high division starting to be explored at t=200. Random exploration, however, generates mostly droplet that do not move and do not divide or dilutes entirely, only rare events lead to more dynamic droplets.

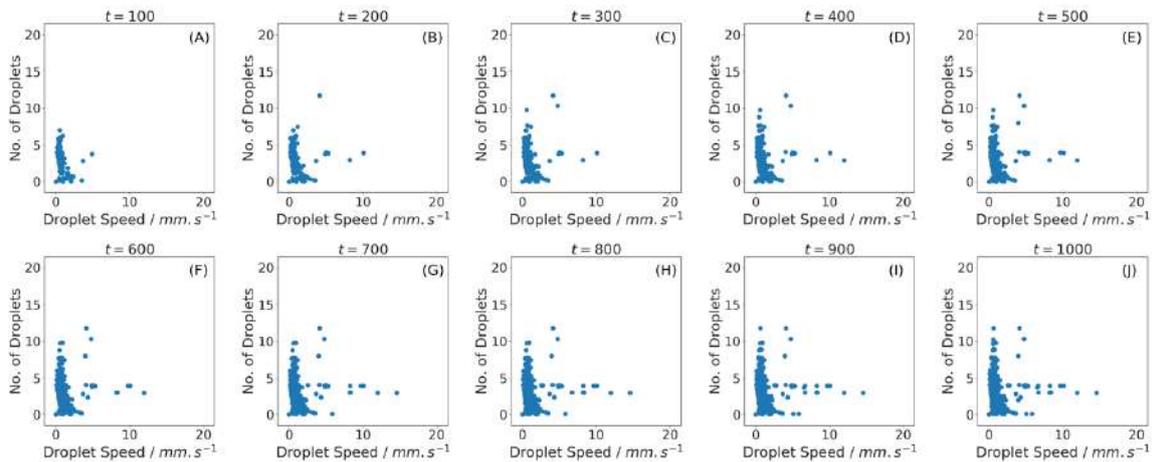

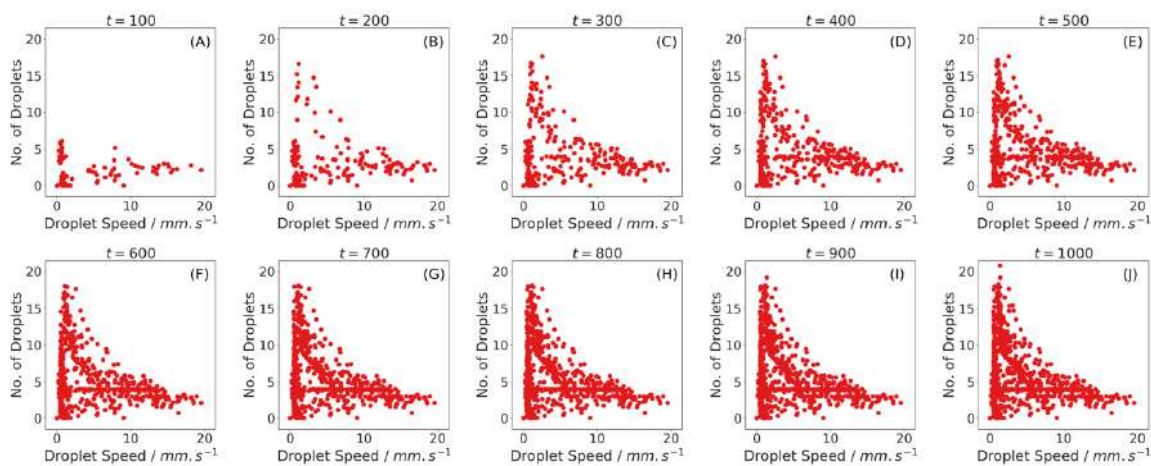

Supplementary Figure 8 – Comparison of the droplet behaviour observed at various iteration of the algorithm exploration procedure, here every 100 steps. Top is for random exploration and bottom is for CA. The CA method very quickly generates more observation that random, at t=300 the difference CA already explored most of the reachable space.

The difference between the random and CA algorithms is that the former focuses on the parameters space while the latter focus explicitly on the observation space, in our case, droplet speed and division. However, in their respective space of interest, they generate targets at random. For example, in the space of oil formulation, the distribution of experiments performed is uniform.

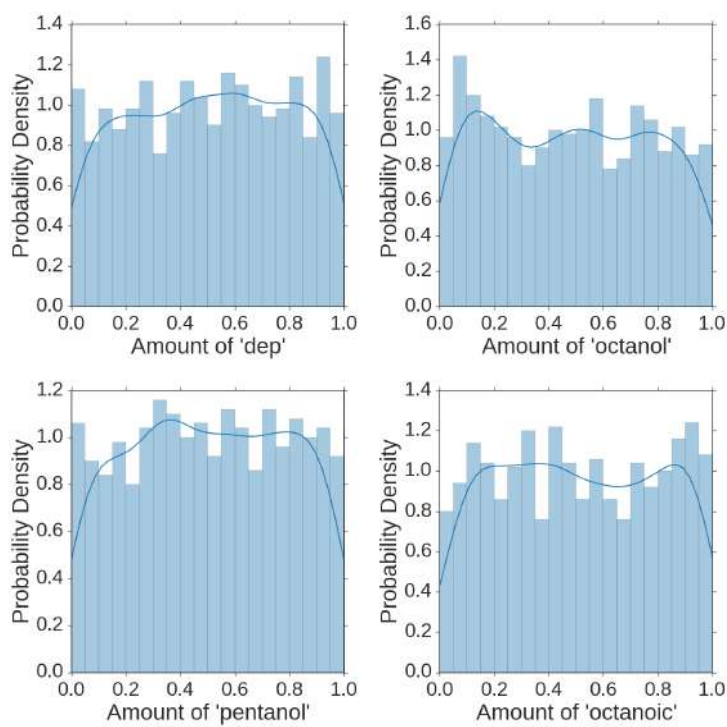

Supplementary Figure 9 – The distribution of experiments performed by the random algorithm in the recipe space – notice the uniform spread across all oils.

As seen above in Supplementary Figure 8, the distribution of the quantity of oils selected to be in each oil mixture by the random method is uniform. Similarly, as seen in Supplementary Figure 9, the CA algorithm selects experimental targets randomly in the goal space.

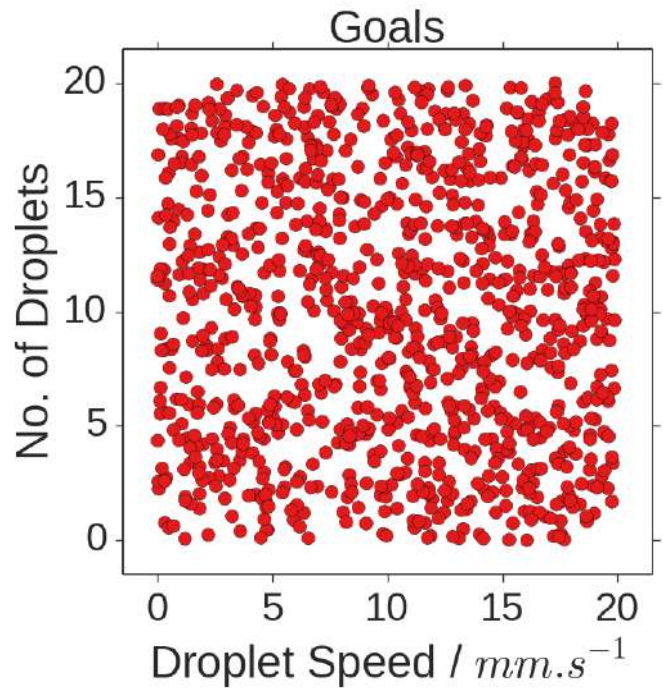

Supplementary Figure 10 - The distribution of experimental targets generated by the CA algorithm in the goals space – notice the uniform spread across the whole goal space.

The distribution of goals (the targeted observation generated in terms of droplet speed and average number of droplets) is distributed uniformly in the output space. The algorithm then, given the past observations it has made of the system, selects the best experimental parameters (oil quantities) to try and achieve these targets. The powerful idea of the CA is that it drives the exploration towards areas that are hard to explore and discover by chance. As a result, the distribution of experiments actually performed in the oil space becomes much narrower, yet leading to a wider variety of droplet behaviours.

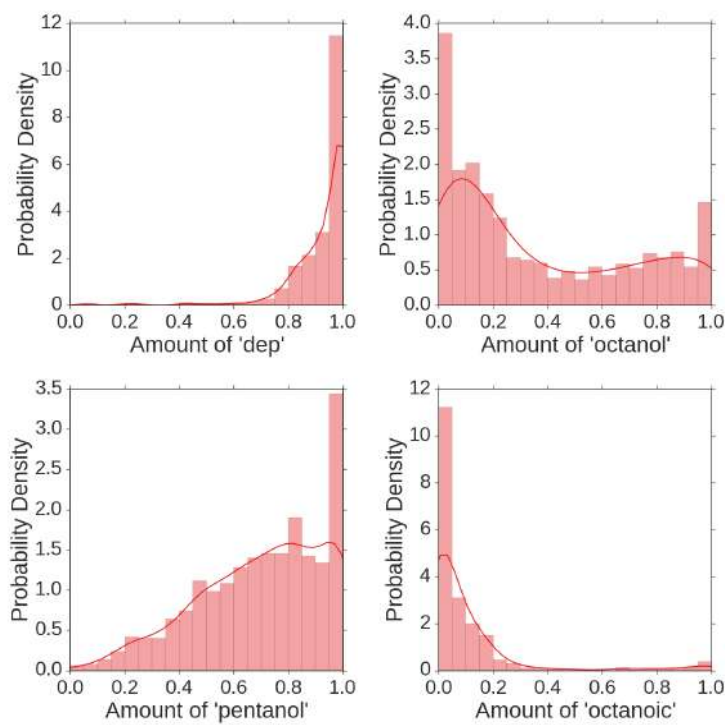

Supplementary Figure 11 - The distribution of experiments performed by the CA algorithm in the recipe space – notice that specific parts of the formulation space are explored in more detail.

The distribution of experiments performed by the CA algorithm is indeed not uniform as shown on Supplementary Figure 10. CA focuses its experimentations on a smaller subspace of the experimental oil formulation space. High levels of DEP and low levels of octanoic acid are comprehensively explored, for example, whilst the reverse case is barely explored at all. This effect is much more visible when we compare the ratio of each oil in the final formulation and between the two methods, as seen on Supplementary Figure 11.

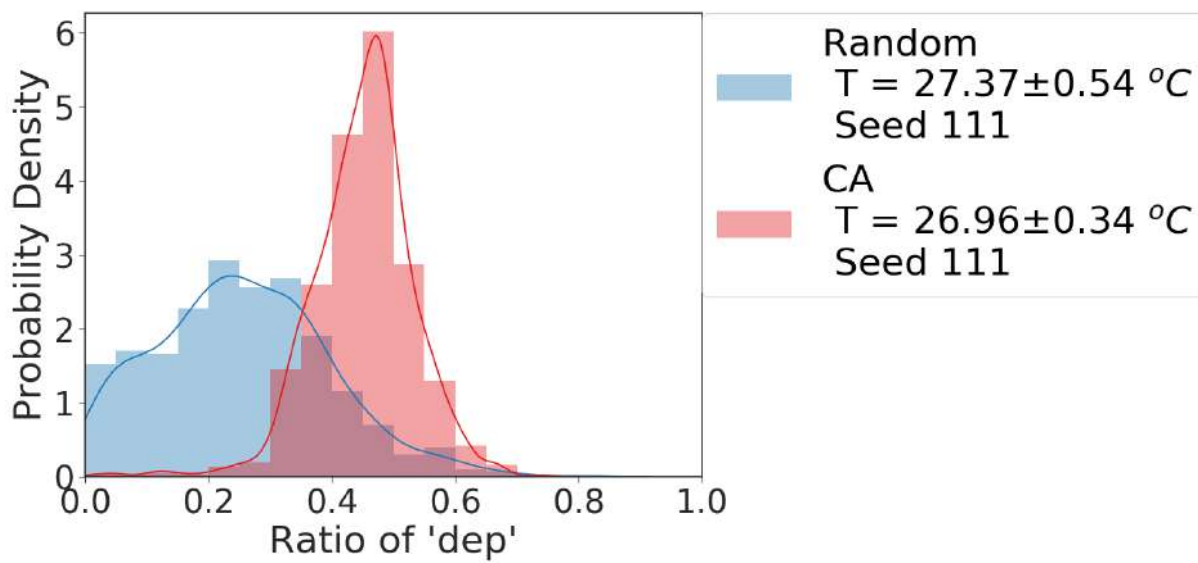

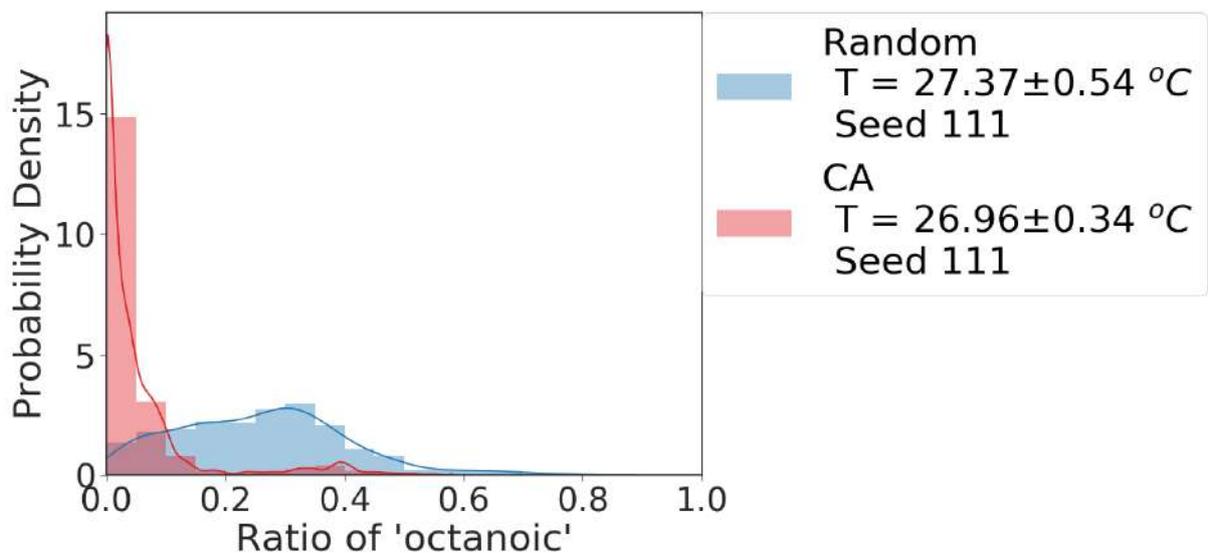

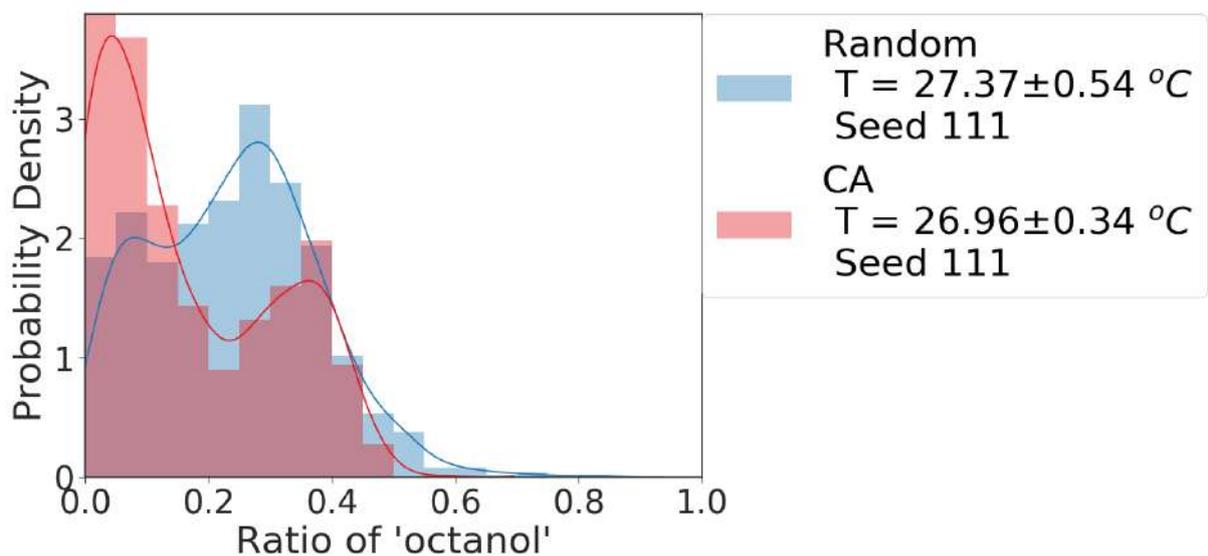

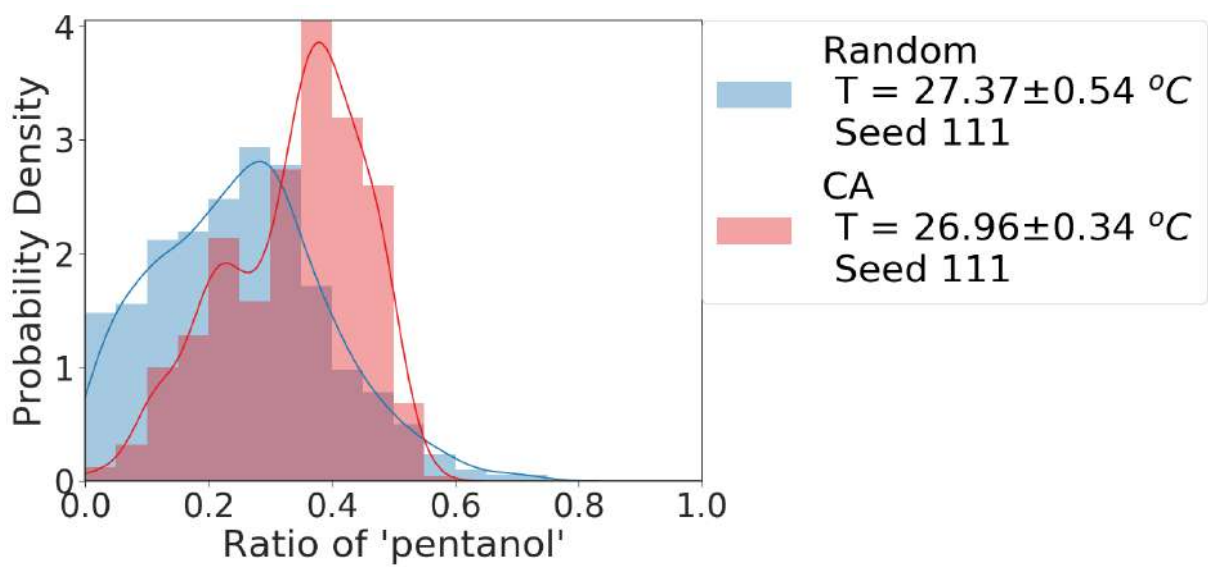

Supplementary Figure 12 – The distribution of the ratios of each oil explored by each algorithm during a 1000-experiment run. Notice how the distribution are different between the two algorithmic approach, the CA focused on a narrower part of the formulation space.

The above plots highlight the fact that a simple change of focus from the experimental (input) space to the observation (output) space can impact significantly the patterns and distribution of experiments performed. Of high interest for this research, the CA algorithm can observe much more varied droplet behaviours despite performing experiments on a smaller subset of the parameter space. This illustrates that random screening and high-throughput methodologies that purely focus on the experimental space are less efficient search methods for unknown systems. This issue is referred to as the redundancy problem in artificial intelligence, machine learning and robotics, specifically in sensory-motor exploration problems.

To go further, we can observe on Supplementary Figure 12 how the different distribution of experiments from each method translates in terms of droplet's physical properties, such as dynamic viscosity, density and surface tension. This in turn can inform us about what properties lead to different droplet behaviours.

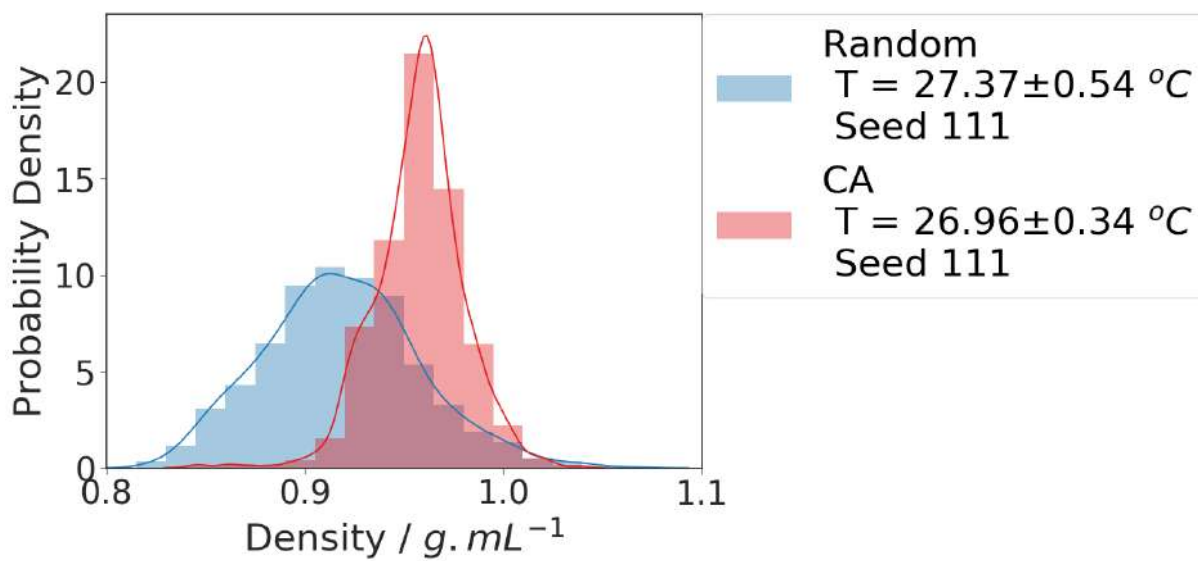

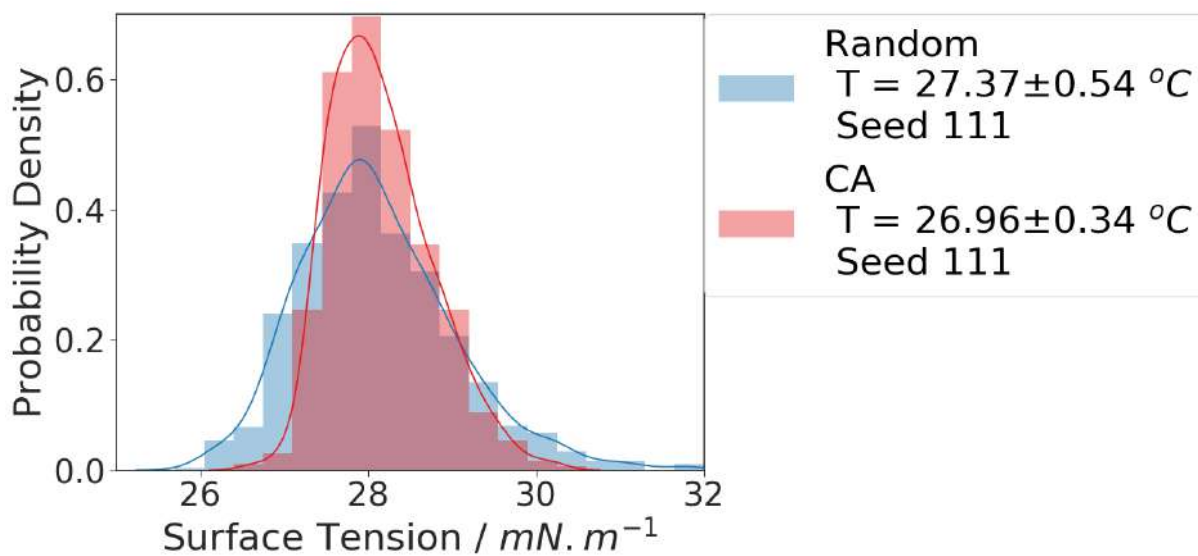

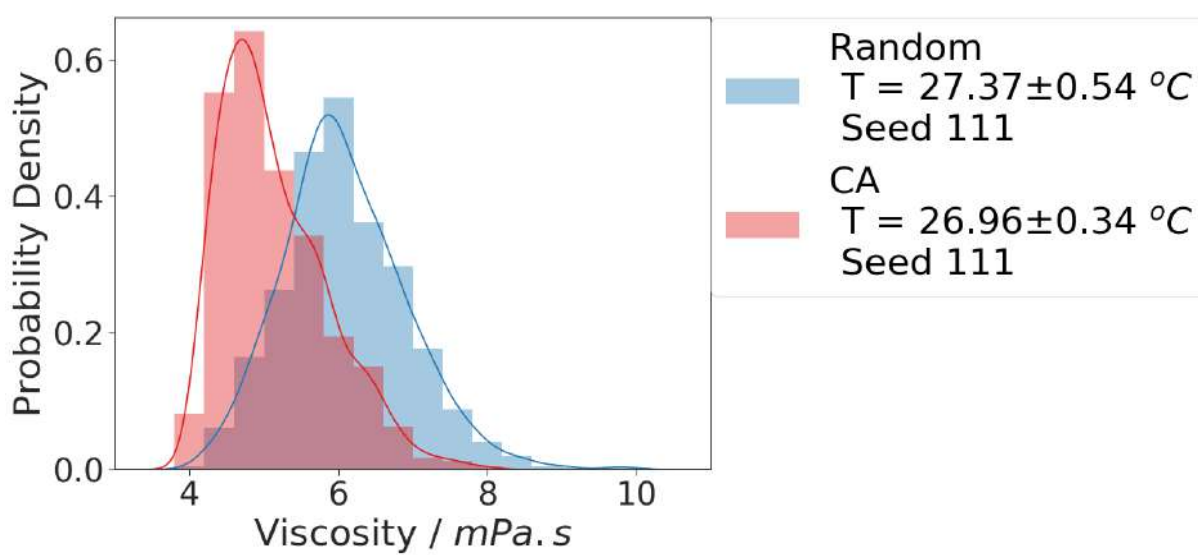

Supplementary Figure 13 – The distribution of experiments performed in terms of the resulting formulation densities, surface tensions and dynamic viscosities, for both the random and CA algorithms.

Interestingly, CA generated droplets of higher density and lower viscosity, indicating that these are the areas with the richest array of droplet behaviours. Viscosity, density and surface tension were computed following the method described in [38].

Finally, we compare other droplet metrics. For example, as seen in Supplementary Figure 13, CA generates droplets that travel around the petri dish much more. The most extreme of these values are likely to be where high division and speed are occurring, allowing impressively large path lengths over the 90s experiment.

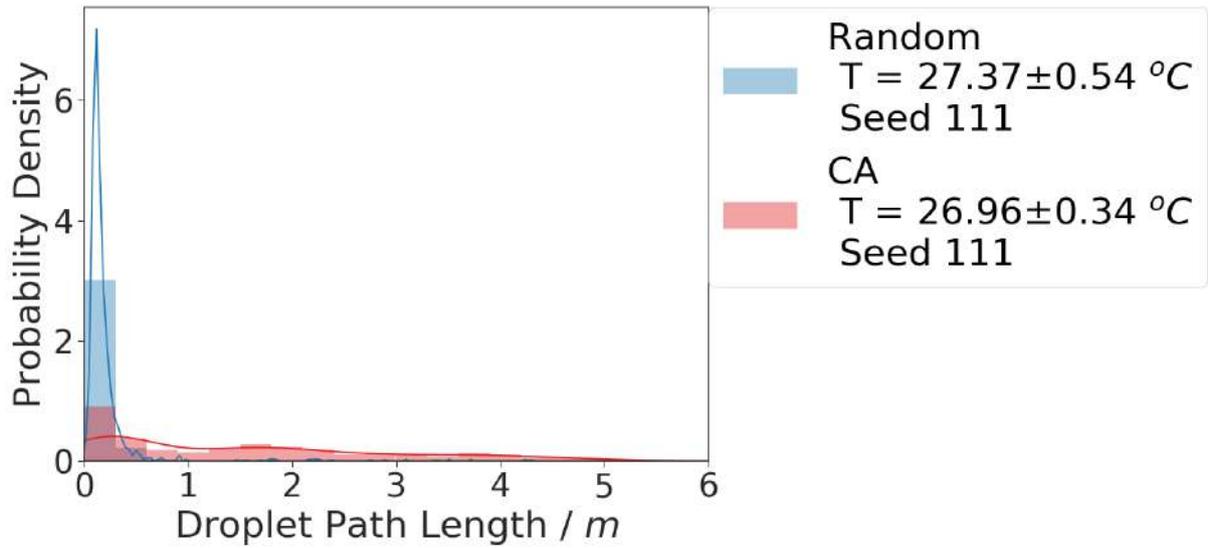

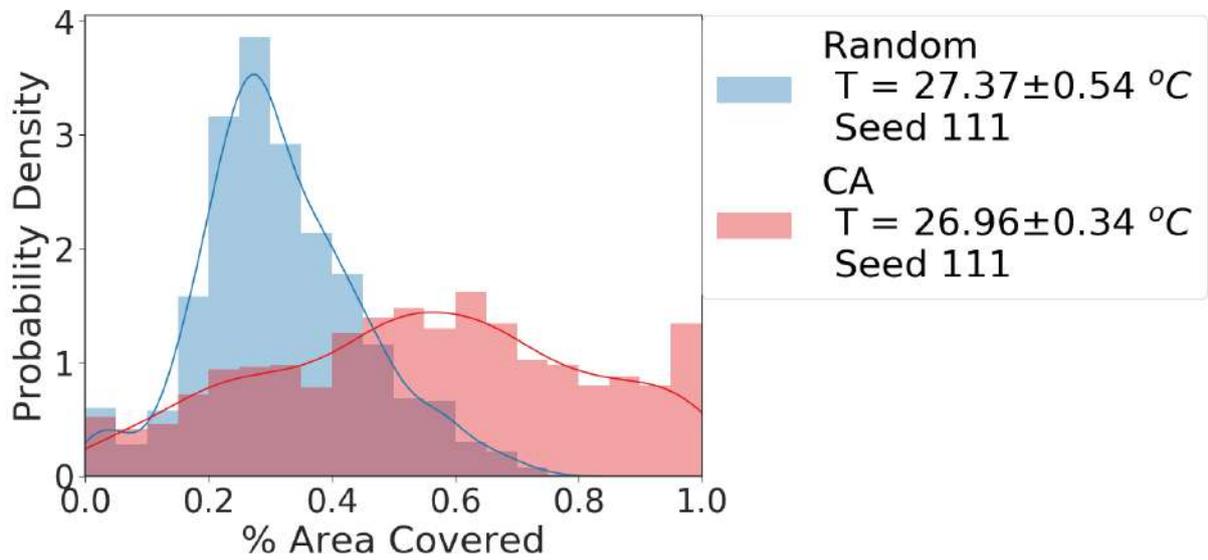

Supplementary Figure 14 – Distribution of the total droplet path length and fraction of the dish area covered for the random and CA algorithms. Path length and fraction of the dish area covered properties are described in section 2.1.5.4 titled Droplet Metrics Measured.

CA also produces droplets that tends to be slightly smaller as per their perceived diameter from the top (camera) view, see Supplementary Figure 14.

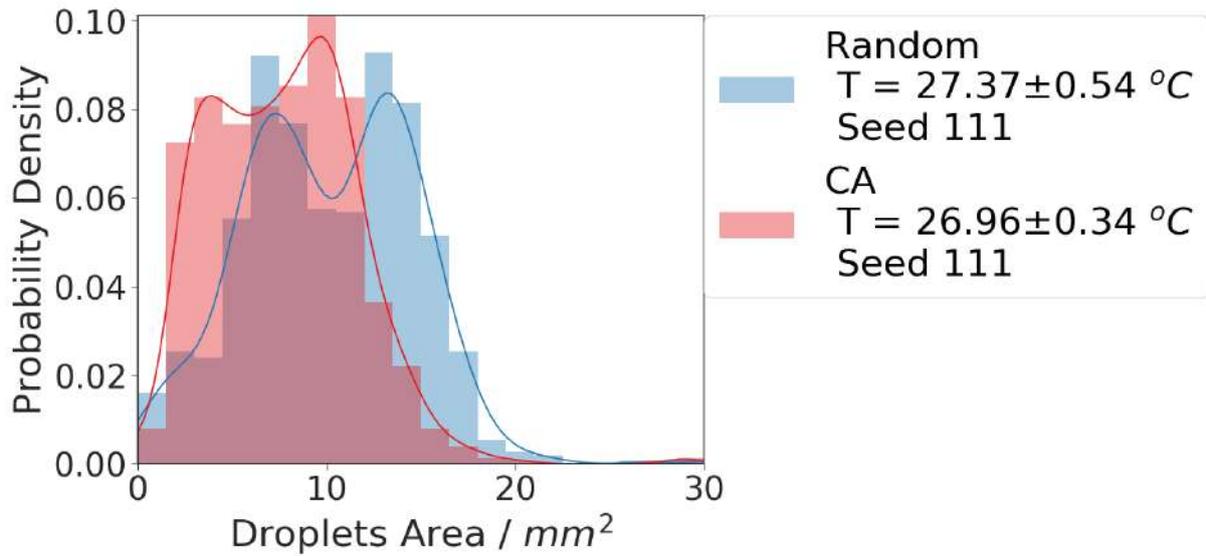

Supplementary Figure 15 - Distribution of the droplet visible area for experiments generated using the random and CA algorithms. The droplet visible area property is described in section 2.1.5.4 titled Droplet Metrics Measured.

## 1.4 Temperature Comparison

The same comparisons as made in section 1.2 "Algorithm Comparison" can be made for experiments using the same algorithm but at different temperatures. As shown in the main manuscript, the CA algorithm, by providing us with a more accurate picture of our droplet system capabilities, enabled the discovery of a temperature effect. This was discovered because the envelope of the observed droplet behaviours was consistently and reproducibly changing as the room temperature changed, something that we were not able to observable when performing random parameters experiments. The run with seed 111 for AC26 condition and the run with seed 212 for AC22 condition are used for the following analysis.

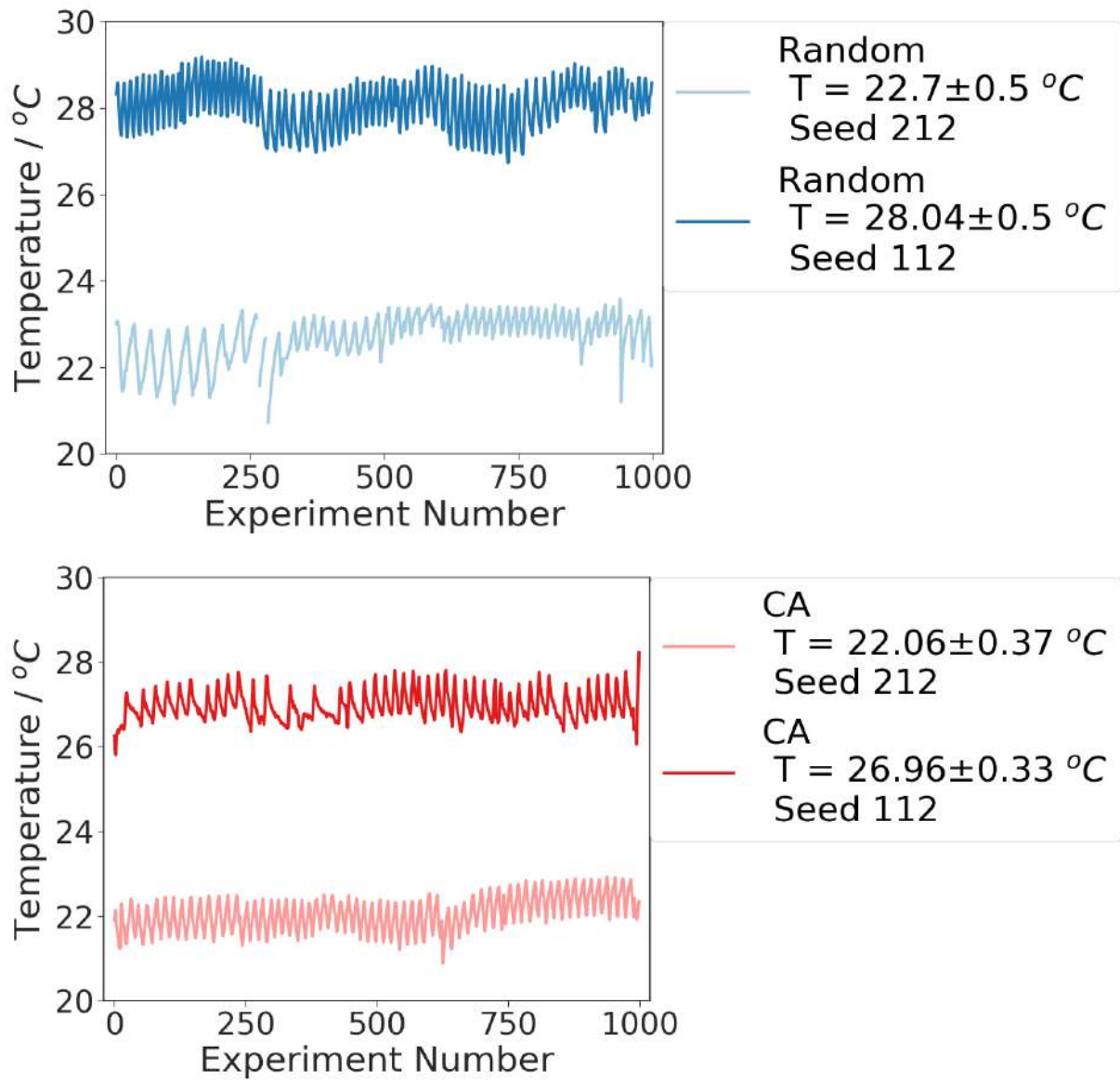

Supplementary Figure 16 – The recorded temperatures for the random and CA algorithms at AC22 and AC26.

Supplementary Figure 15 confirms that the experiments were indeed performed at significantly different temperatures, with a gap of ca. 5°C between the 'cold' (AC22) and the 'hot' (AC26) conditions. Supplementary Figure 16 compares the observations of droplet behaviours between methods and conditions.

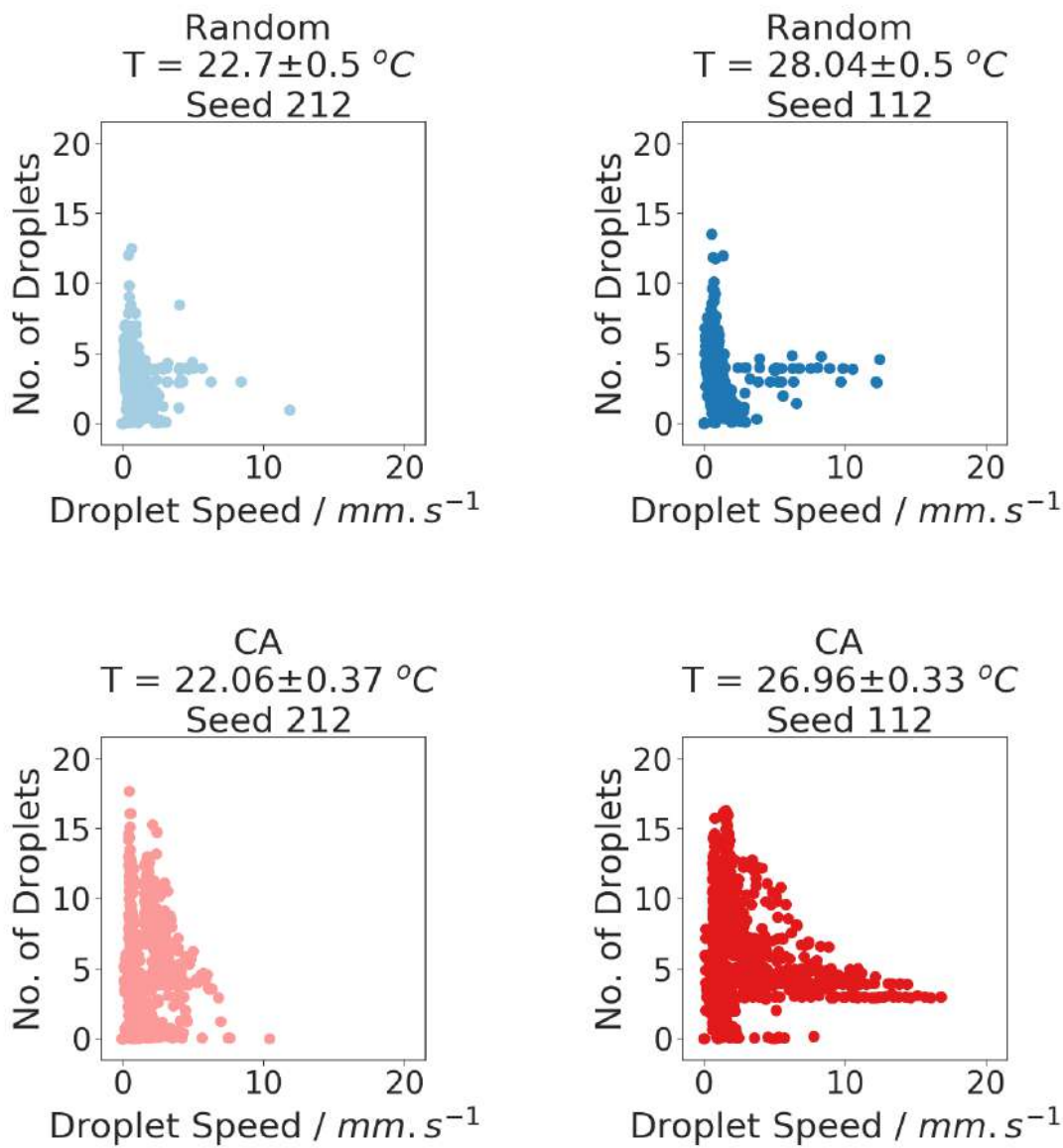

Supplementary Figure 17 - Comparison of the observed droplet behaviours after 1000 individual experiments for random (top) and CA (bottom) at low (left) and high (right) temperature. Each scatter dot represents the average speed and number of droplets for a single 90 seconds droplet experiment.

There are three information to take home from Supplementary Figure 16. First, the CA algorithm consistently generates more varied observations than random experiments, under both AC22 and AC26 conditions. Second, there is no visually significant difference between the observations made under the 'cold' and 'hot' conditions with random parameter search. Third, there is a significant visual difference of observations between cold' and 'hot' conditions with the CA algorithm, with a large number of experiments producing droplets with speed above 5 mm.s$^{-1}$, as well as experiments that divide in more than 10 droplets. These differences of exploration are also visible on density plots as shown in Supplementary Figure 17.

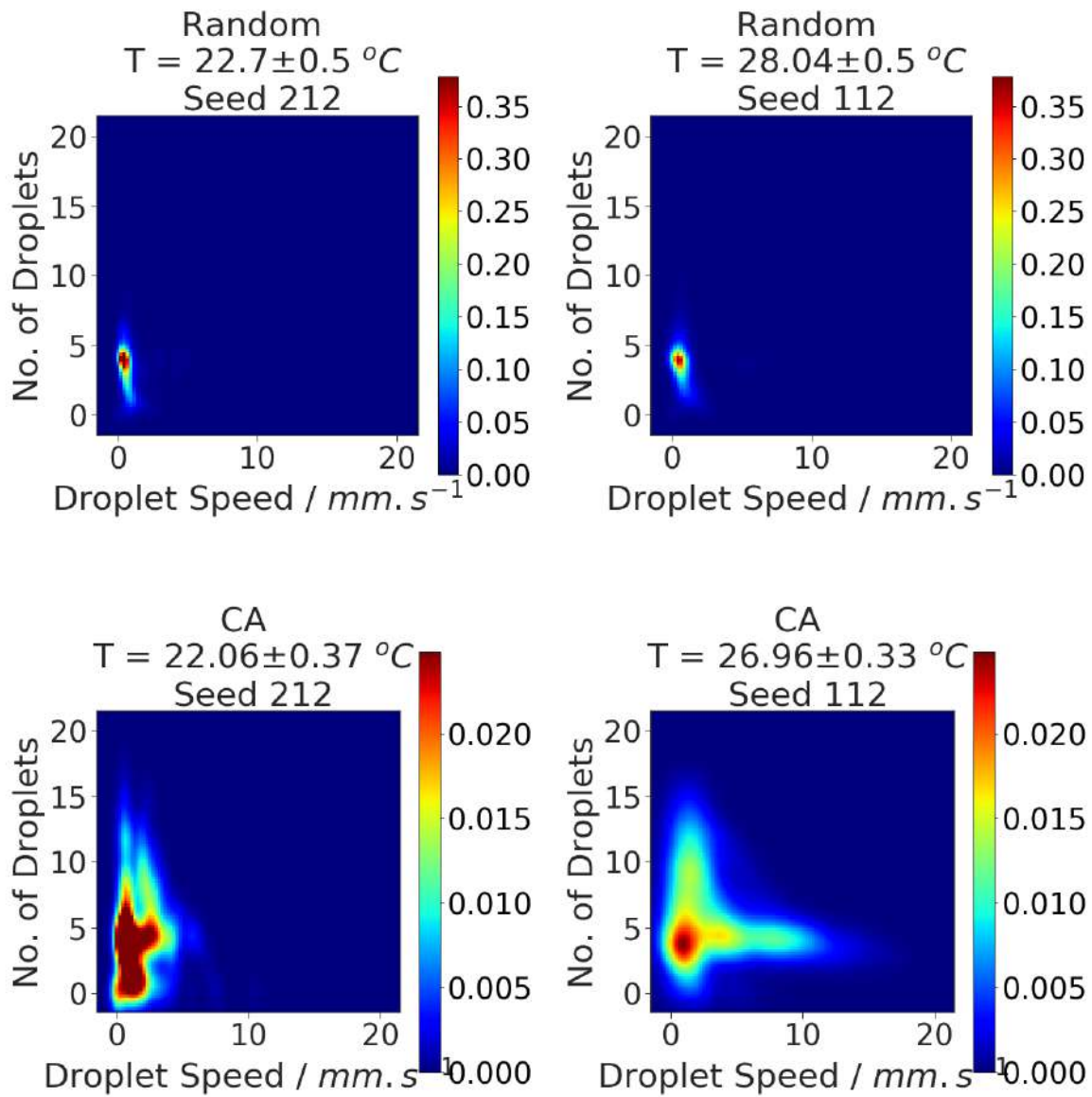

Supplementary Figure 18 - Comparison of the observed droplet behaviours after 1000 individual experiments for random (top) and CA (bottom) at low (left) and high (right) temperatures.

The density of observations is similar for random parameter search at both temperatures, but expands more in the speed dimension at higher temperature for the CA method.

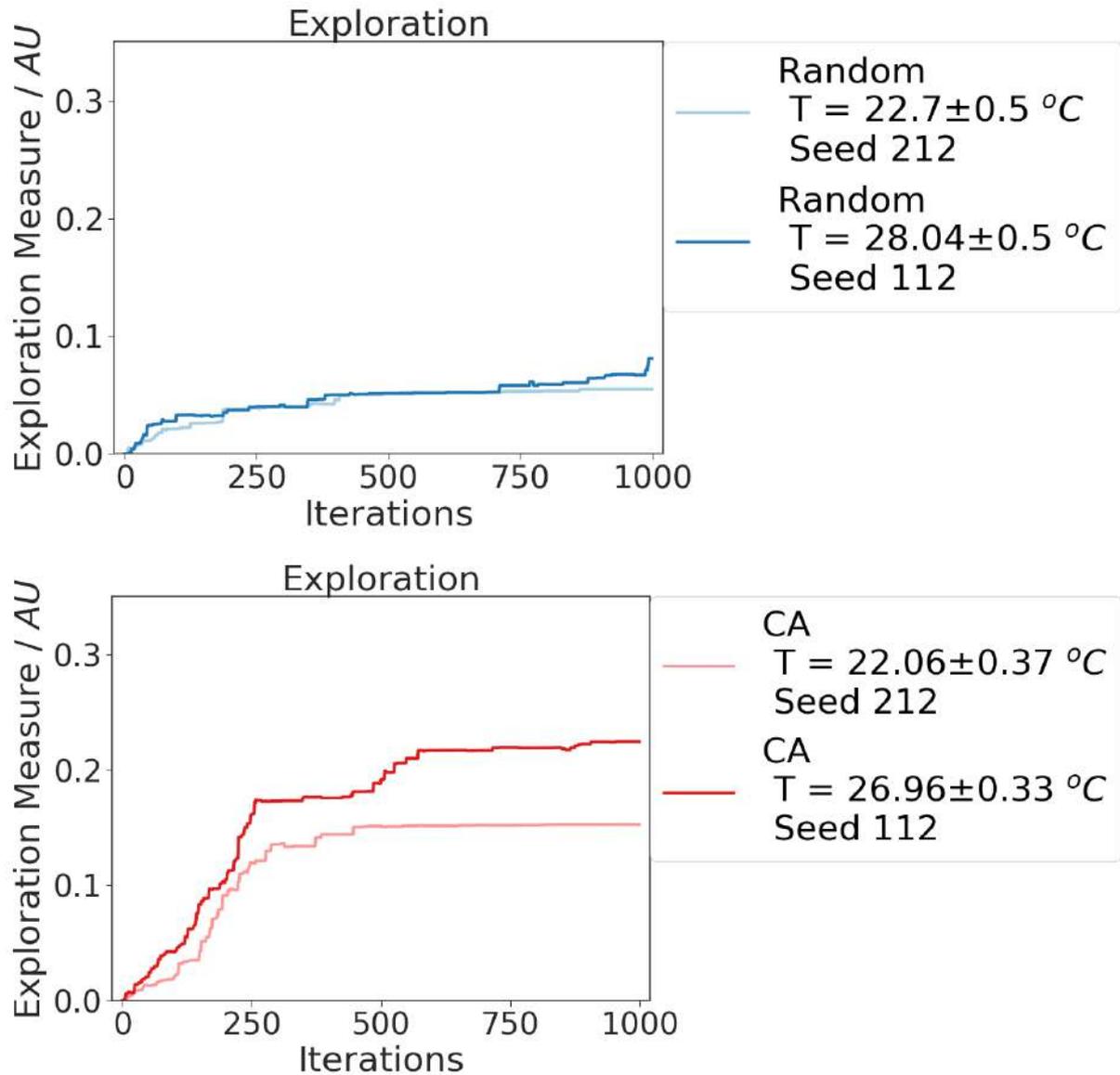

Supplementary Figure 19 - The evolution of the exploration measure throughout experiments for the random (top) and CA (bottom) algorithms, at low and high temperature.

The exploration metrics confirm this effect as shown on Supplementary Figure 18. There is little difference of exploration at both temperatures for random experiments, while with CA the higher temperatures allow for the exploration of more varied droplet behaviours. The effect is especially visible on the distribution of speeds observed as shown on Supplementary Figure 19. For random, both distribution look alike, while for the CA a distribution with a longer tail is observed at higher temperature, extending beyond 5mm.s$^{-1}$, above which very few experiments were observed from the lower temperature.

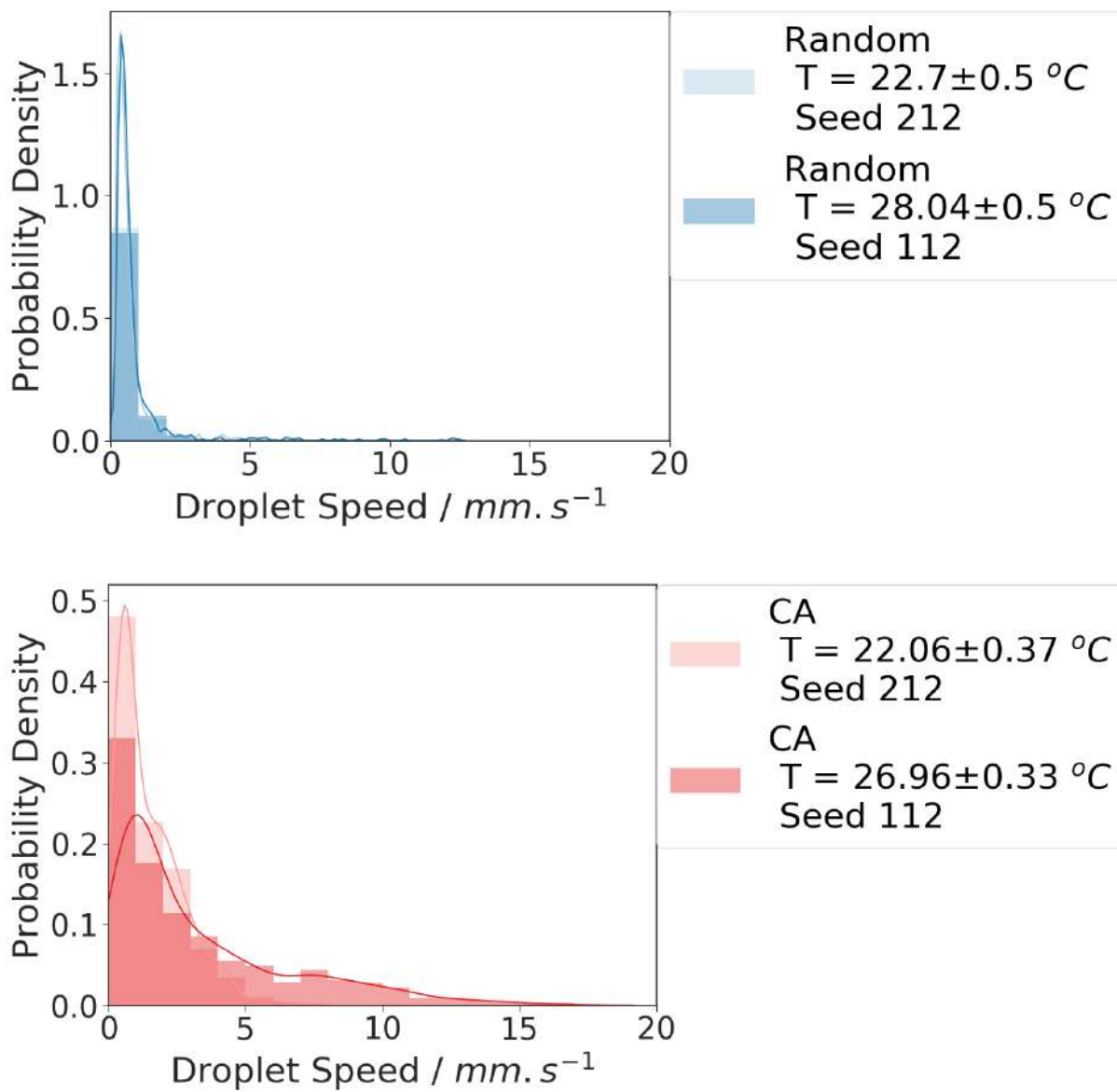

Supplementary Figure 20 - Comparison of the distribution of the average droplet speed for the random (top) and CA (bottom) algorithms at high and low temperature.

This difference of observations due only to a slight change in temperature was only observable with the CA algorithm, which raised our interest and initiated further study on the temperature relationship as described in the main manuscript and detailed further in this document.

Interestingly, although the algorithms were not informed of the change of temperature in the room, the small change in temperature we applied impacted the experiments selected and performed by the CA algorithm (and not from the random). For example, Supplementary Figure 20 shows the distribution of the ratio of 1-pentanol in droplet formulations which illustrates the change in exploration patterns from the CA algorithm induced by the change in temperature.

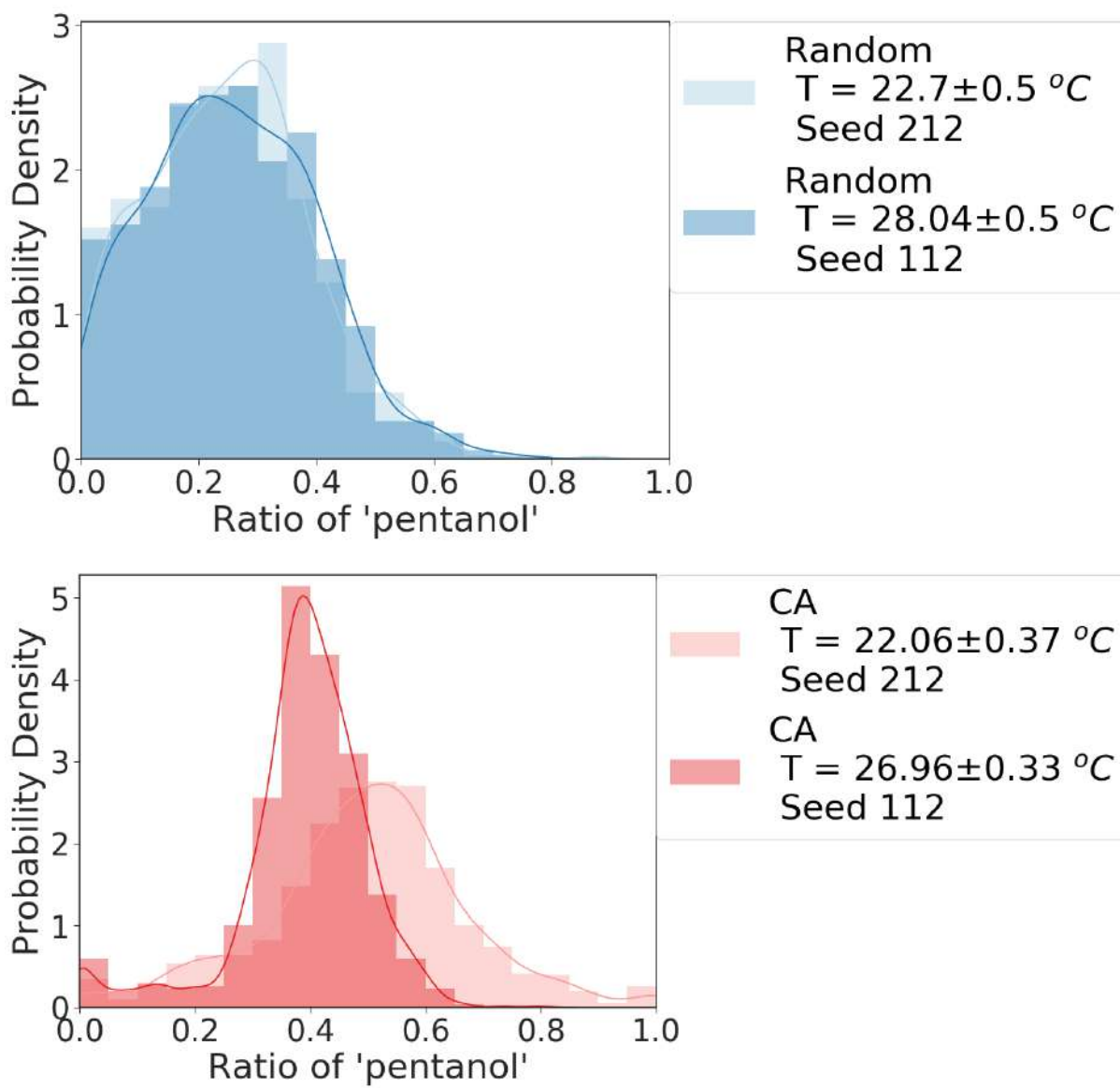

Supplementary Figure 21 – The distribution of the ratios of pentanol in droplet formulation tested during a 1000-experiment run compared between each algorithm (top random, bottom CA) at low and high temperature.

This result indicates that to cover most of the output space (speed / division), the CA had to employ different strategies at different temperatures. In this case, adding less pentanol to the droplet at higher temperature. This is an adaptation that the CA algorithm made without having any direct information about the temperature in the room, but rather learned from the direct closed-loop observation of the system and without any prior information. We know from our chemical analysis that pentanol is one of the key elements driving the motion and division of droplets and its dissolution rate is significantly impacted by temperature. We can hypothesize here that the algorithm increased the amount of pentanol at lower temperatures to compensate for the effect of temperature that reduces the rate of dissolution at lower temperatures. The platform achieved this adaptation without direct knowledge of the chemistry of the system or the current temperature in the room. Only the history of experiments done previously in that particular run were used to define the next experiments (see section 2.2.2 "Curious Algorithm: Random Goal Exploration" for more details).

## 1.5 Going Further

The code used to generate the above plots can be found in the *figures* folder of the associated Github repository: https://github.com/croningp/dropfactory_analysis/tree/master/figures/SI

The interested reader can perform the same kind of analysis for all repeats, methods and conditions. All the plots shown in this section have been generated for all experimental runs and for each method, they can be found at:

https://github.com/croningp/dropfactory_analysis/releases/download/SI/detailed_plots.zip

In the file containing additional plots, the random and CA plots are respectively referred to as *random_params* and *random_goal*, and each run is denoted with the seed number used for the random number generator. We used the following seed: [110, 111, 112] for AC26 experiments and [210, 211, 212] for AC22 experiments, respectively when the air conditioning in the room at 22°C and 26°C.

## 1.6 25 Recipe Temperature Screen

To investigate the effect of temperature on various droplet formulations, 25 recipes were selected for further investigation. These were selected to have a wide range in composition and behaviours, especially speed and division. We ran 140 repeats of each recipe across the range of temperatures accessible in the room (17-30°C), representing 3500 additional droplet experiments. The recipes chosen and a summary of the behaviours observed are shown in Supplementary Figure 21.

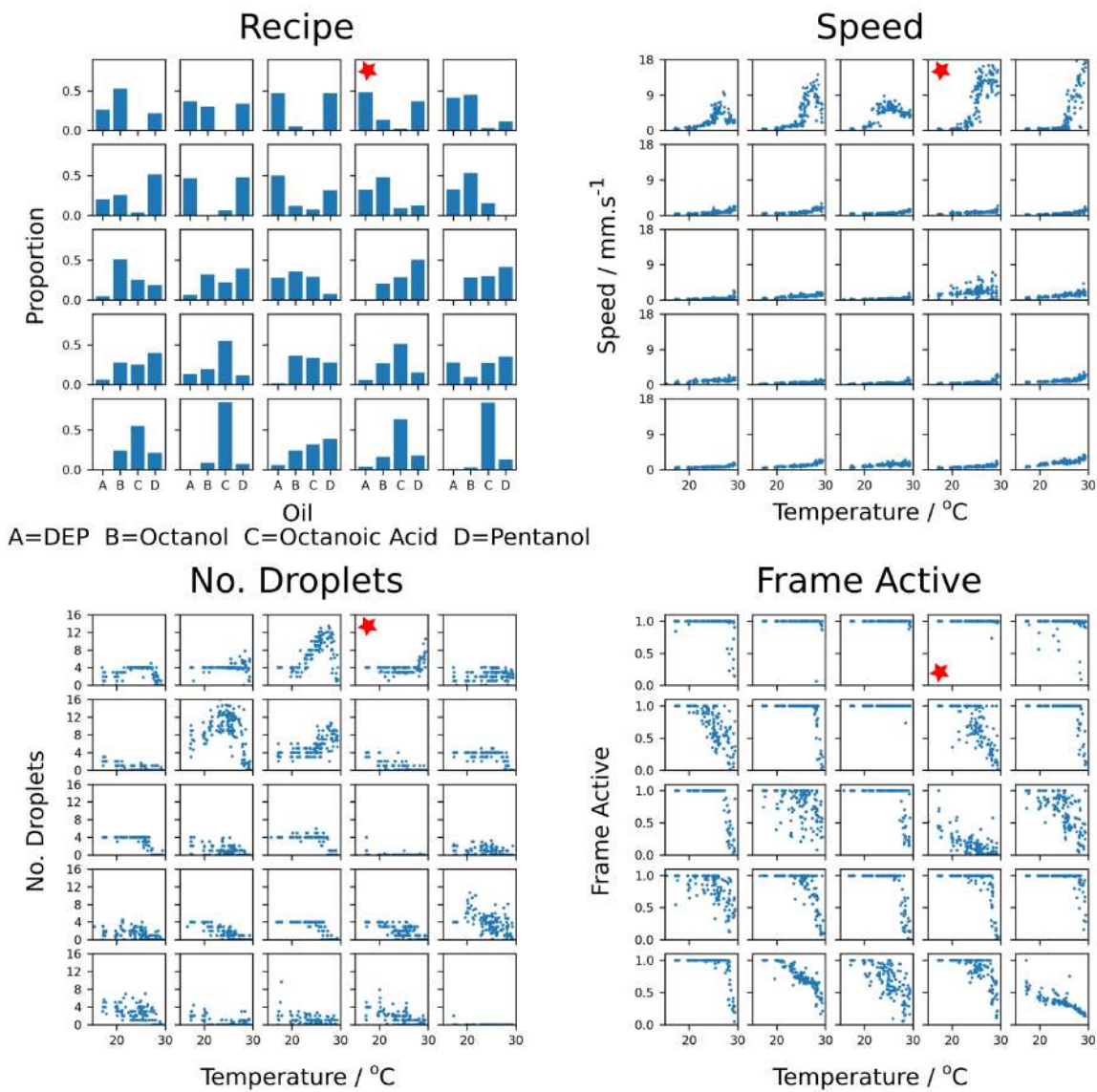

Supplementary Figure 22 – The results of a set of 90-second experiments undertaken at a range of temperatures (17-30°C) for 25 recipes. The red star indicates the recipe taken forward for further studies (recipe 4).

As can be seen from Supplementary Figure 21, there is a large amount of variation in the effect of temperature on droplet behaviour for different recipes. It can be seen that whilst almost all recipes display increasing average speeds with increasing temperatures, this effect is far more pronounced for a small number of recipes, such as numbers 1-5 (top row). Not only are huge increases in speed observed for temperature changes of only a few degrees Celsius for some recipes, these increases are seen to be non-linear and with different onset temperatures. Furthermore, some of them appear to have an optimal temperature for high speed (numbers 1-3), represented by a peak in the speed plot, with higher temperatures impeding the fastest movement. Others still (e.g. 5) seem to still have an upward temperature-speed trend at the maximum sampled temperature of around 30 °C.

There are similar interesting trends in the number of droplets parameter. This quantifies the number of droplets present at the end of the 90-second experiment. Many recipes, especially at lower temperatures, clearly have a trend for having 4 droplets still present at the end of the experiment. It is a general trend for less droplets to be present at the end of the experiment for higher temperatures – along with a reduction in the number of frames with an active droplet – indicating that higher temperatures can often lead to the dissolution or deactivation of droplets. It is interesting to note that recipes 1-2 both have either droplet division or death at the highest temperatures – probably a factor in the decrease in average speed at these temperatures. For recipe 3, however, droplet division and speed do not appear to be incompatible.

There are many interesting comparisons to be made between these recipes, and it is our hope that in future work we can try to investigate these differences further and understand why small compositional changes can lead to very different temperature-behaviour effects. For example, recipes 3+7 are quite similar compositionally, and yet they behave very differently. Whilst they both divide into many droplets at higher temperatures, this only occurs at lower temperatures for recipe 7. More interestingly, recipe 3 displays very high speeds at higher temperature whilst recipe 7 barely moves at any temperature. Recipes 4+8 and 5+9 are also similar in composition but hugely different in the average speeds observed. These examples serve to highlight how, with only small changes to a 4-constituent composition, significant and unpredictable effects on the droplet behaviour can be observed.

For the detailed chemical and physical analysis performed and presented in the main manuscript and in section 1.10, 1.12, and 1.13 of this document, recipe 4 (highlighted with a red star on Supplementary Figure 21) was selected because of its strong speed to temperature sensitivity and a weak impact of temperature on the division of the droplets (the droplets do not divide within the range of temperature accessible for our experiments). Recipe 4 is composed of 1.2% octanoic acid, 46.7% DEP, 5.4% 1-octanol, and 46.7% 1-pentanol.

## 1.7 Droplets as Temperature Sensors

The code to replicate this analysis can be found at: https://github.com/croningp/dropfactory_analysis/tree/master/figures/temperature_prediction

Having observed the significant effect small temperature changes could have on droplet behaviours, we asked if we could deduce the temperature in the room by simply observing our droplets. For each video and recipe, we computed 10 different droplet behavioural measures (for details see section 2.1.5.4 titled "Droplet Metrics Measured") to be used as features to train a machine learning regression model taking as inputs a droplet video and predicting as output the temperature in the room.

We trained a support vector regressor (SVR) for each of the 25 recipes tested and for every possible number and combination of features. A combination of the covered dish area and droplet distance travelled was shown to be a very accurate temperature predictor for recipe 4 (see section 1.5 titled "25 Recipe Temperature Screen").

The prediction accuracy on 140 individual droplet experiments is shown in Supplementary Figure 22 and illustrates that by solely analysing the motion of droplets we can accurately predict the temperature in the room. The residual distribution, modelled as a Gaussian distribution, has a mean of 0.05 (indicating a prediction error of only 0.05°C on average) and a standard deviation of 0.66 (indicating small spread of errors around the mean), demonstrating the high predictability of the temperature in the room from the sole observation of droplet motion. The Pearson correlation coefficient between measured and predicted temperature is $r = 0.98$ ($p < 1e - 95$). The parameters of the SVR are $C = 1000$ and $gamma = 0.1$ that have been estimated using 10-fold cross-validation. The final predictions shown on Supplementary Figure 22 were computed using a leave-one-out methodology such that the predicted experiment was never part of the training set.

The fact that such a complex system can be used as a temperature sensor is both surprising and remarkable, underlining the reproducibility of these observations when the experiments are carried out with sufficient control, as enabled by the robotic platform.

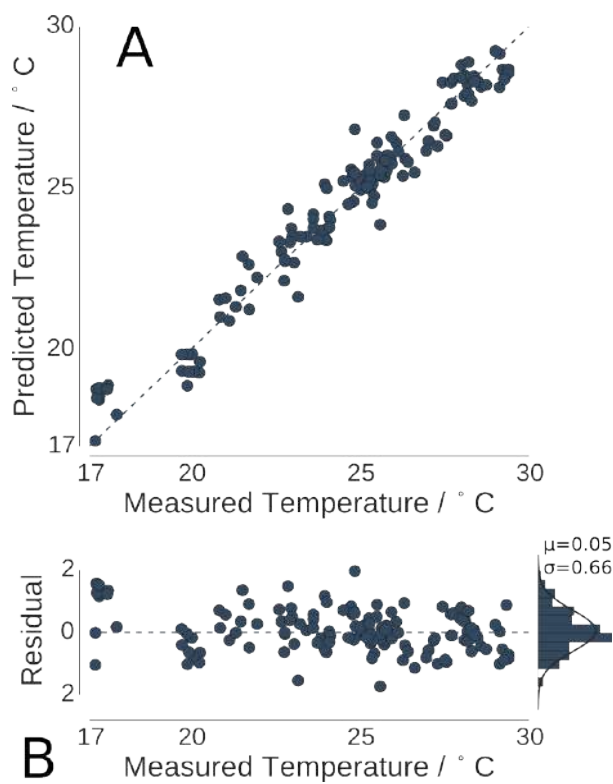

Supplementary Figure 23 - Measured vs predicted temperature of 140 droplet experiments based uniquely from their video. A support vector regressor was trained on 140 droplet experiments using only the information about the covered dish area and the total distance travelled by the droplet droplets. The predicted temperature match and correlate strongly with to the measured values (r=0.98, p<1e-95), with a residual error contained within ±2°C and a mean error of 0.05°C with a standard deviation of 0.66.

## 1.8  Temperature Controlled Dye Release

The code to replicate this analysis can be found at: https://github.com/croningp/dropfactory_analysis/tree/master/figures/dye_release

To investigate the potential for such a system to be used for temperature-controlled chemical release, for example drug release, we wanted to test whether a chemical could be incorporated into the droplet for release.

For ease of analysis, we chose the dye molecule methylene blue – thus allowing quantification of dye release via our droplet tracking software (see section 2.1.5 titled "Droplet Tracking"). As such, recipe 4 (see section 1.5 titled "25 Recipe Temperature Screen") was dyed with both Sudan Black B (at the standard 0.5 gL$^{-1}$, to enable droplet tracking throughout) and methylene blue (1.25 gL$^{-1}$). Methylene blue was chosen as it was known to possess both water solubility and solubility in our oil mixture. Thus, with the room air-conditioning set at either the minimum (17°C) or maximum (28°C) values (see Supplementary Figure 23), 20 repeats of a 5-minute experiment were undertaken, with 4 × 4μL droplets placed in each. Over time, as the methylene blue was released from the droplets into the aqueous phase, the aqueous phase turned blue.

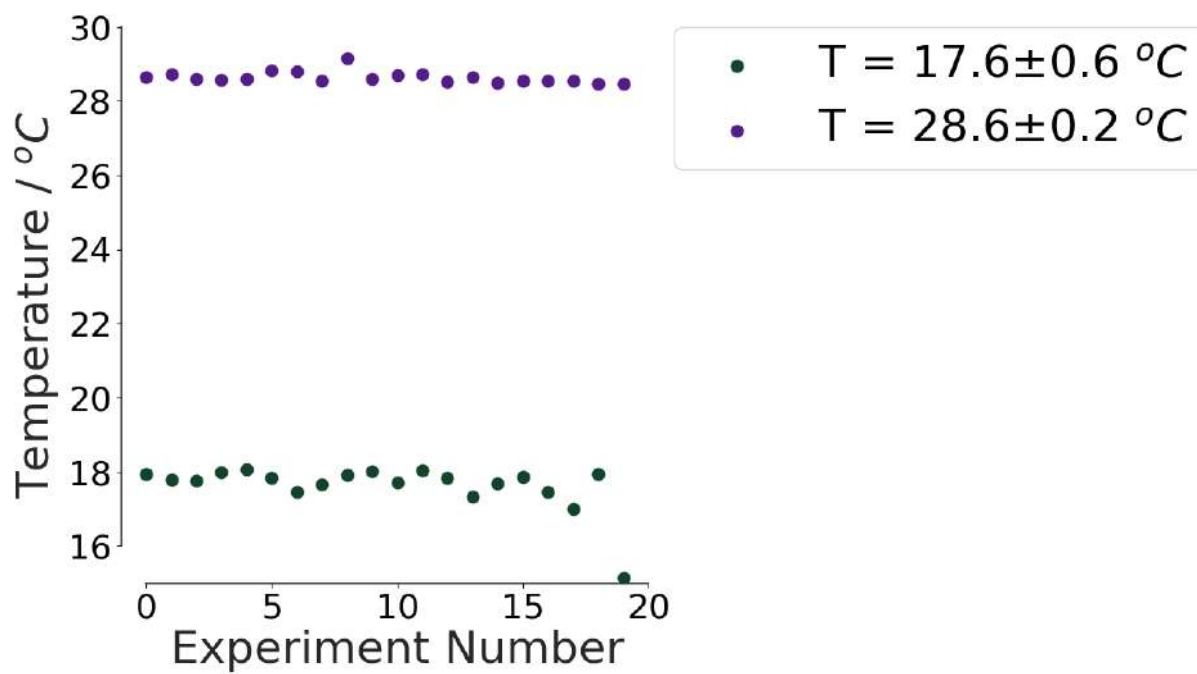

Supplementary Figure 24 - Recorded temperatures for each of the 20 repeats of the dye release experiments.

We used image analysis to quantify the blueness of the aqueous phase through time for each condition. Using OpenCV,[39] for each frame of the video we extracted first the pixels that were both part of the tracking arena and not part of a droplet (see section 2.1.5.1 for tracking arena description and section 2.1.5.2 for droplet detection description). The selected pixel RGB values were transformed in the HSV colour space. We considered a pixel of the aqueous phase as dyed if the intensity of its hue value was above 80 (range is between 0 and 255). This threshold separates the white background from the blue colour of the methylene blue dye and was defined after careful observation of the evolution of the hue channel over time for many videos (see Supplementary Figure 24). Our final measure of dye release is the ratio of dyed pixel to the total number of pixel in the arena that are not part of a droplet. We subsequently fitted the data with a sigmoid function.

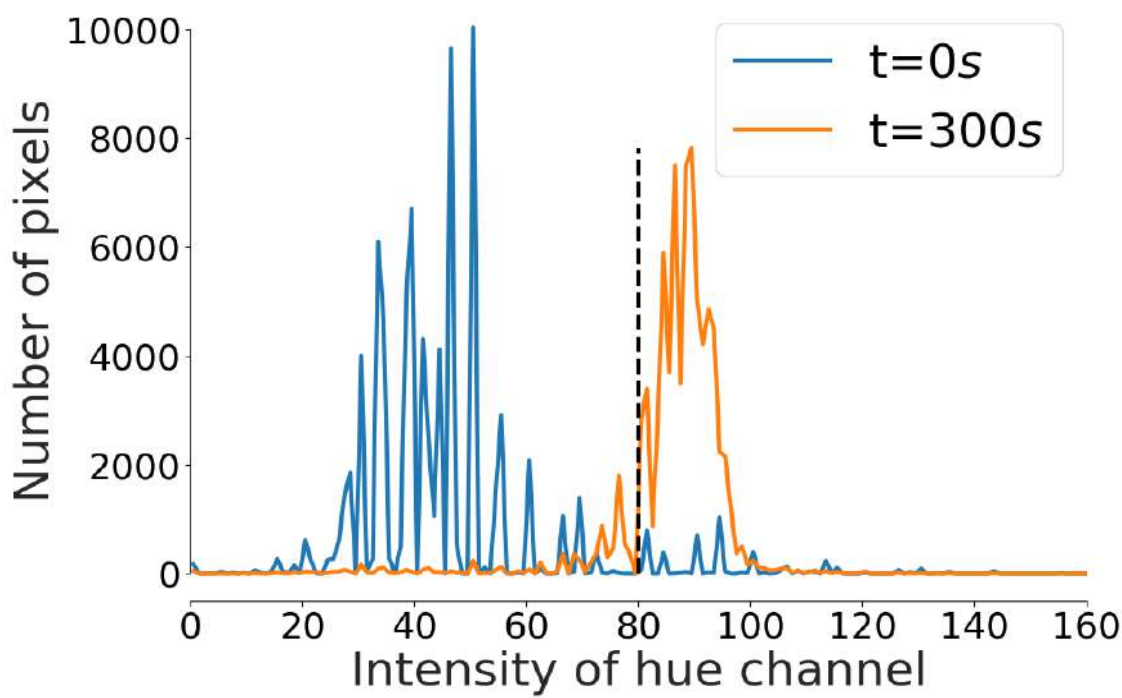

Supplementary Figure 25 - Histogram of the hue channel pixel distribution (pixel that are not part of a droplet and inside the tracking arena) at the start an on experiment (blue) and at the end of a 5 minutes experiment (orange) – for experiment number 17 in the cold condition at a measured temperature of 17°C. At the start of an experiment there is no dye released in the aqueous phase and the distribution represents the background hue channel intensity of the petri dish. At the end of the experiment, the aqueous phase is mostly blue, and the distribution of hue intensity has shifted to the right. Our measure of dye release is the ratio of the number of pixel that are above the threshold of 80 in intensity on the hue channel (dashed black line) over the total number of pixels.

As can be seen from Supplementary Figure 25 and see Supplementary Movie 6 (https://youtu.be/zOURJEnbmV4) the dye is released far more rapidly at 28.6 ±0.2°C than at 17.6 ±0.6°C. At the lower temperature, there is a significant dormant period of around 50s before any blue dye release is observed. Once the dye is released in the lower temperature case, it is released at a significantly slower rate than at high temperature. Thus, the peak blue count at low temperature is only reached after around 250s, compared to around 100s at higher temperature, a factor of ca. 2.5. This is a proof of concept that droplets such as these could be used as containers for active molecules – be they dyes, drugs or catalysts, which release their cargo differently depending on the temperature. This was done using a single recipe and tested with a single cargo and a whole optimisation process could be undertaken for a specific system of interest to further differentiate the release at low and high temperature - potentially enabling a binary release/no-release scenario. This was achieved with a temperature difference of only approximately 10°C, and further heating or cooling could also achieve even more differentiated results.

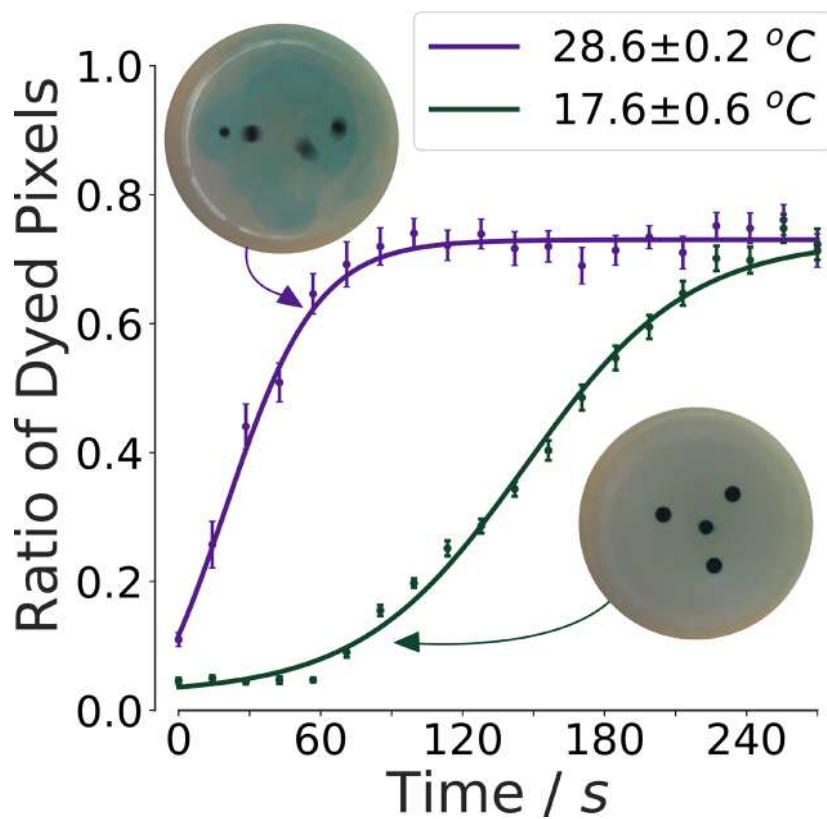

Supplementary Figure 26 - A plot of the measured ratio of pixels dyed blue against time at two different temperatures. The ratio of pixels dyed blue is a proxy for the amount of methylene blue dye released from the droplets during the experiment. Note that the value becomes constant / decreases over time, due to a reduction in blueness as the dye is dispersed homogenously in 3D – see Supplementary Movie 6 at https://youtu.be/zOURJEnbmV4.

## 1.9 15 Minute Experiments

Given the above, we chose one recipe of interest for further analysis to study the causes behind these drastic variations in droplet behaviour. The recipe chosen was recipe 4 (designated with a red star in Supplementary Figure 21), which is composed of 47.9% DEP, 13.5% 1-octanol, 1.9% octanoic acid, and 36.7% 1-pentanol. This recipe shows a steep and sudden increase in the average droplet speed at around 25°C, whilst displaying barely any movement below 22°C and little change in its average speed in the 27-30°C range. Recipe 4 also consistently maintains 3-4 droplets throughout the experiment at all except the highest temperatures and droplets remain active throughout the 90 second duration of our experiments, which removes droplet division or inactivity as extra factors to account for. Supplementary Movie 4 (https://youtu.be/zhTeDofB6mk) shows how droplets made with this recipe behave at a range of temperatures. Whilst there was logic behind our decision to study this droplet recipe in more detail, we could equally have chosen a number of other droplet formulation to study.

To study recipe 4 and analyse the mechanisms behind the droplet temperature dependant motion, we performed longer experiments to observe the system until it reaches an equilibrium. 15 minutes appeared to be an adequate duration. A key point for these 15-minute experiments is that droplet behaviours are measured over time, rather than just having an overall / average metric for a whole experiment as we did for the algorithmic exploration. It is indeed possible to track droplet behaviour throughout the course of the experiment using a chunking method, that is by running our analysis on smaller section of a video (e.g. chunks of 2 seconds) iteratively over the course of an experiment. Supplementary Figure 26 shows such data for one experiment for the speed and number of droplets metrics. Each point represents the average over a window of 2 seconds sliding with a time step of 1 second.

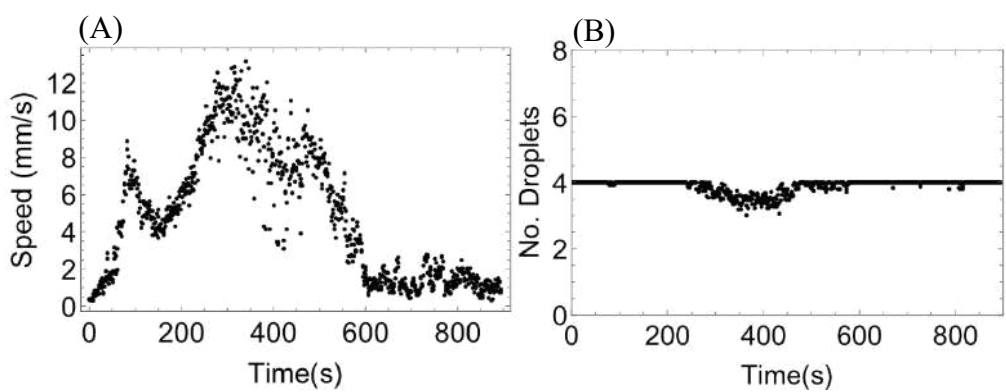

Supplementary Figure 27 – (A) shows average droplet speed vs. time and (B) shows a number of droplets vs. time for a single experiment of recipe 4 performed at 28.74 °C. The two peaks in the droplet speed vs. time plot is a specific feature of the chosen recipe that was never reported before. For this experiment, the number of droplets vs. time stays constant over the course of the experiment, 4 droplets are placed in the beginning and they remain present for the full 15 minutes of the experiment (except occasional period when droplets leave the tracking area due to high movement).

For this 15-minute experiment, two high droplet speed peaks are clearly observed with a period of slower speed in between followed by very little movement after around 600 seconds. The number of droplets is seen to be consistently 4, except for the period 240-480s were a slightly lower average number of droplets is detected. This period corresponds to the second peak of high droplet speed during which sometime droplets exit the tracking arena by going close to the dish walls. As a result, the tracking algorithm can perceive the number of droplet as oscillating between 3 and 4.

We have repeated such 15-minutes experiment 72 times at a wide range of temperature ranging from 20°C to 30°C. Supplementary Figure 27 shows the aggregated time-speed-temperature profile for all experiment for recipe 4.

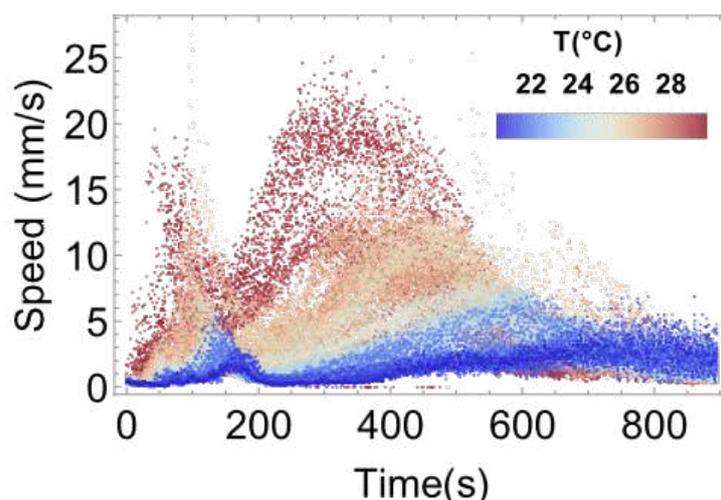

Supplementary Figure 28 – Average droplet speed vs time over all 15-minute experiments performed within the temperature range 20-30°C. Droplet speed is analysed in chunks of 2s with 1s overlap. The colour represents the temperature at which an experiment was performed, from 20°C (blue) to 30°C (red). The highest the temperature the faster the droplets move and the earlier they initiate movements.

Several trends can be observed from Supplementary Figure 27. First, the speed peaks occur earlier and with increasing speeds as the temperature increases. This is true for both the first and second peak. In particular, at the lowest temperatures the second speed peak is very broad, and the droplets are still moving at the end of the 15-minute experiment. The change in maximum speed due to temperature is surprising – rising from <5 mms$^{-1}$ at around 22°C to as much as 25 mms$^{-1}$ at around 28°C, a 5x speed increase for just a ca. 6°C increase. It is also interesting to note that preliminary experiments with other fast-moving recipes do not display these two movement peaks. This phenomenon has, to the best of our knowledge, not been reported before - making this droplet recipe quite unique in its time-speed profile.

## 1.10 Generating the temperature-time phase diagram

The code to replicate this analysis can be found in the form of a *Mathematica 11.3* (Wolfram Ltd.) notebook here:

https://github.com/croningp/dropfactory_analysis/releases/download/SI/droplet_motion_analysis.nb

The two distinct high-speed droplet peaks described above (in section 1.8) highlight the existence of different phases of droplet motion, as well as a significant impact of temperature on each phase initiation time and magnitude. To investigate further this phenomenon, we utilise the time-dependent droplet coordinate data to find inflection points in the droplet's acceleration profile (representing transition stages between different phases of motion) with the aim to create a time-temperature dependent phase diagram for this specific droplet composition.

After the completion of image processing and feature extraction to generate metrics such as the average speed and number of droplets (see section 2.1.5), time-dependent coordinate data of individual droplets were collected. We used *Mathematica 11.3* (Wolfram Ltd.) to further analyse and quantify the droplet behaviour. A complete summary of workflow to extract the temperature-time phase diagram from the droplet trajectory data is shown in Supplementary Figure 28.

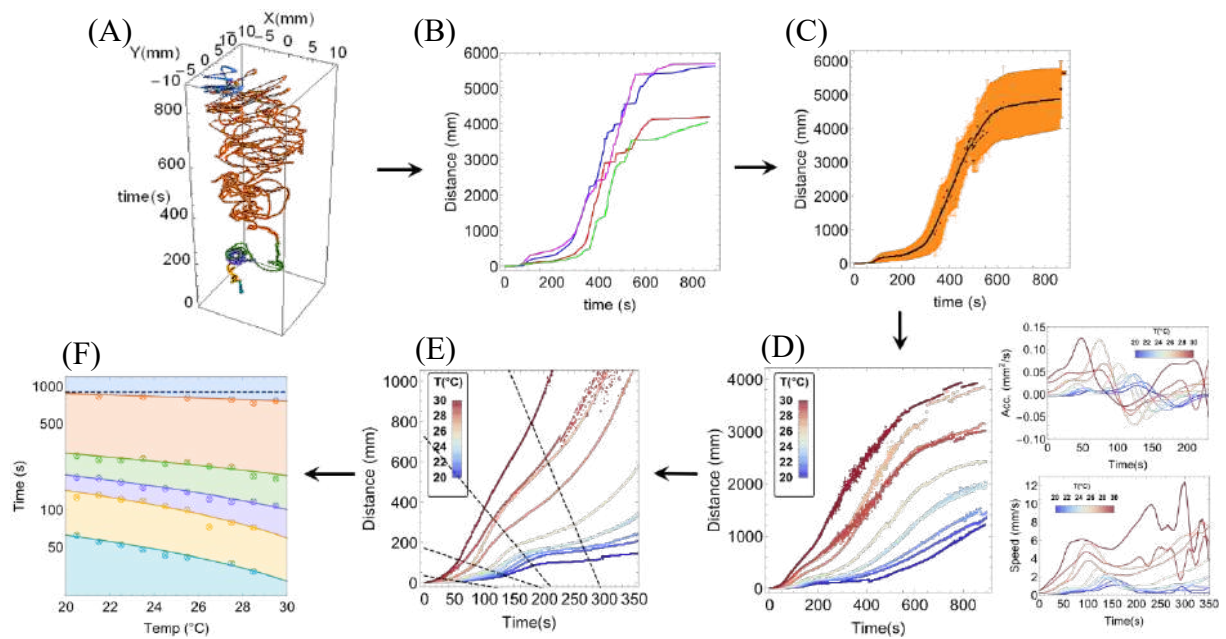

Supplementary Figure 29 - A summary of the workflow used in the preparation of the droplet motion temperature-time phase diagram. (A) Extraction of individual droplet trajectories from an experiment. (B) Estimation of the distance travelled by each droplet in an experiment. (C) Estimation of the average distance travelled by all droplets in an experiment. (D) Binning individual experiments by temperature intervals of 1°C in the range 20-30°C and estimating average speed and acceleration of the droplets vs. time for each bin. (E) Identification of different phases by extracting maxima, minima and inflection points in speed and acceleration vs. time data. (F) Fitting of the phases transition times through temperature and time to generate a continuous temperature-time phase diagram.

For simplicity, the two-dimensional trajectories (defined by X-Y coordinates) of each recognized droplet were stored in separate files. Our image analysis creates many trajectory files for each experiment due to collision events between droplets in the course of an experiment. After a collision, droplet tag numbers are updated, and each droplet is considered as a new droplet and assigned a new droplet number.

Based on the assumption that four droplets were present in each experiment, we started our analysis by connecting these different trajectory files to create complete trajectory for each of the four droplets in each experiment. We specifically searched for pairs of files separated by the shortest times by comparing last time point of one file with the first time point from the other file. Based on this scheme and by combining all the data files, we obtained four complete X-Y trajectory files for each experiment, one for each droplet.

Based on these data, we calculate the cumulated distance travelled by each droplet (Euclidean distance), an example of which is shown in Supplementary Figure 29A. Sudden jumps in the distance data correspond to deviations due to droplet collisions, misdetections, or droplets leaving the tracking arena. These errors will be averaged out as we agglomerate data from many repeats of each experiment. For each experiment, we average the cumulated data of each individual droplet as shown in Supplementary Figure 29B. Several transitions and non-linearities can be observed, which are an indication that different stages of droplet motion with variable speeds and acceleration occur in the timescale of 900 seconds.

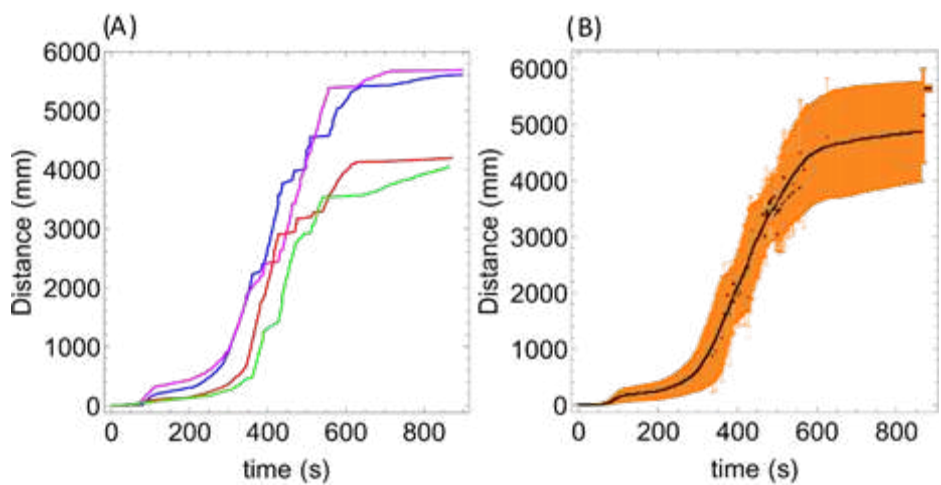

Supplementary Figure 30 – Cumulated distance travelled by droplets during a droplet experiment. (A) Cumulated distance (Euclidean distance) travelled for each of the four droplets in an experiment performed at 27.04 °C. (B) Mean of the cumulated distance over all four droplets (black data points) together with standard deviation plotted in orange

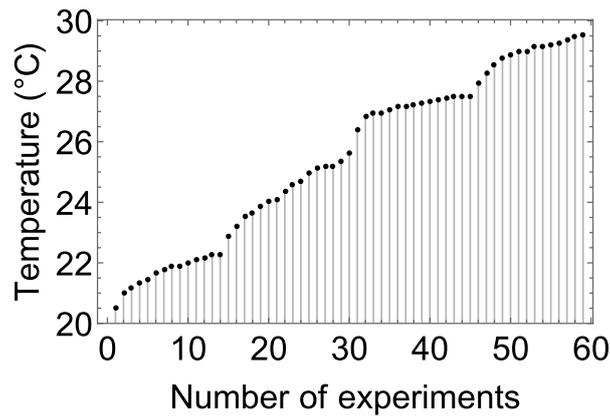

Supplementary Figure 31 – A plot illustrating the spread of temperatures of the 59 experiments used for phase diagram preparation.

The same analysis was performed on 59 experiments performed at different temperatures ranging between 20-30°C. After calculating mean cumulated distance for each experiment, the data were binned together by temperature intervals of 1°C and 2°C (e.g. 20-21°C, 21-22°C, …, 29-30°C). Supplementary Figure 31 shows the mean cumulated distance and standard deviation for different temperature intervals. A first observation is that the total cumulated distance travelled by droplets after 900 seconds increases monotonically with temperature. At higher temperature ranges (26-28°C and 28-30°C) the droplets eventually start decelerating as can be seen from the decrease in the slope of the cumulated distance curve between 600 and 900 seconds.

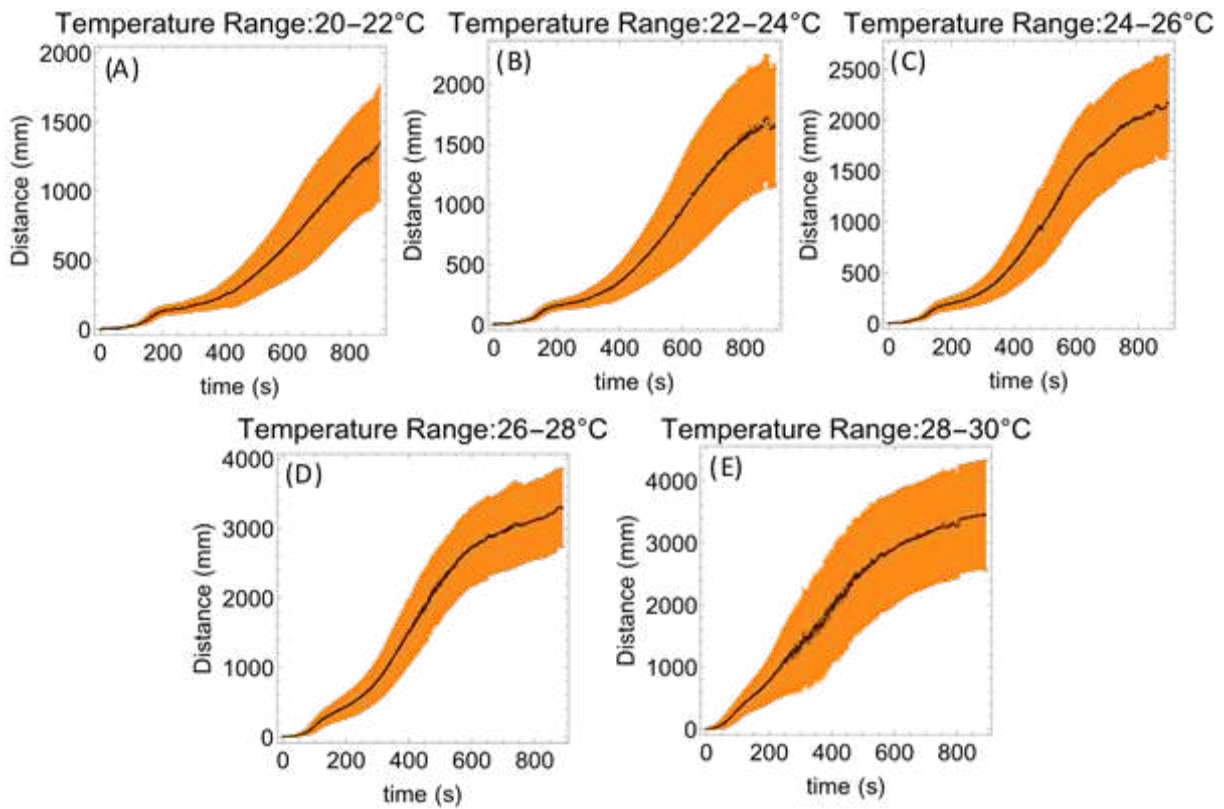

Supplementary Figure 32 – Cumulated distance travelled by droplets for all experiments binned in different temperature intervals. (A-E) respectively shows intervals of 20-22°C, 22-24°C, 24-26°C, 26-28°C, 28-30°C.

From the averaged travelled distance data, we can compute the speed and acceleration of droplets vs time which can be used to estimate maxima, minima and inflection points in the droplet motion in time. The speed and acceleration vs. time were estimated from the forward difference from averaged distance data and together with smoothing using moving average and low pass filter (see Mathematica Notebook for details). Supplementary Figure 32 shows the average speed and acceleration of droplets *vs.* time at temperature intervals binned by 1°C increment. Out of 900 seconds experiment, speed and acceleration plots are only shown up to 300 seconds. After these timescales, due to multiple collisions among droplets and between droplets and walls, the droplet speed and acceleration shows erratic peaks which cannot be used in quantification. Both speed and acceleration plots show multiple peaks which drift monotonically in certain directions with temperature. In the average speed *vs.* time plot, we observe a single peak prior to 250 seconds. The peak position shifts linearly, and its magnitude increases rapidly with increasing temperature. Similarly, in the average acceleration *vs.* time plot, multiple peaks were observed directly indicating different phases in droplet motion. At t ≤ 50s the first peak appears, which indicates an initiation period where the droplets first start to show early fluctuations.

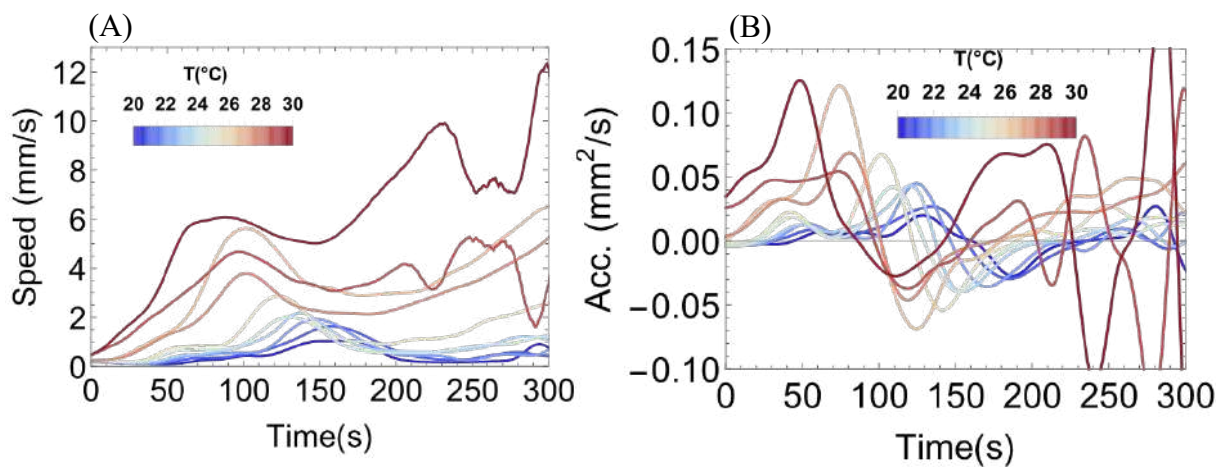

Supplementary Figure 33- Estimated average speed and acceleration of droplets from the cumulative displacement data. (A) shows speed vs time at different temperature intervals between 20-30 °C. The average speed increases monotonically with temperature up to ca. 200 seconds, then there are strong fluctuations in speed due to rebounding and collision events from the walls and pair of droplets. (B) shows acceleration vs time at different temperature intervals between 20-30 °C. There is a clear trend of acceleration at different temperatures up to ca. 200 seconds and similar fluctuations can be seen later as in speed vs time plot.

Based on the analysis of speed and acceleration vs. time data, we partitioned the droplet motion into several phases by finding the positions of maxima, minima in the acceleration profiles of droplets and cut-off criteria in the speed data. Six different phases could be distinguished from this analysis and over the timespan of 900 seconds, for each phase of motion the criteria of selection are described in Supplementary Table 1 along with a description of the characteristic droplet motion.

Supplementary Table 1 – Description of each of the phases P1-P6 and the criteria used to calculate the phase transition times.

| Phase | Criteria | Description |
|---|---|---|
| **Initiation (P1)** | Between start of the experiment and the 1st maxima in the acceleration vs. time data. | Initiation stage is defined between start of the experiment and minute fluctuations which corresponds to first maxima in acceleration vs time plot. |
| **Fluctuation (P2)** | Between 1st maxima and 2nd maxima in acceleration vs. time data. | Fluctuation stage is defined as first observed fluctuations before strong directional motion can be observed. |
| **Irregular (P3)** | Between 2nd maxima and 1st minima in acceleration vs. time data. | Irregular motion is defined by region when droplet starts moving erratically after initial fluctuations till slows down entering deceleration phase. The first peak in the speed vs. time data shows the time when the irregular phase is at its peak. |
| **Deceleration (P4)** | Between 1st minima and 3rd maxima in acceleration vs. time data. | Deceleration stage starts when the droplets starts slowing down before speeding up again. |
| **Continuous (P5)** | Between 3rd maxima and saturation criteria (speed < 1.3mm/s) | Continuous stage is defined when droplet speed starts increasing again after the deceleration stage till it finally starts slowing down. |
| **Saturation (P6)** | Between saturation criteria (speed < 1.3mm/s) and end of the experiment (900s) | The saturation stage is defined when the droplet speed after the continuous motion falls down below a certain threshold (in this case 1.3 mm/s). |

A temperature-time phase diagram was then created by calculating the intercept between cumulative distance travelled plots and linearly-fitted transition times. Supplementary Figure 33A and B show fitted lines (black dotted) on cumulative distance data which partitions the droplet motion into different phases. Supplementary Figure 33C shows the extracted temperature-time phase diagram with each coloured region indicating an individual stage of droplet motion. Initiation, fluctuation, irregular, deceleration, continuous and saturation phases are respectively shown in blue, yellow, purple, green, orange and blue. The different phases of droplet motion clearly show strong temperature dependence. In particular, each phase is initiated earlier with higher temperatures, which is also evidenced in Supplementary Movie 5.

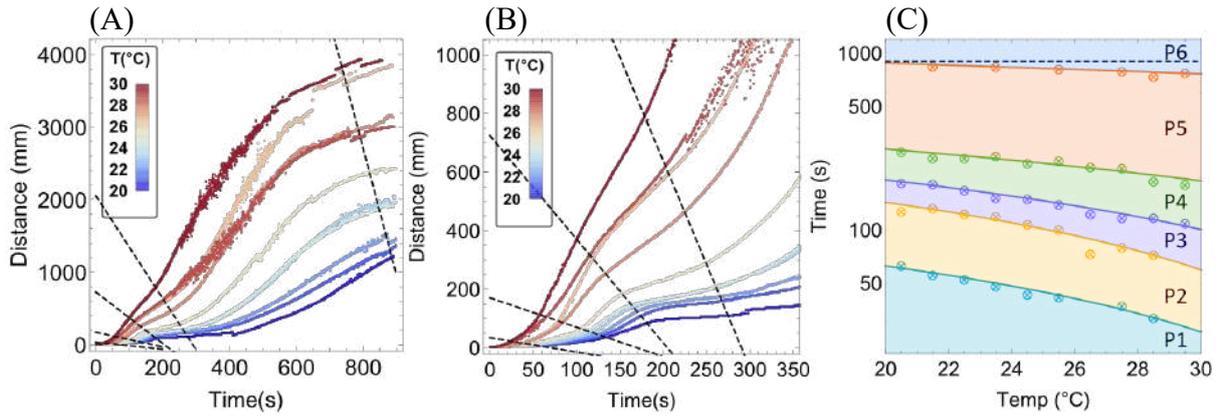

Supplementary Figure 34 – Temperature-time dependence on droplet behaviour. (A) shows mean distance travelled by droplet up to 900 seconds at different temperature bins. The dotted lines show the transition between different stages of droplet motion. (B) shows zoomed view of early stages of droplet motion up to 350 seconds. (C) Temperature-Time phase diagram showing different stages of droplet motion, with time plotted on a log scale for clarity. The different stages P1-6, as mentioned, are initiation, fluctuation, irregular, deceleration, continuous and saturation.

Finally, as a means to visualize the different phases of droplet motion along with the droplet motion, we plotted the trajectory of a single droplet in a 3D plot and applied the same colour scheme as the phase diagram to the trajectory for better visualization. Supplementary Figure 34 shows single droplet trajectories at 21.44°C and 27.39°C, for each plot the X-Y axis represents the position of the droplet and the Z axis represents time.

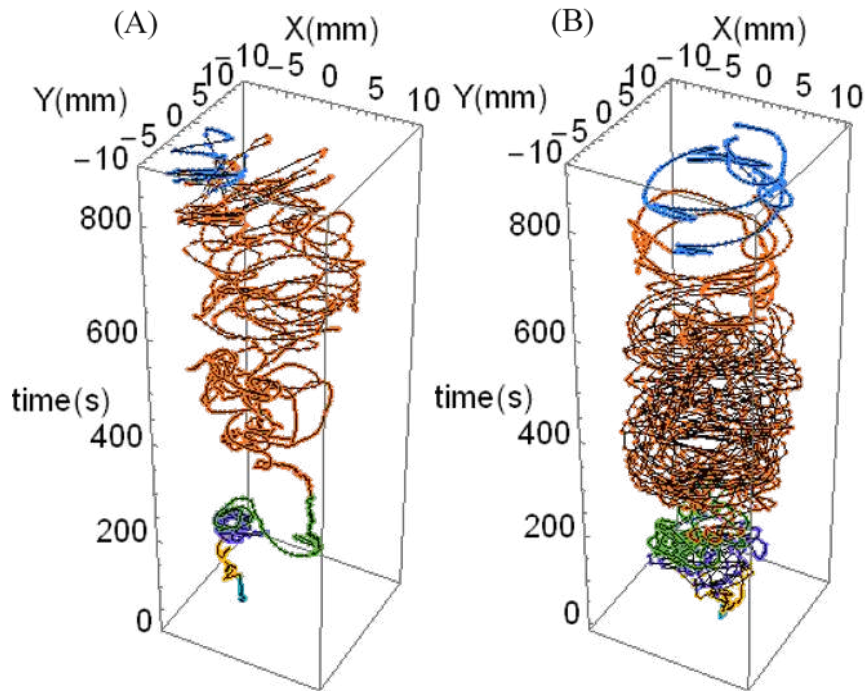

Supplementary Figure 35 – Single droplet trajectories plotted in 3D at 21.44°C (A) and 27.39°C (B). X and Y axis represent droplet position in mm from the centre of the petri dish and Z is for time ranging from 0 to 900 seconds in time steps of 0.25 seconds. The different colours correspond to different phases of droplet motion as used in the phase diagram of Supplementary Figure 33.

# 1.11 ¹H NMR Oil Dissolution Analysis

The code to replicate this analysis can be found in the form of a *Mathematica 11* (Wolfram Ltd.) notebook here:

https://github.com/croningp/dropfactory_analysis/releases/download/SI/droplet_motion_analysis.nb

Based on the analysis of NMR spectroscopy data (see details in section 2.3.2 titled "¹H NMR Spectroscopy Experiments"), we have estimated time-dependent concentration profiles of pentanol, octanol, octanoic acid, DEP and ethanol in the aqueous phase at two temperatures, 22°C and 28°C. When regulated to 22°C, the temperature at the experimental location was recorded at 22.4 ±0.2°C. When regulated to 28°C, the temperature at the experimental location was recorded at 27.7 ±0.2°C. The NMR data for the four oils and ethanol (produced via the hydrolysis of DEP) is shown in Supplementary Figure 35. As expected, oil concentrations in the aqueous phase increase with time.

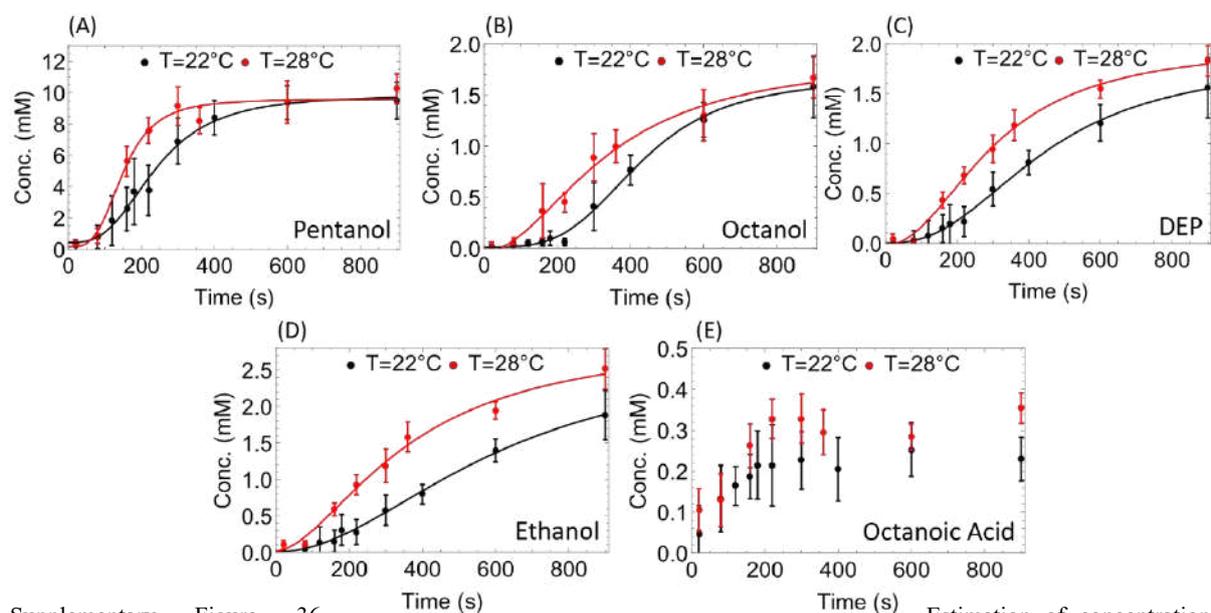

Supplementary Figure 36 - Estimation of concentrations of droplet components in the aqueous phase. (A-D) shows fitted concentration data vs. time for pentanol, octanol, DEP and ethanol at 22 °C and 28 °C. (E) shows the concertation of octanoic acid vs. time estimated from NMR data.

To investigate the dependence of droplet motion on oil dissolution, we first modelled the rates of oil dissolution into the aqueous phase ($C_{i,aq}$) by fitting the NMR data with a four-parameter logistic (4PL) function,

$$C_{i,aq} = d_i + \frac{a_i - d_i}{1 + (t/c_i)^{b_i}}$$

where the parameters $a_i$, $b_i$, $c_i$, $d_i$ corresponds to each oil $i$. The fitting parameters correspond to the minimum asymptote ($a_i$), Hill's slope ($b_i$, describing the steepness of curve at point $c_i$), inflection point ($c_i$), and maximum asymptote ($d_i$). Using *NonLinearModelFit* routine in *Mathematica 11.3* (Wolfram Ltd.), we fitted the experimentally determined concentration of pentanol, octanol, DEP, and ethanol. Due to the low proportion of octanoic acid in the recipe, and its rapid dissolution, its oil concentration profile was not suitable for fitting in this manner. The fitting parameters are shown in in Supplementary Table 2 at both 22°C and 28°C.

Supplementary Table 2 – Oil concentration in the aqueous phase fitting parameters at 22°C (left) and 28°C (right)

| Oil | Low T (22°C) | High T (28°C) |
|---|---|---|
| Pentanol | $a \to 0.43877$<br>$b \to 2.96298$<br>$c \to 239.825$<br>$d \to 9.88366$ | $a \to 0.18178$<br>$b \to 3.43907$<br>$c \to 149.829$<br>$d \to 9.59902$ |
| Octanol | $a \to 0.01566$<br>$b \to 3.60849$<br>$c \to 427.262$<br>$d \to 1.66740$ | $a \to -0.03002$<br>$b \to 1.94483$<br>$c \to 339.909$<br>$d \to 1.87132$ |
| DEP | $a \to 0.00140$<br>$b \to 2.44643$<br>$c \to 444.168$<br>$d \to 1.82304$ | $a \to -0.01605$<br>$b \to 2.07579$<br>$c \to 304.351$<br>$d \to 1.98727$ |
| Ethanol | $a \to 0.00758$<br>$b \to 1.99447$<br>$c \to 605.574$<br>$d \to 2.74819$ | $a \to 0.02029$<br>$b \to 1.74354$<br>$c \to 363.369$<br>$d \to 2.94498$ |

Supplementary Figure 35 shows the fitted curves at 22°C and 28°C for pentanol, octanol, DEP, and ethanol. The NMR data shows that pentanol dissolves faster than the other oils in the aqueous phase and that dissolution of all oils occurs faster at higher temperature. The characteristic timescale can be defined from the inflection point ($c_i$). For all oils, there is a large decrease in the characteristic time scale at higher temperature, which illustrates the faster kinetics of oil dissolution into the aqueous phase. The characteristic time scale for pentanol dissolution reduces from 239 seconds to 149 seconds with an increase in temperature. Alongside the increased pentanol dissolution, the rates of dissolution of the other oils also increase. Similar to pentanol, the characteristic timescale of octanol reduces from 427 seconds to 339 seconds, DEP from 444 seconds to 304 seconds and ethanol from 605 seconds to 363 seconds with the increase in temperature.

## 1.12 Associating physical and chemical analysis

The code to replicate this analysis can be found in the form of a *Mathematica 11* (Wolfram Ltd.) notebook here:

https://github.com/croningp/dropfactory_analysis/releases/download/SI/droplet_motion_analysis.nb

Having extracted the different phases of droplet motion in section 1.9 and modelled the dissolution of oils from the droplets into the aqueous phase in section 1.10, we can now utilise these together (see Supplementary Figure 36 and main text Figure 5). The aim was to find correlations between the different phases of motion and the oil dissolution rate profiles for the oils and hence better understand which oils drive which phase of motion.

As can be seen from Supplementary Figure 36, there appears to be some definite correlations between oil dissolution and the times of the different phases of droplet motion. For example, the irregular-deceleration phase transition (purple-green) is very closely correlated to the time of the peak pentanol dissolution rate. During the fluctuation phase, weak and unstable pentanol concentration gradients begin to form in the aqueous phase, leading to the fluctuating motion. Following this, as pentanol dissolution reaches its peak, the concentration gradient fluctuations become stronger leading to the irregular phase of droplet motion. This stage is an outcome of short-term symmetry breaking due to the formation of strong asymmetric oil gradients around the droplet-water interface. Formation of these gradients leads to a net force on the droplet due to asymmetric interfacial tension. At this stage pentanol dissolution is too rapid for the concentration gradients to be stable hence only irregular, rather than continuous, motion is observed.

When the rate of pentanol dissolution begins to decrease the droplets begin to slow down in the deceleration phase of motion. This occurs earlier with increasing temperature, and this phase transition is seen to be closely correlated to the time at which the pentanol dissolution rate begins to decrease. As the rate of pentanol dissolution is now decreasing, and there is already a substantial pentanol concentration in the aqueous phase, the pentanol concentration gradients are weaker at this time. It is not until the dissolution of other oils begins to dominate that the continuous phase of motion is observed. At the higher temperature, the pentanol dissolution rate is lower than the octanol + DEP + ethanol dissolution rate after 277 seconds, and the octanol + DEP + ethanol dissolution rates reach their maximum over pentanol dissolution after 405 seconds. Both of these are seen to correlate well with the acceleration and peak speed of the continuous phase of motion, implying that it is primarily these oils, and not pentanol, that causes the continuous phase of motion. These oils can set up new concentration gradients that drive the droplet via the Marangoni effect. As their dissolution is more gradual, sustained symmetry breakage can occur in both the aqueous and oil phases, as discussed in the main manuscript.

At the end of the continuous phase, the aqueous phase is almost saturated with all oils, reducing the dissolution rates of octanol, DEP and ethanol. This leads to the saturation phase of motion, in which the droplets again decelerate. As oil dissolution is so limited at this time there are no concentration or surface tension gradients to drive droplet motion. This saturation stage is more prominent and earlier at higher temperature as the increased oil dissolution rates earlier in the experiment lead to earlier saturation of the aqueous phase.

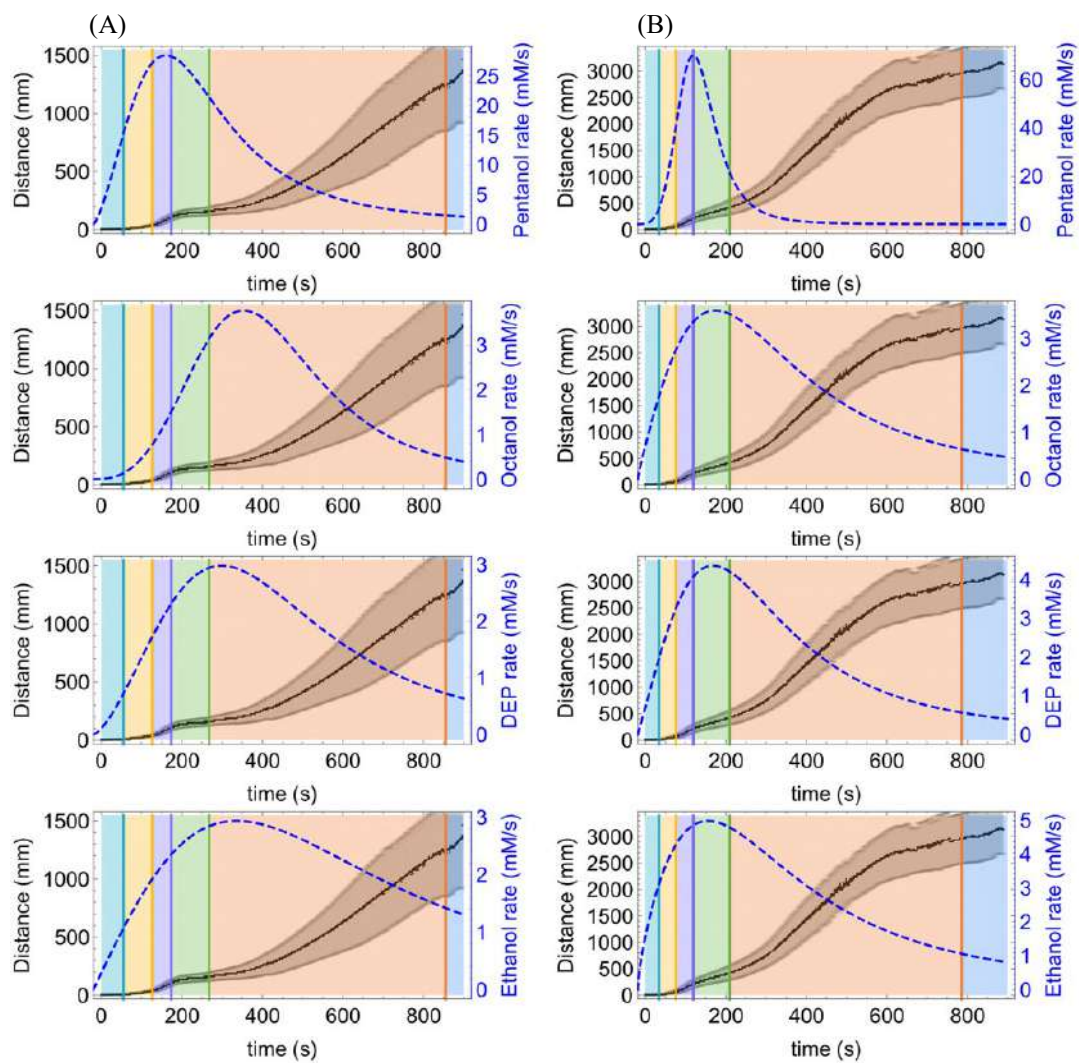

Supplementary Figure 37 - Comparison between rates of oil dissolution estimated from NMR experiments and cumulated distance travelled at 22°C (column A) and 28°C (column B). From top to bottom, the rows correspond to pentanol, octanol, DEP, and ethanol. On each plot, the black line shows the average cumulated distance moved by droplets in the experiments as described in section 1.9 (scale on left axis). The blue dashed line shows the dissolution rates of each oil as modelled in section 1.10 (scale on right axis). The background of the plot is coloured according to each phase of the droplet motion as described and extracted in section 1.9.

## 1.13 Additional Experiments to Probe the System

To probe our oil-in-water droplet system further, a range of experiments were undertaken to explore the effect of changing various parameters on the droplet behaviour. We subsequently present the following experiments to observe the effect on the time-dependant droplet motion:

- Variation of the pH of the surfactant containing aqueous phase
- Variation of the octanol-pentanol ratio in the droplet recipe
- Variation of the alcohol chain length in place of the pentanol component
- Variation of the number of droplets placed in the petri dish

Finally, DLS experiments are presented that were performed to check for the formation of micelles in the aqueous phase and possible change of size during an experiment.

### 1.13.1 Vary Aqueous Phase pH

The aqueous phase pH was varied by first preparing 20mM TTAB$_{(aq)}$, then adding the estimated NaOH$_{(s)}$ required for the desired pH before fine tuning with dilute NaOH$_{(aq)}$. The pH was then confirmed to be ±0.1 pH units (pH = 10, 13) and ±0.5 pH units (pH = 7) using a pH meter.

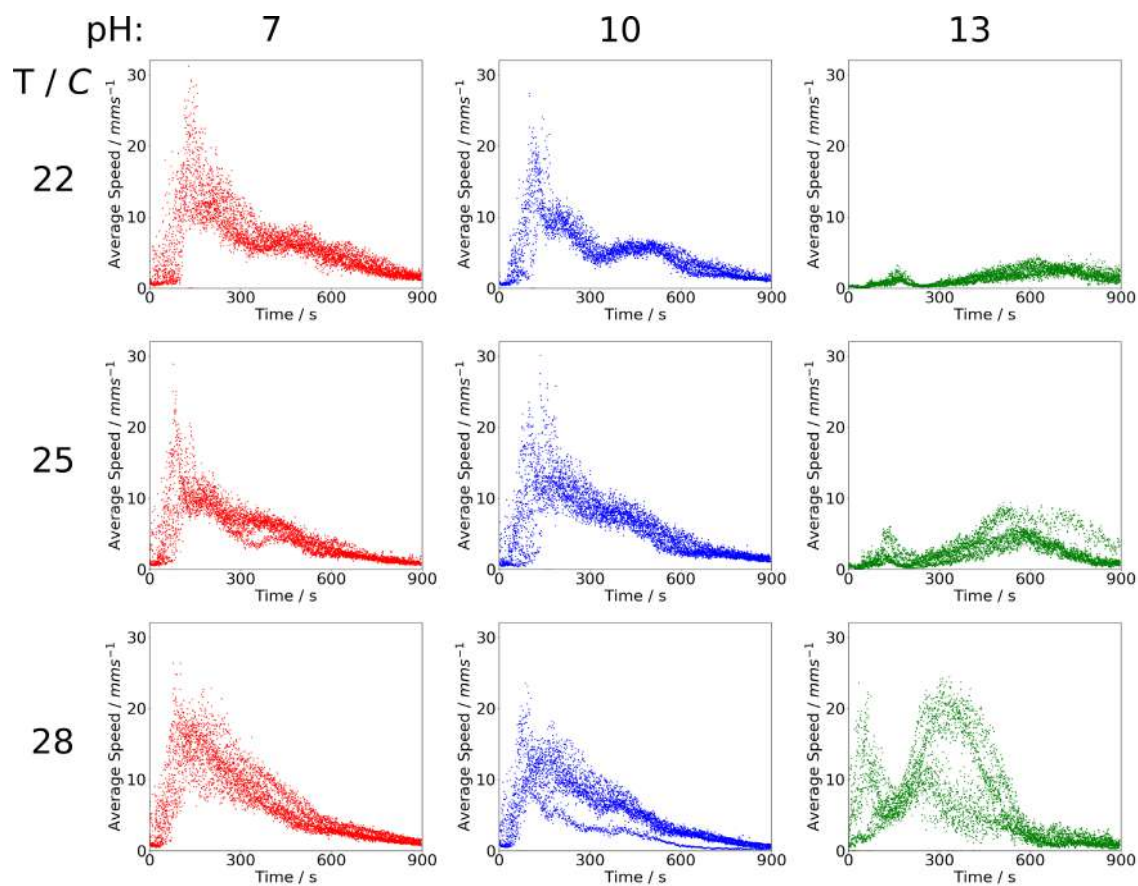

Supplementary Figure 38 - The variation in oil droplet speed as the temperature and aqueous phase pH are varied. Each plot shows the average speed measured every second in a 900 s experiment for 6 experimental repeats.

As can be seen from Supplementary Figure 37 and as discussed in the main text, changing the pH has a significant impact on the oil droplet behaviour. As the pH is decreased the second movement peak, corresponding to the continuous phase of motion, occurs earlier and with lower maximum speeds. Indeed, it is increasingly hard to justify this as a second peak as the pH is decreased.

Both lowering the pH and lowering the temperature is expected to significantly decrease the rate of DEP hydrolysis. We confirmed the change in oil dissolution by performing solvent suppressed $^1$H NMR spectroscopic as shown in Supplementary Figure 38. These experiments confirm that ethanol is only produced at high pH and is increasingly produced at higher temperature. This trend is mirrored in the level of DEP dissolution. Conversely, the dissolved octanol concentration is seen to vary little with the change in pH. Thus, this implies that octanol dissolution is not the primary cause of the large second movement peak at pH 13 and it is in fact DEP hydrolysis / ethanol. Interestingly, we only see a significant second movement peak at pH 13 and this is also the only time when we see DEP hydrolysis.

The first movement peak (corresponding to fluctuation and irregular motion) is also seen to vary in maximum speed and time with pH. At pH 7, the maximum speed is high, with this peak occurring earlier at higher temperature. At pH 13, the peak is supressed and later at 22°C, and this is reflected in the low level of pentanol dissolution at 120 seconds.

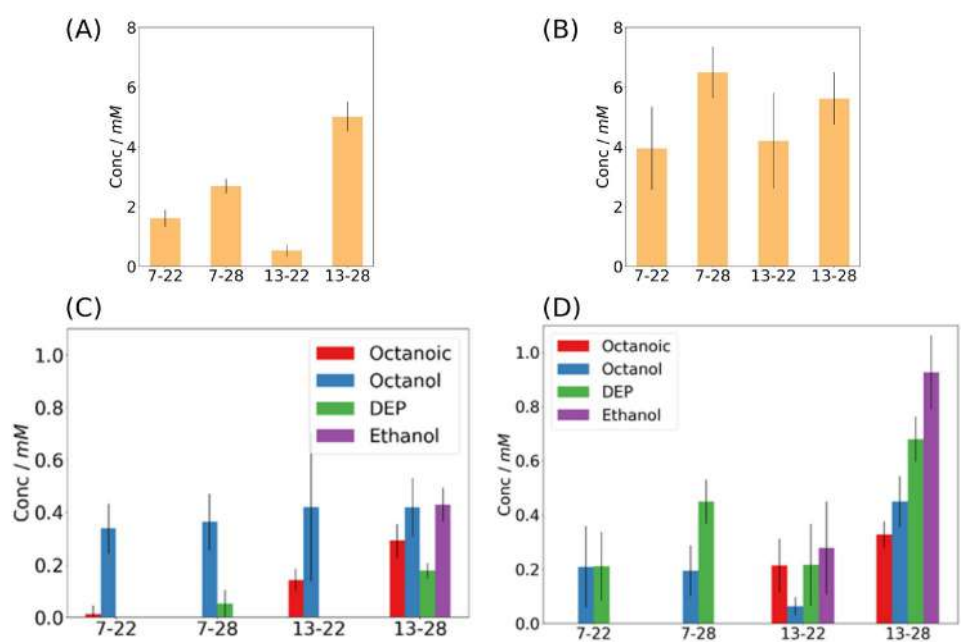

Supplementary Figure 39 – The variation in pentanol (A,B) and other oil (C,D) dissolution level as temperature and aqueous phase pH is varied. (A+C) are for samples 120 s after droplet placement (B+D) plots for samples 220 s after droplet placement. The x-axis labels refer to the pH and temperature, e.g. 7-22 is pH 7 and 22°C.

### 1.13.2 Vary Octanol-Pentanol Proportion

To study the effect of small recipe variations on droplet behaviour and gain further insight into the system, a series of experiments were undertaken in which the octanol and pentanol proportions were varied in 2% increments from their level in the focus recipe (Supplementary Figure 39).

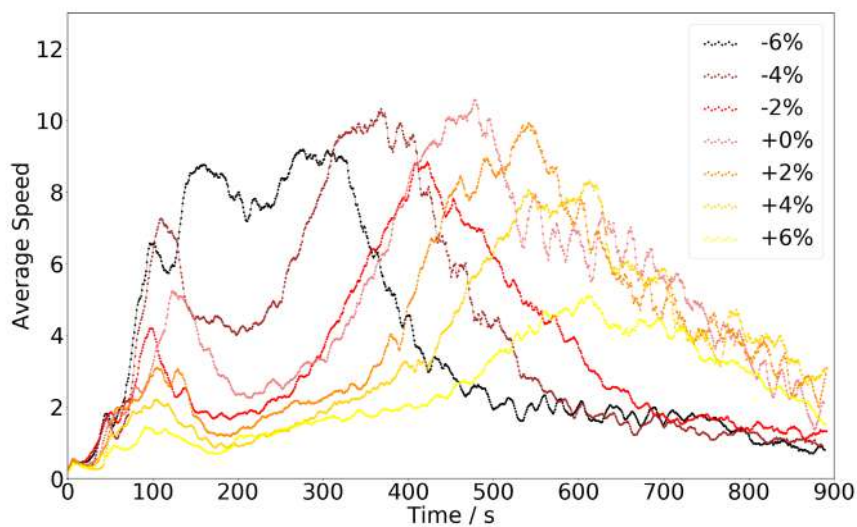

Supplementary Figure 40 - A plot of the 10s moving average of the droplet speed against time for a 15-minute experiment, averaged over 6 repeats, for oil droplets composed of recipe 4 with small changes in the proportion of the recipe of each of the alcohols octanol and pentanol. The legend keys refer to the % change of octanol, the reverse value is true for pentanol, i.e. black is -6% octanol +6% pentanol compared to recipe 4.

Looking first at the later peak (at ca. 150-700s), corresponding to the continuous mode of motion, there is a very clear trend for increased octanol content causing the movement to occur later. It is thought that this second movement peak is primarily due to DEP hydrolysis and dissolution, and we hypothesise that this effect is due to the decreased movement and mixing in the early stages with increasing octanol proportion. Hence, with low levels of mixing for higher levels of octanol, DEP hydrolysis is slowed, and this second peak occurs later and with less intensity. It is interesting to compare the 0% and -2% octanol dynamics. The 0% peak corresponding to irregular motion is unexpectedly higher than the -2% peak, and the same ordering is observed for the continuous peak. Thus, it appears that in this case the increased motion in the early stages for 0% then leads to increased motion later on due to the increased mixing. For all the other mixtures tested, the trend is as expected with increasing pentanol proportion leading to both peaks becoming faster and earlier. This together provides further evidence that it is DEP, and not octanol, that is the primary influencing factor for the continuous phase of motion.

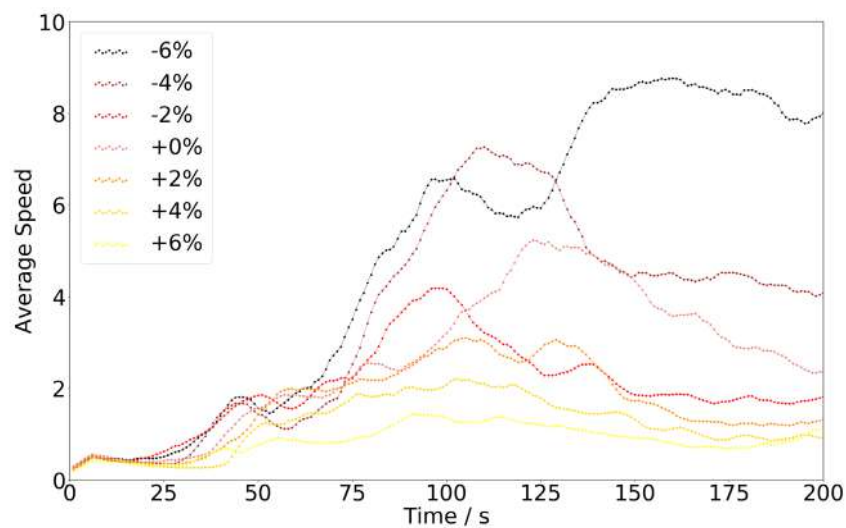

Supplementary Figure 41 - A plot of the 10s moving average of the average droplet speed against time for the first 200s of a 15-minute experiment, averaged over 6 repeats, for oil droplets composed of recipe 4 with small changes in the proportion of the recipe of each of the alcohol octanol and pentanol. The legend keys refer to the % change of octanol, the reverse value is true for pentanol, i.e. black is -6% octanol +6% pentanol compared to recipe 4.

Looking now at the early stages of droplet motion (see detailed view in Supplementary Figure 40), we see a clear trend, where increasing the proportion of pentanol and decreasing the proportion of octanol has an effect on both of the initial movement peaks (corresponding to the fluctuation and irregular forms of motion). Even with only these small formulation changes, the fluctuation peak (at ca. 50s) shows a trend for moving slightly earlier and displaying slightly higher speeds. Similarly, the irregular motion peak (at ca. 100s) becomes much more pronounced with higher proportions of pentanol. Reversely, when the pentanol proportion is decreased by only 4-6%, this first motion peak is significantly suppressed. This confirms the direct link between pentanol / octanol proportion and these initial phases of movement, with an increased pentanol proportion leading to earlier and stronger droplet motion. Whilst the differences between the average movement speeds at 50 s is only small, the trend is strong and the standard deviation in the average speed values at this time are low (Supplementary Figure 41).

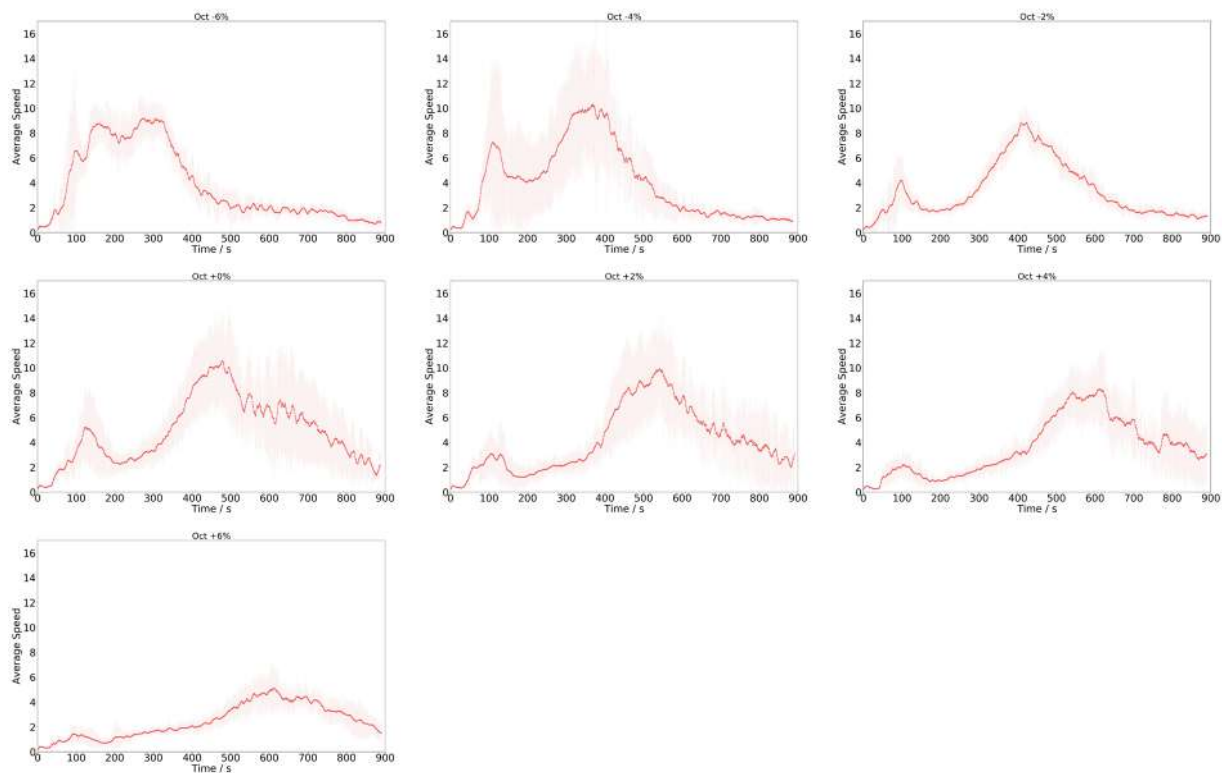

Supplementary Figure 42 - A plot of the 10s moving average of the average droplet speed against time for a 15-minute experiment, averaged over 6 repeats, for oil droplets composed of recipe 4 with small changes in the proportion of the recipe of each of the alcohol octanol and pentanol. Shaded areas show standard deviation.

### 1.13.3 Vary Alcohol Chain Length

Following this, a study was undertaken in which the pentanol in the formulation (which contributes 36.7% by volume) in replaced with other straight chain primary alcohols of varying chain length, the results for which are shown in Supplementary Figure 42.

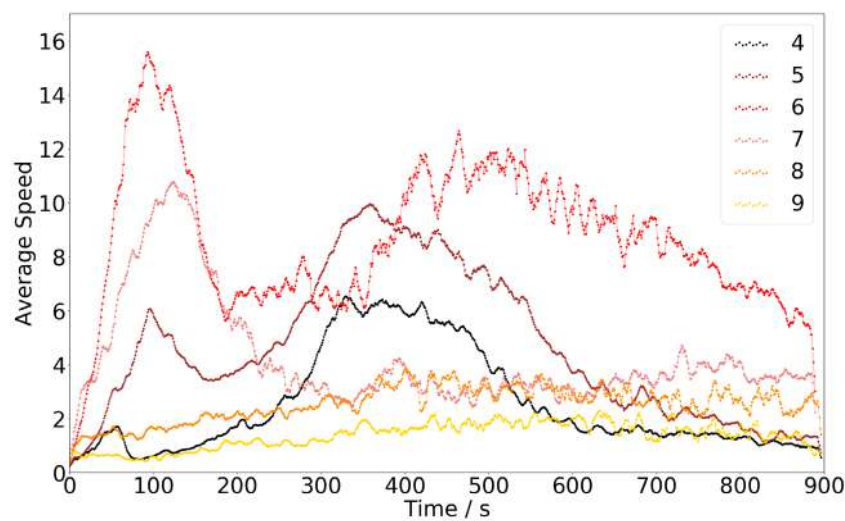

Supplementary Figure 43 – A plot of the 10s moving average of the average droplet speed against time for a 15-minute experiment, averaged over 6 repeats, for recipe 4 oil droplets with various straight chain primary alcohols used in the place of pentanol in the formulation. The number in the key refers to the alcohol used, for example, 7 = 1-heptanol used in the place of 1-pentanol.

As can be seen from Supplementary Figure 42, there are very clear differences in the time-speed droplet profile as the alcohol that replaces pentanol is varied. It appears that there is an optimum alcohol chain length for high speed in the first movement peak, with hexanol, heptanol and pentanol displaying the fastest movement. For butanol there is a small peak whilst for octanol and nonanol this first speed peak is not observed. It is proposed that this is due to a trade-off between the dissolution rate of the alcohol and the impact of the alcohol on the interfacial tension and hence Marangoni instabilities occurring on the droplet. The more hydrophobic alcohols (octanol and nonanol) are hypothesised not to dissolve in sufficient quantities in this timeframe to influence the interfacial tension and droplet motion. Conversely, butanol, despite presumably dissolving rapidly to significant levels, has a limited effect in these early stages. This implies that it has a more limited effect on the interfacial tension, potentially due to the dissolution being too rapid and symmetric for interfacial tension imbalances to be set up. Hexanol appears to represent the perfect chain length for this recipe, giving both significant dissolution and having an effect on the interfacial tension.

It is very interesting that even in the octanol case, in which octanol is used in the place of pentanol, as well as the octanol already in the formulation, no second peak (continuous phase of motion) is observed. This is a surprising result, and appears to confirm that the second, continuous movement peak is not directly due to the longer chain alcohol dissolving, adding weight to the theory that the second peak is primarily due to DEP hydrolysis. This continuous phase peak is increasingly dominant for butanol, pentanol and hexanol, and not present for heptanol or above. It is hypothesised that the increased hydrophobicity of the oil phase for heptanol and above reduces phase mixing and hence the DEP hydrolysis that causes this continuous motion. For butanol, pentanol and hexanol the size on the continuous motion peak mirrors the trend of the fluctuation-irregular peak, again implying that increased dynamics and mixing in the early phases impacts the continuous phase due to increased phase mixing.

### 1.13.4 Vary Number of Droplets

A study was undertaken in which the number of droplets placed was varied. In all cases, a symmetric placement pattern was used, starting with a droplet at the centre of the dish and then subsequent droplets placed in a symmetrical pattern around this first droplet (line, triangle, square, pentagon). Eight repeats were undertaken for placing 3, 4, 5 and 6 droplets, the results for which are shown in Supplementary Figure 43.

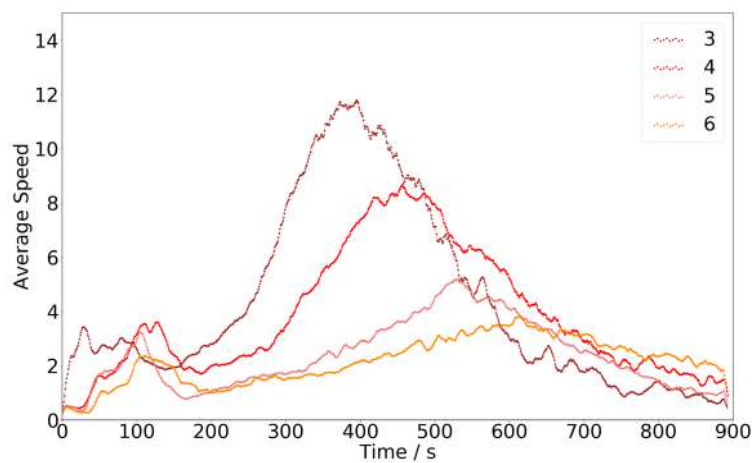

Supplementary Figure 44 - A plot of the 10s moving average of the average droplet speed against time for a 15-minute experiment, averaged over 8 repeats, for recipe 4 where 3, 4, 5 or 6 droplets are placed (indicated in the legend). Note for the 6-droplet case 2 outliers are removed from the average, as in these cases many droplets were placed onto one another, leading to only 1 to 2 remaining active droplets that were bigger in size hence resulting in very different droplet dynamics.

As can be seen from Supplementary Figure 43, there is a clear trend for both movement peaks to occur later and with lower maximum speeds as more droplets are placed. As more and more droplets are placed, the net oil dissolution from each droplet should be reduced, due to Le Chatelier's principle. Thus, there will be a reduced effect of the oils on the surface tension and reduced surface tension gradients and slower and later oil droplet movement. This again confirms the direct relationship between oil dissolution and oil motion. The same data can be seen on Supplementary Figure 44 with the standard deviation around the average values.

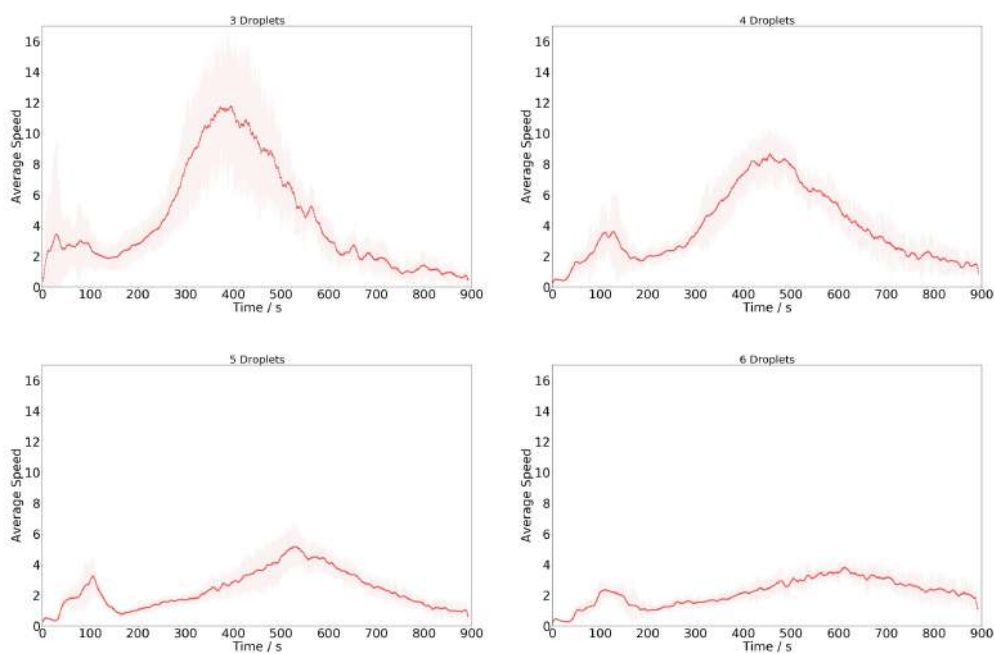

Supplementary Figure 45 - A plot of the 10s moving average of the average droplet speed against time for a 15-minute experiment, averaged over 8 repeats, for recipe 4 where 3, 4, 5 or 6 droplets are placed. Note for the 6-droplet case 2 outliers

are removed from the average, as in these cases many droplets were placed onto one another, leading to only 1-2 remaining active droplets and very different droplet dynamics. Shaded areas show standard deviation.

## 1.14 DLS Experiments

Dynamic light scattering (DLS) measurements were performed on a Malvern Instruments Zetasizer Nano–ZS spectrometer fitted with a 633 nm laser. Samples were analysed at 25 °C on the same day as sample preparation, with a 10-minute equilibration time. For each sample the analysis was repeated 10 times, with each analysis representing the mean value of 20 runs with a 20 seconds scan time per run. At least 5 separate samples were analysed in this way at each timepoint.

To study the effect of oil dissolution on the aqueous phase supramolecular assemblies, a range of samples were prepared in a similar manner as for the $^1$H NMR samples (without $D_2O$ addition) and analysed via DLS. Samples were collected at an experimental temperature of 28.0±2.0°C.

As can be seen from Supplementary Figure 45, for fresh aqueous phase micelles were observed to have a hydrodynamic diameter of 5.29±0.28 nm (0 s datapoint). After 200 s, these micelles were observed to have shrunk to around 4.71±0.23 nm. At this stage, only pentanol has dissolved to a significant level and the droplet has been undergoing fluctuating and irregular motion. Indeed, pentanol dissolution in TTAB solutions above the CMC is known to have the effect of decreasing the micelle size *and* reduce the free surfactant concentration via reducing the TTAB aggregation number.[40] After 500 seconds (5.60±0.13 nm) and 900 seconds (5.40±0.20 nm) the hydrodynamic diameter of these micelles has risen to a value close to / slightly above the original value. During this period, the droplets have been undergoing continuous motion and octanol, DEP and ethanol have been dissolving at their highest rate. Whilst the exact relationship between these findings and the droplet motion mechanisms are unknown, there is a noticeable difference between the fluctuation / irregular phase of motion (decreasing micelle size) and the continuous phase of motion (increasing micelle size).

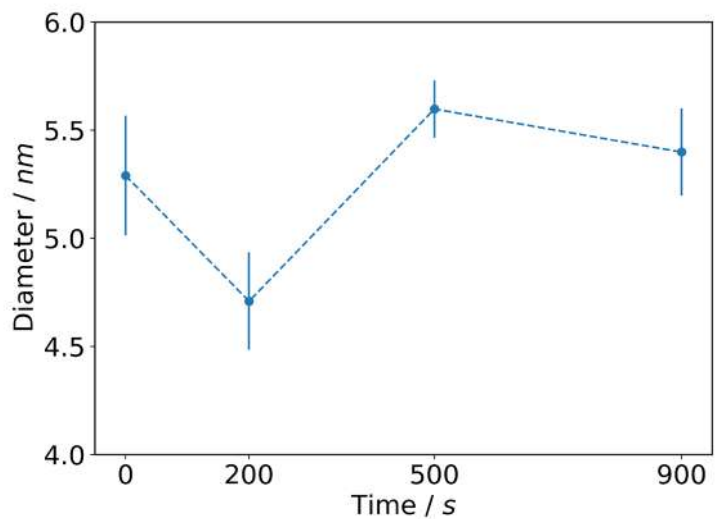

Supplementary Figure 46 – A chart illustrating how the hydrodynamic diameter of micelles in the aqueous phase changes, as measure by DLS.

To confirm these trends for more extreme samples, three further model aqueous phases were prepared in which 16 µL, 32 µL or 3.5 mL of the focus recipe was stirred for 10 minutes with 3.5 mL of the aqueous phase before analysis by DLS. Our intention was that such extreme cases would magnify the trends observed above. Results from these samples are detailed in Supplementary Table 3. As can be seen, the aqueous phase assembly size increases with the increasing level of oil dissolution.

Supplementary Table 3 – A table illustrating the hydrodynamic diameters measured for samples in which 16 µL (A), 32 µL (B) or 3.5 mL (C) of the focus recipe are stirred for 10 minutes with 3.5 mL of the aqueous phase.

|  | A | B | C |
|---|---|---|---|
| **Hydrodynamic Diameter / nm** | 5.81 ± 0.22 | 25.3 ± 0.96 | 618 ± 120 |

## 2 Supplementary Materials and Methods

In this section, we describe the robotic platform in detail and link to the available software and hardware open-access resources. We then detail the standard operating procedures used to prepare the reagents. Finally, we detail both the implementation of the algorithms and the analysis with appropriate links to the corresponding software open resources that are made available.

The information regarding to the platform, algorithm and data analysis are distributed across the following GitHub repositories:

- the robotic platform: https://github.com/croningp/dropfactory

- the algorithms & running experiments: https://github.com/croningp/dropfactory_exploration

- the analysis of experiments: https://github.com/croningp/dropfactory_analysis

- the code to track and analyse droplet's video: https://github.com/croningp/chemobot_tools

- the framework code to control Arduino boards from Python:

    - https://github.com/croningp/commanduino

    - https://github.com/croningp/Arduino-CommandTools

    - https://github.com/croningp/Arduino-CommandHandler

- the library to control the tricontinent pumps used to handle liquids on the platform: https://github.com/croningp/pycont

- the modular linear actuator design used for the syringe and filling and cleaning stations: https://github.com/croningp/ModularSyringeDriver

- the modified explauto library used in this work: https://github.com/jgrizou/explauto, which was derived from https://github.com/flowersteam/explauto

- a Python file manipulation library used across some repository work: https://github.com/jgrizou/filetools

## 2.1 Robotic Platform: Dropfactory

In this study, the robotic platform, called Dropfactory, is only a tool for querying information on the system in study in the real world and in a fully automated fashion. Because we aimed to compare algorithms, we needed to design and build a tailored robot enabling high-throughput of highly reproducible experiments. Previous platforms could undertake around 50 experiments per working day, we needed a throughput of around 300 experiments per working day for our comparative study to be achievable in reasonable time.

Supplementary Movie 1 (https://youtu.be/bY5OoRBJkf0) shows Dropfactory in operation.

In the following, we first explain the principle and conceptual shift in the design of Dropfactory compared to other platforms and later details each of the working stations and their respective role and action sequences.

### 2.1.1 Principles

An oil-in-water experiment consists of placing small oil droplets (made from an arbitrary mixture of oils) at the surface of an aqueous medium (made from a mixture of aqueous phases), we then need to video record the droplet movements and analyse them. To be able to run such experiments continuously, the platform must be able to mix, sample, clean and dry both the oil and aqueous phases. To increase efficiency, compare to previous platforms, we moved from a sequential to a parallel mode of operation.

Previous robotic platforms were based on an open-source 3D printer and all the tools (syringe, filling and cleaning tubing) were situated on a single moving head - this means only one action could be performed at the same time. To reach a throughput of 300 experiments a day, different stages of the preparation of an experiment have to be performed in parallel. Dropfactory is designed to do so.

The platform is organized as a little factory, enabling us to fully parallelize all required operations (mixing, droplet placing, recording, cleaning, drying). To do so we designed the platform around two 'turntables' - called Geneva wheels - with specialized workstations positioned all around the wheel. Supplementary Figure 46 illustrates this principle.

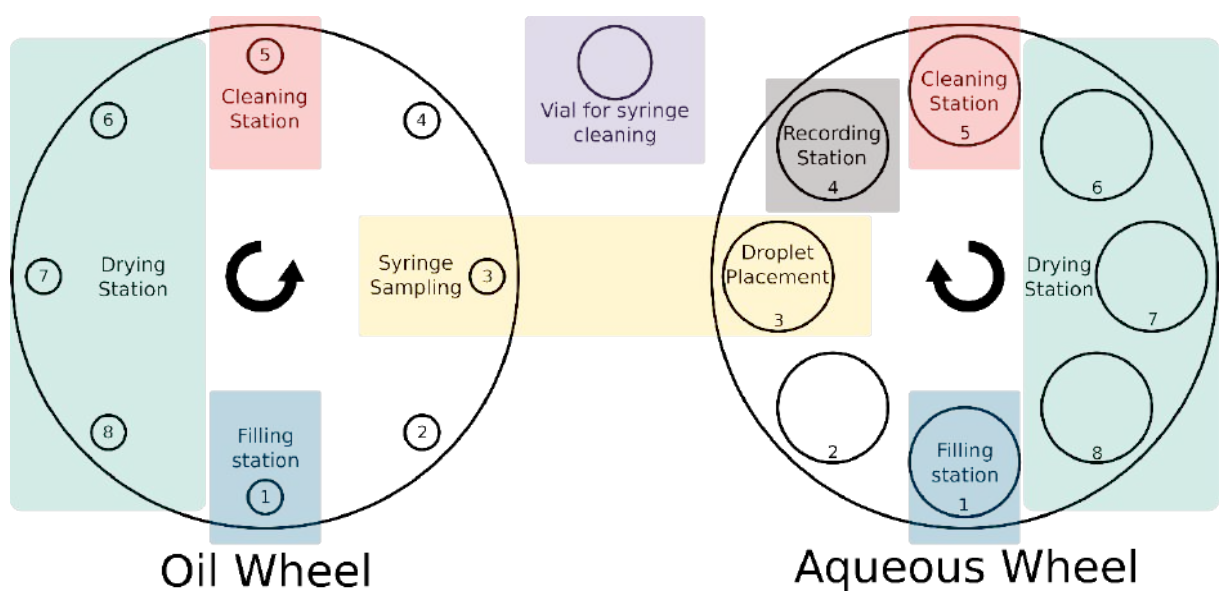

Supplementary Figure 47 - Conceptual design of Dropfactory. The oil and aqueous phase are handled separately on two turntables that can move the containers (oil vials or petri dish) from one fixed working station to the next. At each position some specific actions are performed on the containers.

This design allows the movement of oil vials and petri dishes between specialized working stations, rather than requiring the tools to move to those containers. As a result, Dropfactory is:

1. **Robust**, because there are far fewer moving parts. Each working station performs only its specific task at the location it was designed to work. This also means less tubing and wires moving around.

2. **Easier to maintain**, because all working stations are clearly separated, both physically and digitally. Thus, identifying a bugs or mechanical failures is quick and fixing them is easier.
3. **Fast**, because while an oil mixture is prepared, a previous one is being cleaned, another three are being dried, and another is sampled by a syringe to be placed on a previously filled petri dish. At the same time, another petri dish is being filled, one is being cleaned, three are drying, and one contains droplets in motion that are being recorded under a camera.

Dropfactory is able to record 1 experiment of 1min30sec every 1min51sec, faster by a factor of 6 versus previous sequential platform. Thanks to its robustness, the platform was consistently running for months in the lab recording more than 30,000 droplet experiments.

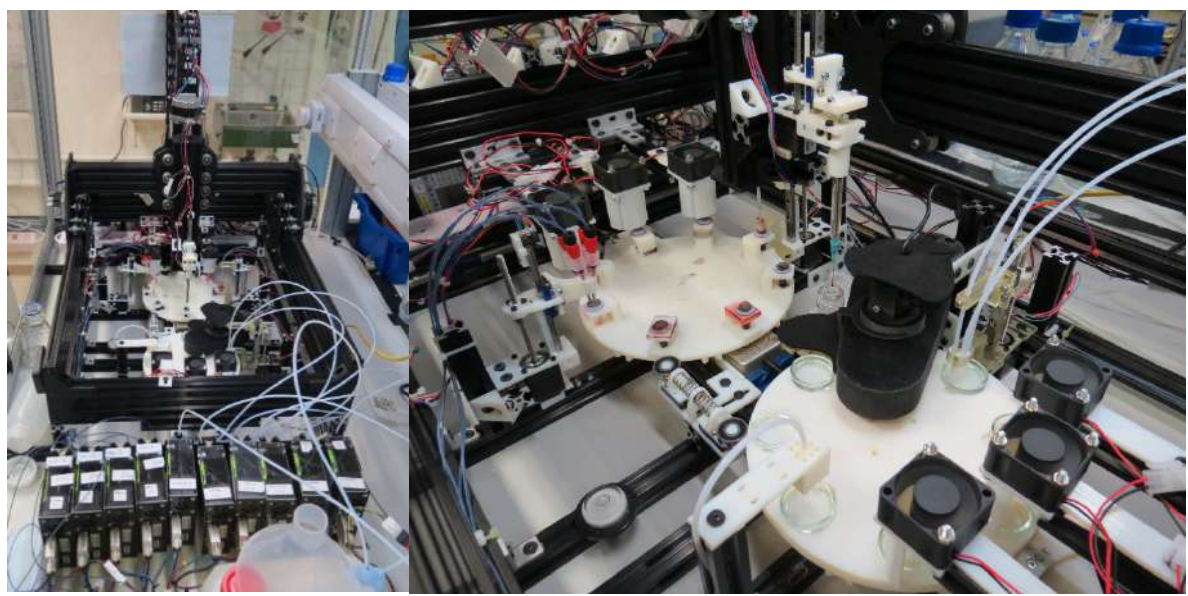

Supplementary Figure 48 - Left: photo of the platform from the side, showing the liquid handling pumps in the foreground, and the XYZ CNC frame behind. Right: The inside of the XYZ frame containing the two Geneva wheels with the filling, cleaning and drying stations. The syringe system for oil droplet sampling and placement is mounted on the XYZ head enabling the sampling of the oil mixture from one wheel and the placing of droplets on the petri dish in the other wheel caring the aqueous phase filled petri dishes.

Supplementary Figure 47 shows the platform as installed in our lab fumehood. We can clearly see the two Geneva wheels, the XYZ arm and a set of pumps for oil, aqueous and cleaning reagents. One wheel handles the oil cycle (filling a vial with a mixture, mixing, sampling droplet, cleaning, drying), the other handles the aqueous cycle (filling a petri dish with aqueous phase, adding droplets, video recording of droplet behaviour, cleaning, drying). Each time the wheel turns the respective vials/dishes found themselves under a different working station. Finally, a syringe mounted on a XYZ CNC platform makes the bridge between the two wheels by sampling a mixed oil mixture from the oil wheel and generating the droplets on the petri dish on the aqueous wheel, cleaning the syringe between each experiment.

We describe next the robotic frame, the oil wheel, the aqueous wheel, the syringe system, the pumps and some monitoring instruments, as well as what function they perform and where to find all necessary hardware and software information.

### 2.1.2 Platforms and Working Stations Design and Procedures

Most of the information described below and especially the details of the materials can be found on the following repository: https://github.com/croningp/dropfactory

We designed Dropfactory around three mechanisms:

1. **A XYZ CNC frame** that provides both the structural frame and the mechanism to move the syringe around as required to sample the oil mixture and place droplets.
2. **Two Geneva wheels**, one for the oils, one for the aqueous phase. They allow the containers to move from one working station to another in a simple and robust way.
3. **Working stations** that perform only one simple task. The vials and dishes are displaced to the working station thanks to the Geneva wheels. Having specialized working stations makes the whole system easier to design, build, and fix. In addition, a lot less cables and tubes will be in motion, reducing again the possibility of failure in the system.

In addition, we made a point to not over-design or over-specify the platform before building it. Rather we left ourselves room for iterating on the platform while building it and as we encountered problems. For example, most working stations have been redesigned 2 or 3 times after receiving real-world feedback from practical experience.

To achieve this flexibility, we based most of our design on 3D printing combined with aluminium profile technology - enabling us to tune the system on site. This also explains why there is no overall 3D design specifying every detail of the platform down to the last millimetre.

Dropfactory is a research platform, it has been conceived with modularity in mind and we encourage the interested reader to adopt a similar approach if they are willing to build their own robot based on our design. We now describe in more details the XYC CNC frame, the Geneva Wheel, the Modular Linear Actuator, the Pumps and each working station.

#### 2.1.2.1 Robotic Frame

Link: https://github.com/croningp/dropfactory/blob/master/doc/cnc_frame.md

The robot frame (Supplementary Figure 48) is based on an OX CNC Machine from OpenBuilds (OX CNC MECHANICAL KIT (500x750mm) + 4 NEMA 23 Stepper). Details are available on OpenBuilds website: http://www.openbuilds.com/builds/openbuilds-ox-cnc-machine.341/. We sourced the kit from a local reseller: http://ooznest.co.uk/OX-CNC-Machine/OX-CNC-Mechanical-Kit

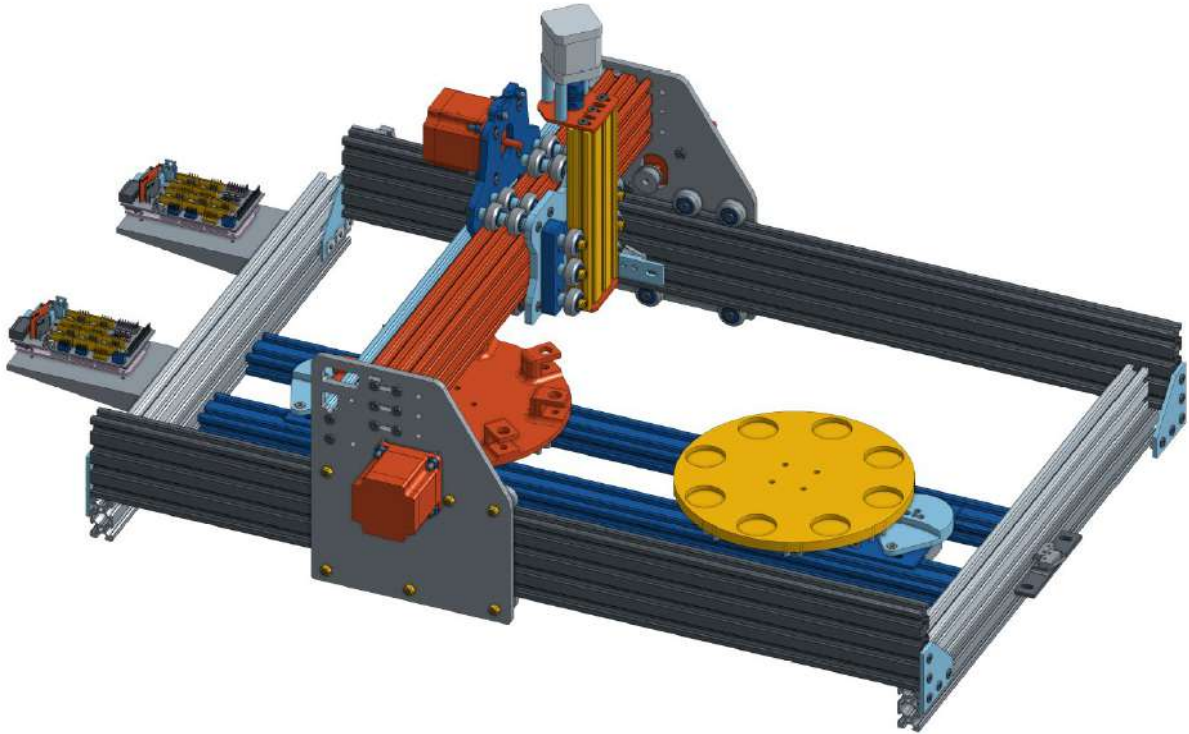

Supplementary Figure 49 - The OX CNC mechanical kit used as the base frame for Dropfactory.

To control the XYZ motion we use two Arduino Mega boards mounted with a Ramps 1.4 shield:

http://ooznest.co.uk/3D-Printer-Electronic-Parts/RAMPS-14-Controller-Board-Premium

Each board can control 5 stepper motors, 3 are needed for the XYZ motion. 1 is needed to control our syringe system, and 3 more to control some of our working stations. The board also controls our temperature and humidity sensor (SHT15 - https://www.sparkfun.com/products/13683) that records the conditions under which each experiment was performed.

The platform is controlled via python on an external computer using our commanduino tool-kits (https://github.com/croningp/commanduino) allowing the quick iteration and prototyping of Arduino based robots:

1. The firmware for both Arduino boards is in the software/arduino folder: https://github.com/croningp/dropfactory/tree/master/software/arduino

2. The corresponding python robot controller is in the software/robot folder: https://github.com/croningp/dropfactory/tree/master/software/robot

The frame design and 3D files are available from Onshape:

https://cad.onshape.com/documents/3aeb7616c1e547bfaae38ba3/w/426b95792e7c48a8b6dd7727/e/af7f485263ee4608affce6e3

This base provides a robust XYZ machine with standard aluminium profile making it an easy platform to customize. We modified this platform to elevate the working area in order to integrate the Geneva wheels and the working stations.

### 2.1.2.2 Geneva wheels

Link: https://github.com/croningp/dropfactory/blob/master/doc/geneva_wheel.md

A Geneva wheel is a mechanism that translates a continuous rotation into a discrete rotary motion. At each discrete wheel position, the wheel is locked into position mechanically - ensuring the location of each station on the wheel by design. A Geneva wheel is made of two parts: the drive wheel and the driven wheel. If the driven wheel has N slots, then one full rotation of the drive wheel advances the driven wheel by 360/N degree.

In our setup N=8 in order to have enough queueing and space to fit all working stations comfortably around the wheel. The drive wheel is rotated by a stepper motor and a mechanical switch is used to measure each full rotation of the drive wheel – corresponding to $1/8^{th}$ of a turn of the driven wheel. Both our Geneva wheels are based on the same design that was made modular to enable us to change only the top plate to tailor it to various vials and dishes. We designed our Geneva wheel system to be compact and compatible in its dimension with the modular aluminium profile system used as the backbone of Dropfacotry, that is by using dimension as multiple of 20mm.

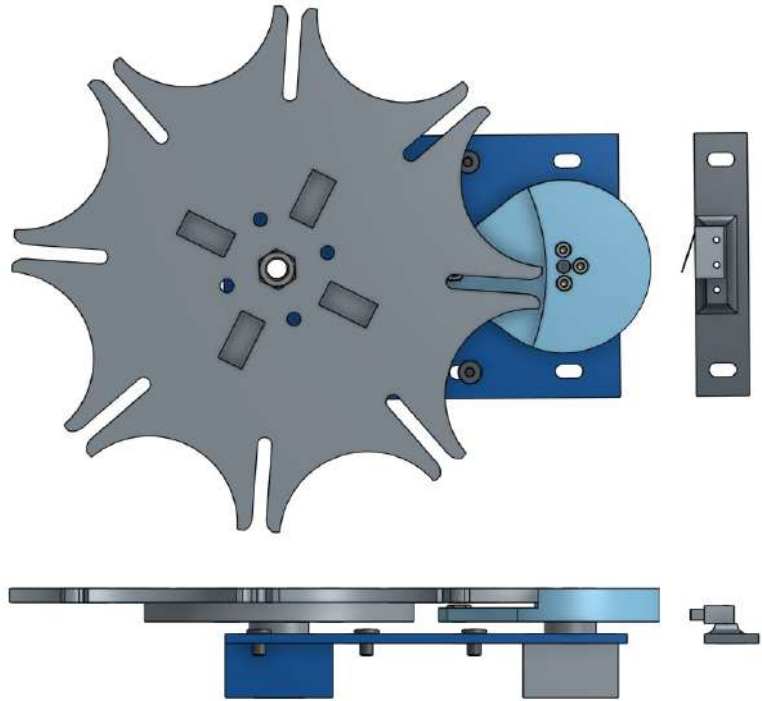

Supplementary Figure 50 - The onshape CAD design files for the drive (light blue) and driven (grey) wheels, the mounting plate (dark blue) and homing sensor (grey, right).

We designed two plateaus, one for handling the oil containers and one for holding the petri dishes.

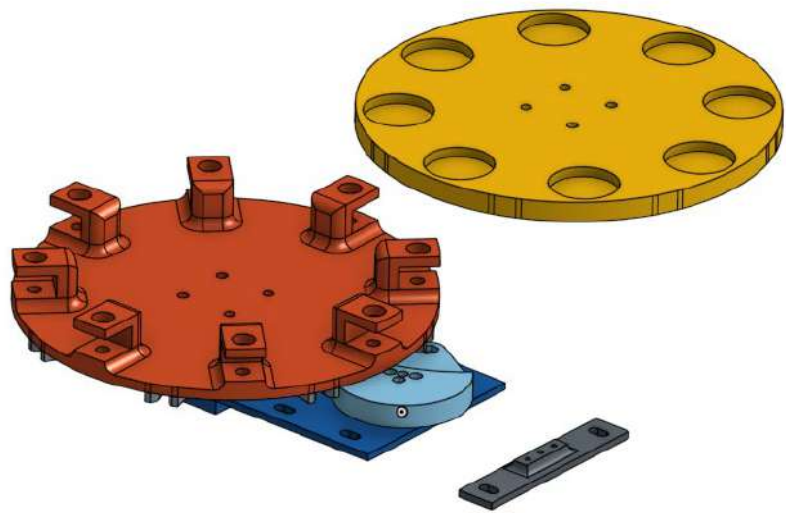

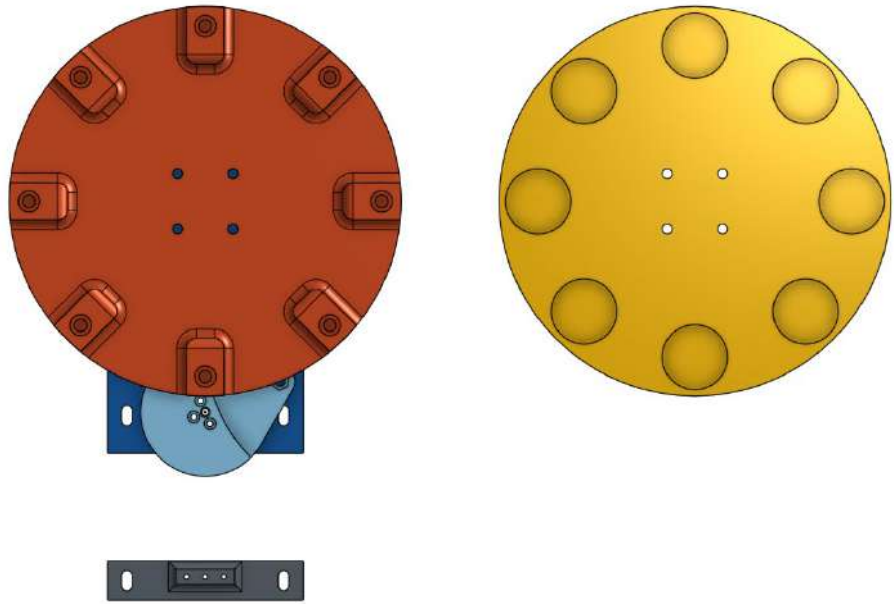

Supplementary Figure 51 - The onshape CAD design files for the two Geneva wheel top-plates used – one for the oil vials (orange) and one for the aqueous petri dishes (yellow).

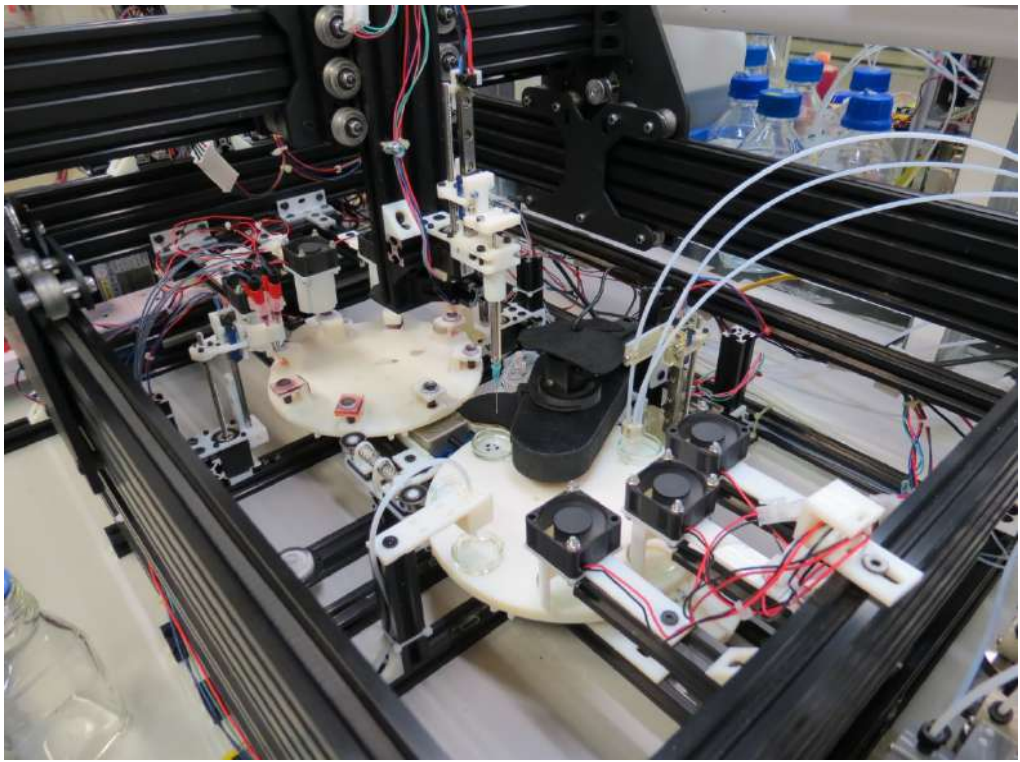

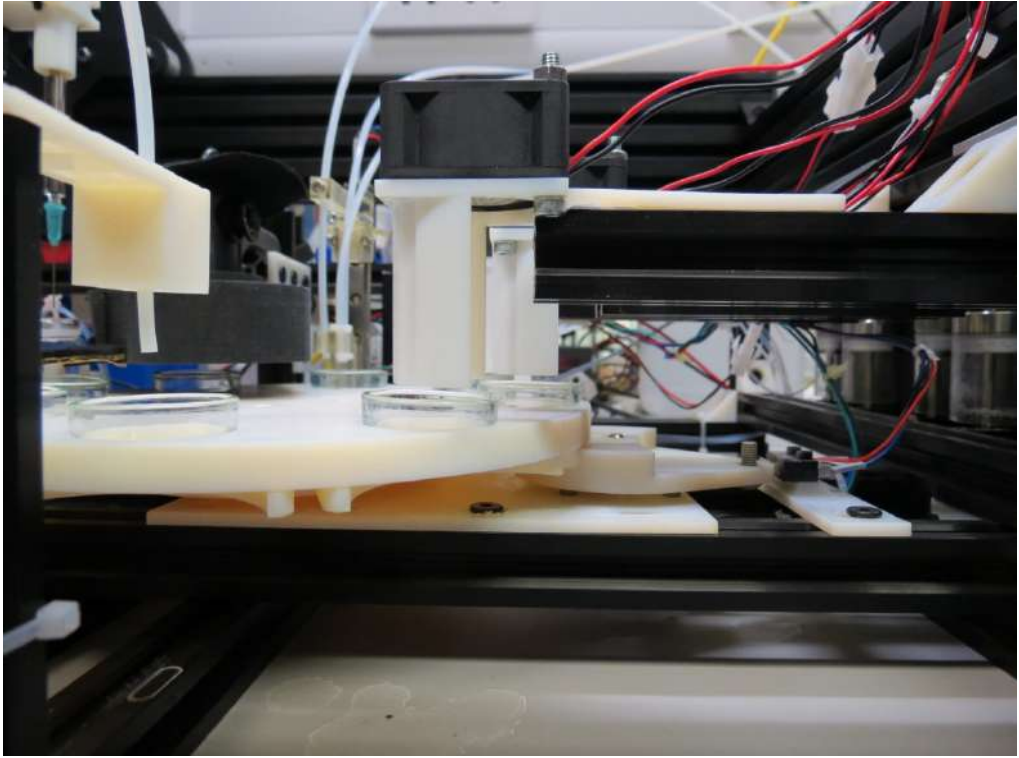

Supplementary Figure 52 – Photographs showing the two Geneva wheels in position on dropfactory

Finally, to improve the smoothness of the motion we designed a stabilizer that comes into contact with both driven wheels and dampens possible jerks in the wheel motion. The stabilizer is made of two bearings on a sliding axis that are pushed against one another and contact with the side of the Geneva wheels, providing a slight damping that is enough to stabilize the motion.

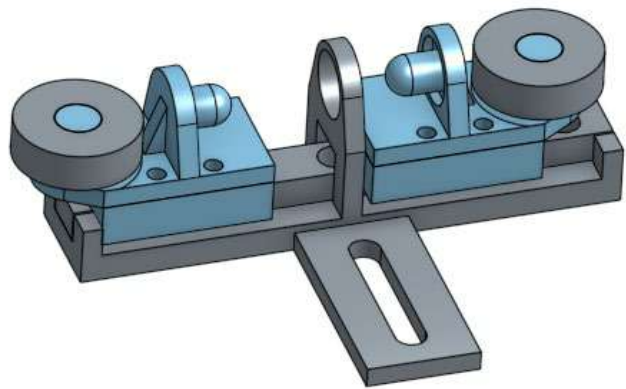

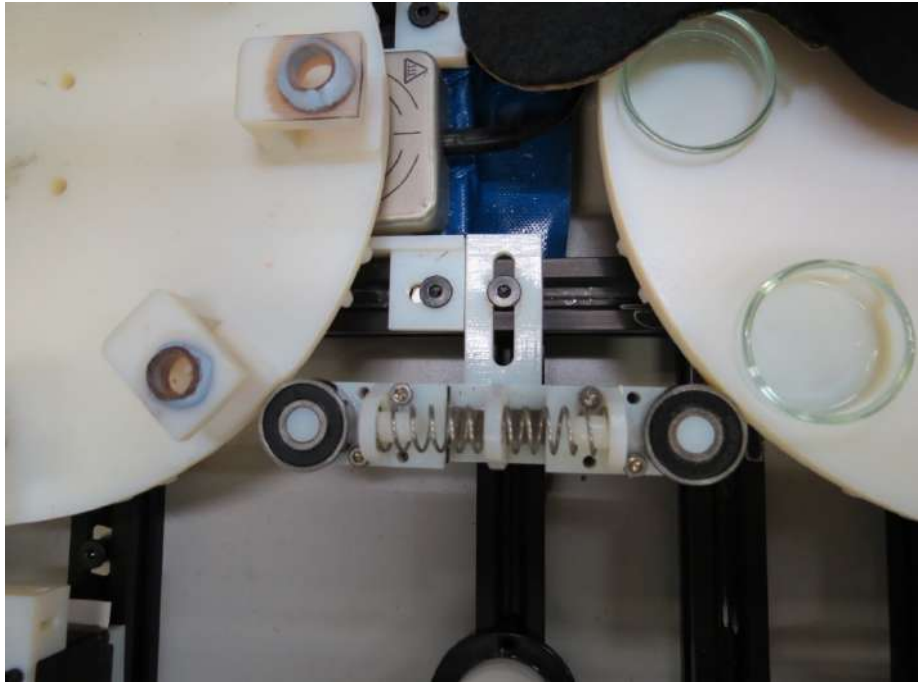

Supplementary Figure 53 - Onshape CAD design file (top) and photograph of the driven wheel stabilizer.

All the 3d stl files are located in the hardware/3d_parts/geneva_wheel folder, along with some visualisation of the parts:

https://github.com/croningp/dropfactory/tree/master/hardware/3d_parts/geneva_wheel

The CAD design files are available on Onshape:

1. Geneva Wheel:
   https://cad.onshape.com/documents/3aeb7616c1e547bfaae38ba3/w/426b95792e7c48a8b6dd7727/e/30b62a18352c4a91b6bc9828
2. Geneva Wheel Stabilizer:
   https://cad.onshape.com/documents/5789121ee4b07256e8184139/w/a0a9bcb1b97b6c43ac68f81e/e/801990910dc3689559c2009a

The code controlling the synchronous rotation of both wheels is in the software/robot/robot.py file, more specifically the *rotate_geneva_wheels()* function. See https://github.com/croningp/dropfactory/blob/master/software/robot/robot.py. That function first makes sure nothing is in the way of the Geneva wheels, then moves the driving stepper motor one full turn, which in turn produces a 1/8 rotation of the plateau. We ensure that the stepper actually does one turn by using a homing switch and raise an error if the stepper does not reach the switch within 30 seconds, indicating that the system got stuck.

### 2.1.2.3 Modular Linear Axis Driver

Link: https://github.com/croningp/dropfactory/blob/master/doc/modular_linear_actuator.md

A full documentation of our small modular linear actuator is available on the following repository: https://github.com/croningp/ModularSyringeDriver

The syringe and most working stations required a small linear axis to be controlled with great precision. The syringe needs to pump and deliver precise amounts, and other working stations need to plunge tubes into the various vials/dishes to extract and clean those containers. This actuator is used multiple times in Dropfactory for the syringe system, the oil filling station, oil cleaning station and the dish cleaning station.

Furthermore, a modular design should allow further applications of the driver in the future. To this end we designed a modular linear driver around a small NEMA8 stepper motor with a threaded rod axis. Both the fixed and moving parts of the module are equipped with a mechanical interface enabling customization of the linear axis to the desired task.

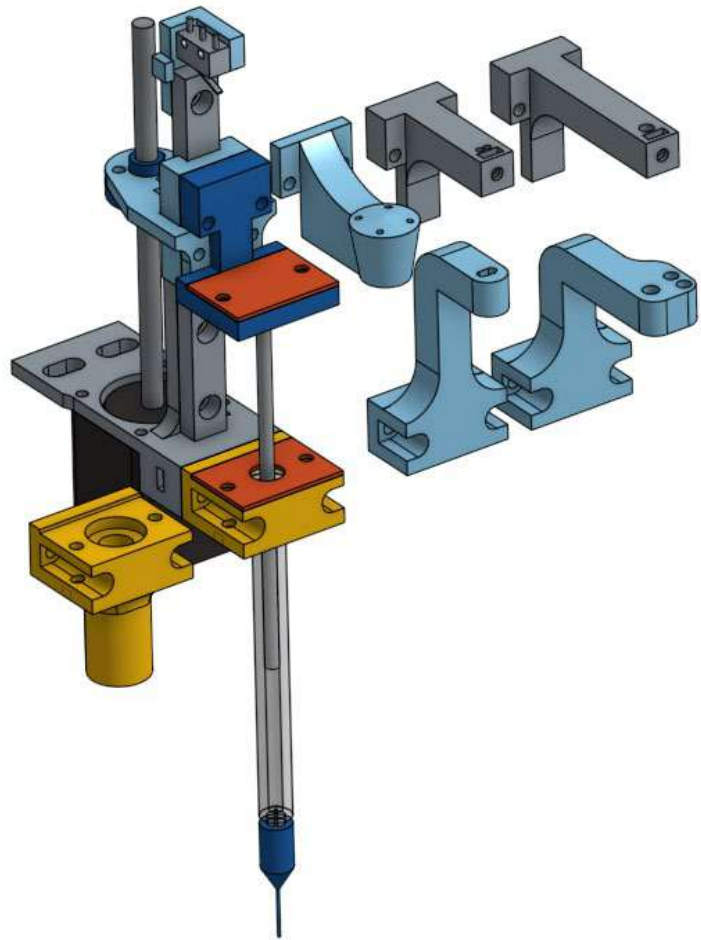

Supplementary Figure 54 - Onshape CAD design file for the small modular linear actuator used several times in the dropfactory platform. Several possible variations of tool attachments are shown.

A full BOM and assembly instructions are available at:

https://github.com/croningp/ModularSyringeDriver/blob/master/hardware/assembly.md

**2.1.2.4 Pumps**

The platform needs to handle various liquids such as oils, aqueous phases and acetone. We used Tricontinent C-Series Syringe Pumps. https://www.tricontinent.com/products/syringe-pumps-and-rotary-valves/c-series-syringe-pumps.html. Dropfactory uses 10 such pumps. Below is a top view of the pumps and the reagent containers:

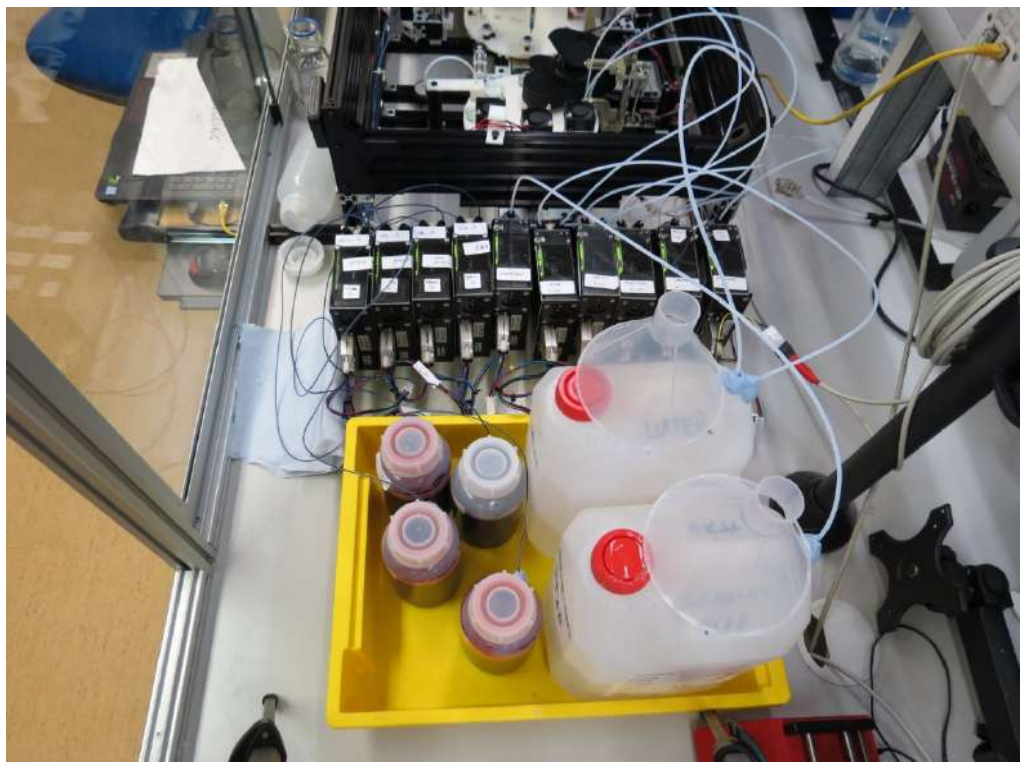

Supplementary Figure 55 – A photograph of the pumps and chemical inputs used for dropfactory.

Full documentation of our python library to control the pumps and how to wire them is available at: https://github.com/croningp/pycont

For Dropfactory, the pumps configuration and control are managed in the software/pump folder: https://github.com/croningp/dropfactory/tree/master/software/pump

We used the following tubing to connect the pumps to containers and dispensing units:

1. 1/8 inch outer diameter tubing and their fitting:
    - http://kinesis.co.uk/products/fittings-tubing/tubing/tubing-tubing-ptfe-1-8-3-2mm-od-x-1-5mm-id-10m-008t32-150-10.html
    - http://kinesis.co.uk/flangeless-fitting-for-1-8-od-tubing-1-4-28-flat-bottom-delrin-etfe-black-yellow-xp-308.html
2. 1/16 inch outer diameter tubing and their fitting:
    - http://kinesis.co.uk/products/fittings-tubing/tubing/tubing-tubing-ptfe-1-16-1-6mm-od-x-1-0mm-id-20m-008t16-100-20.html

o http://kinesis.co.uk/flangeless-fitting-for-1-16-od-tubing-1-4-28-flat-bottom-delrin-etfe-black-blue-xp-208.html

### 2.1.2.5 Oil wheel

The oil wheel has 4 distinct working station types:

1. A **filling station** where oils are mixed in different quantities:
   https://github.com/croningp/dropfactory/blob/master/doc/working_stations/oil_filling.md
2. A **stirring and sampling station** where a magnetic stirrer ensures proper mixing of the oil mixture and the syringe samples the oil mixture for making droplets:
   https://github.com/croningp/dropfactory/blob/master/doc/working_stations/oil_stirring.md
3. A **cleaning station** where oils are sent to waste and containers are cleaned with acetone and water:
   https://github.com/croningp/dropfactory/blob/master/doc/working_stations/oil_cleaning.md
4. Three **drying stations** which blow air on the containers and dry them from the acetone remaining after cleaning.

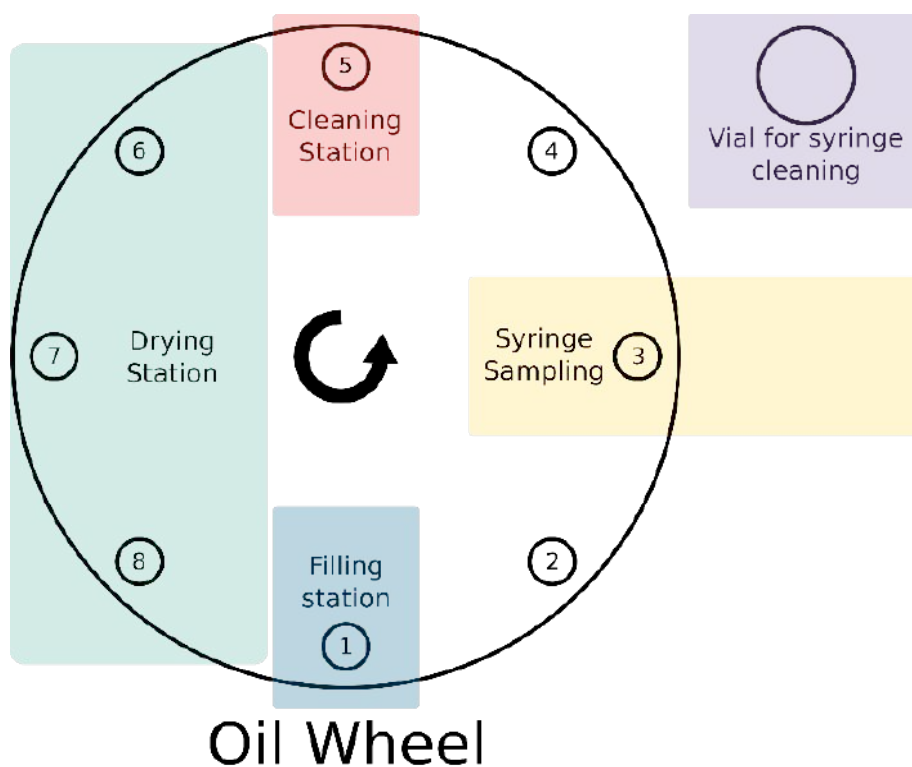

Supplementary Figure 56 - A schematic illustrating the stations located on the oil Geneva wheel.

### 2.1.2.6 Oil Filling station

Link:

https://github.com/croningp/dropfactory/blob/master/doc/working_stations/oil_filling.md

The oil filling station is designed to deliver the 4 oils into a small plastic container in the desired volumes. Because of the narrow diameter of the container, we use our modular linear axis to lower down the 4 oil lines into the container. Because a very small volume is being dispensed, and to increase reproducibility, after the oils have been dispensed the head is lowered down in order to contact the dispensing needle with the surface of the oil mixture to remove any pending droplets that often remain on the dispensing tips.

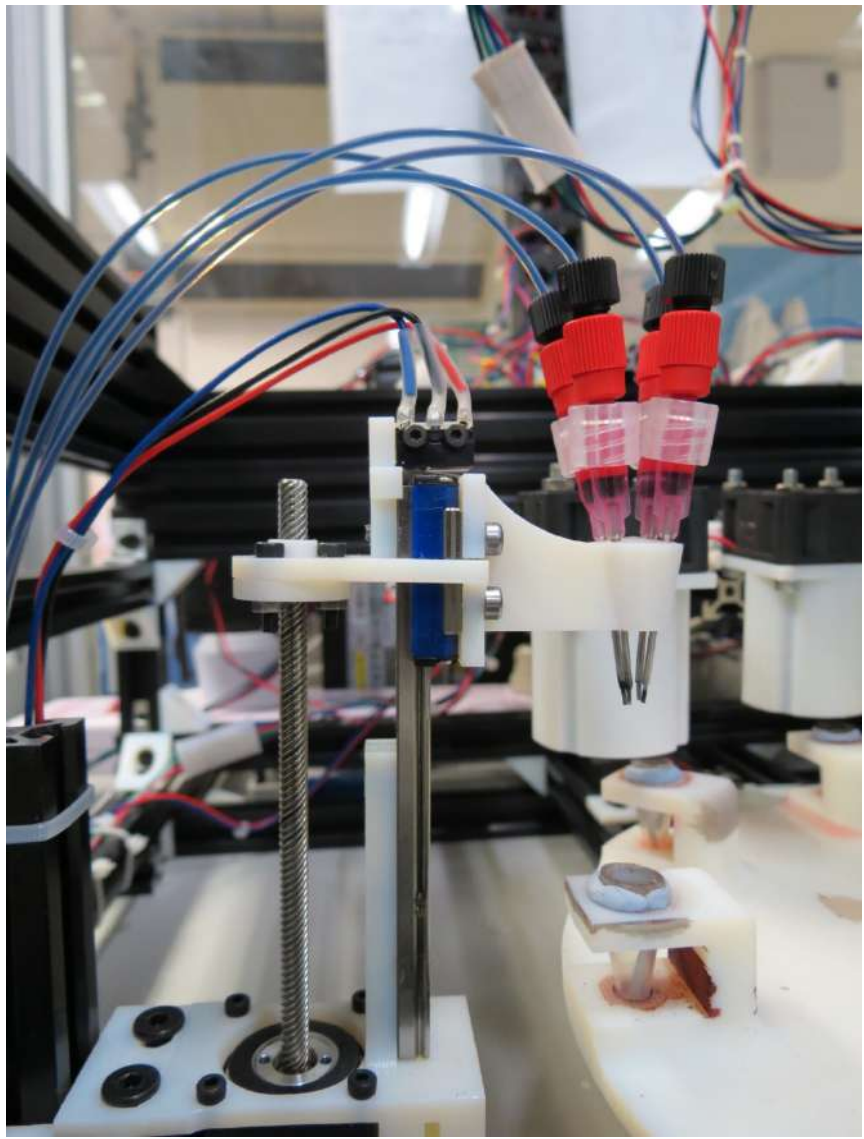

Supplementary Figure 57 – A photograph of the oil filling station. The modular linear axis driver is on the left and the oil wheel on the right.

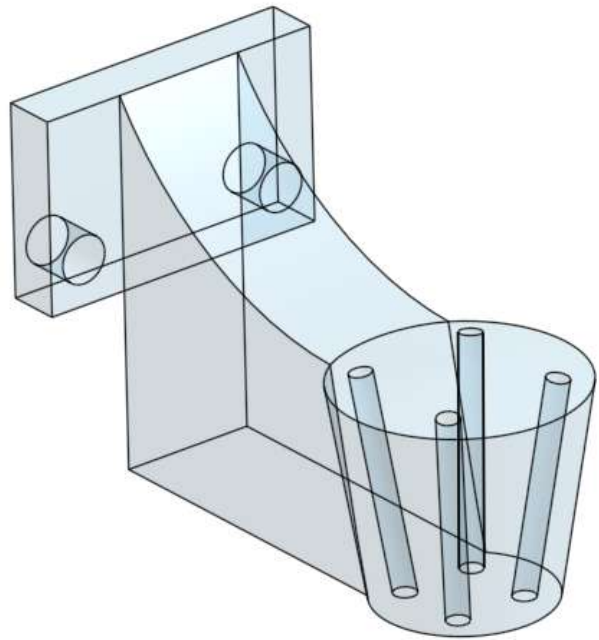

Supplementary Figure 58- The Onshape CAD design file for the modular linear axis driver adaptor for holding the four oil lines.

The filling procedure:

1. The head is lowered down so the tip of the dispensing needles come into the vial
2. The desired oils are dispensed with in the desired ratio to a total volume of 0.5mL
3. The head is lowered down so the tips of the dispensing needles come into contact with the surface of the oil mixture, this is to remove any pending drops remaining on the tips of the droplets
4. The head comes back to its home position

The 3D designs are available online:

1. STL Mount: https://github.com/croningp/dropfactory/blob/master/hardware/3d_parts/oil_filling/oil_filler.stl
2. Modular Actuator: https://github.com/croningp/ModularSyringeDriver
3. Onshape 3D model: https://cad.onshape.com/documents/62d832e8b2dc4f2c03b85d68/w/e45d0051d41b139c7004414d/e/1583bc5599c1a1019a2f3e93

The code managing the oil cleaning working station is in the software/working_station/fill_oil_tube.py file:

https://github.com/croningp/dropfactory/blob/master/software/working_station/fill_oil_tube.py

### 2.1.2.7 Oil Stirring Station

Link:

https://github.com/croningp/dropfactory/blob/master/doc/working_stations/oil_stirring.md

In each oil vial, there is a small magnetic stirrer bead of length 15mm and diameter 1.5mm. At the stirring station (position 3 of the oil wheel) a compact magnetic stirrer ensures that the oils are fully mixed before being sampled by the syringe. Such small magnetic stirrer plates tend to overheat, for this reason a fan has been positioned below the stirrer to regulate the stirrer temperature. The stirring efficiency was checked by adding a dye into some water while being mixed into the vials. In addition, when sampled, the syringe performs additional mechanical mixing by pumping in and out 5 times some oil mixture.

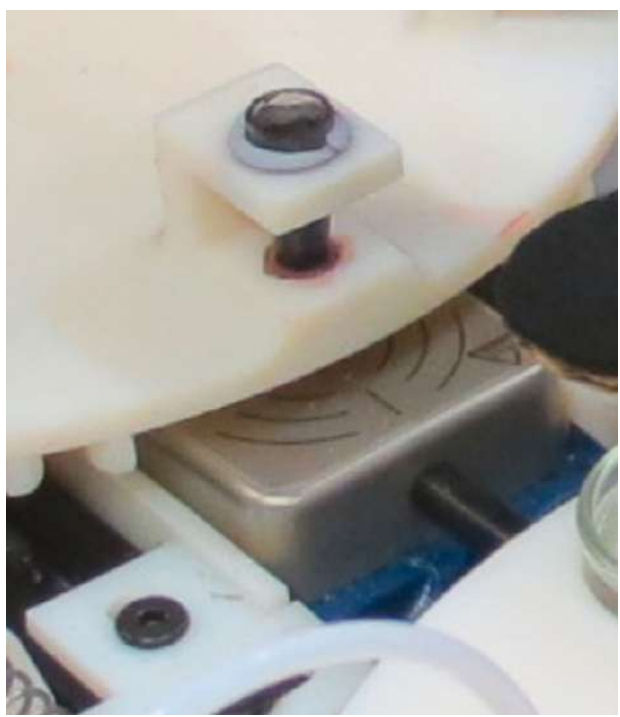

Supplementary Figure 59 - A photograph of the oil stirring station, with the small magnetic stirrer plate (silver) visible below the Geneva wheel.

Materials:

1. Micro stirrer: STIRRER MICRO 07 + UK TELEMODUL 7W, from H+P LABORTECHNIK. Supplied from VWR, catalog number is 442-3116: https://uk.vwr.com/store/product/442-3116/stirrer-micro-07-%2B-uk-telemodul-7w-1-1-items
2. Magnetic stirring bars, micro. Length: 15mm, Diameter: 1.5mm. Supplied from VWR, catalog number is 442-0367: https://uk.vwr.com/store/catalog/product.jsp?catalog_number=442-0367

The 3D design for the supporting mechanical parts can be found at:

1. STL Stirrer Holder:
   https://github.com/croningp/dropfactory/blob/master/hardware/3d_parts/oil_mixing/stirrer_holder.stl
2. STL Stirrer Fan Holder:
   https://github.com/croningp/dropfactory/blob/master/hardware/3d_parts/oil_mixing/stirrer_fan_holder.stl
3. Onshape 3D model:
   https://cad.onshape.com/documents/62d832e8b2dc4f2c03b85d68/w/e45d0051d41b139c7004414d/e/7f2bc6ac687a7f6977a3b478

### 2.1.2.8 Oil Cleaning station

Link:

https://github.com/croningp/dropfactory/blob/master/doc/working_stations/oil_cleaning.md

The cleaning station handles two tubes, one to dispense acetone and one to empty the vial contents to the waste. The waste tube needs to be dipped into the dish, for which we again use our modular linear actuator.

https://github.com/croningp/dropfactory/blob/master/doc/modular_linear_actuator.md

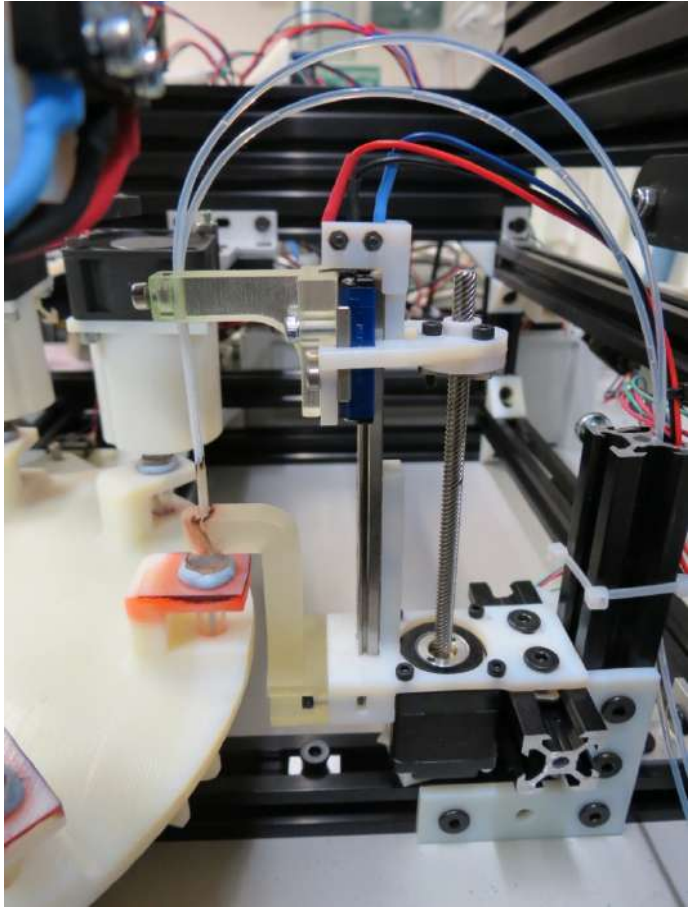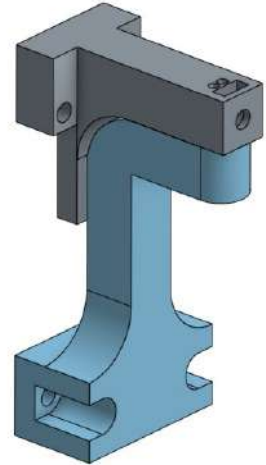

Supplementary Figure 60 - A photograph (left) of the oil cleaning station and the Onshape CAD design file (right) for the tube guide (blue) and mount (grey).

An oil vial is cleaned using the following protocol:

- pump vial contents to waste, 2mL (largely exceeding vial volume)
- repeat 5 times:
    - add 0.7mL of acetone into vial (exceeding oil volume but below total volume available)
    - pump vial content to waste, 2mL (largely exceeding vial volume)

The 3D designs can be found at:

1. STL Mount: https://github.com/croningp/dropfactory/blob/master/hardware/3d_parts/oil_cleaning/oil_cleaning_mount.stl
2. STL Guide: https://github.com/croningp/dropfactory/blob/master/hardware/3d_parts/oil_cleaning/oil_cleaning_guide.stl
3. Modular Actuator: https://github.com/croningp/ModularSyringeDriver
4. Onshape 3D model: https://cad.onshape.com/documents/62d832e8b2dc4f2c03b85d68/w/e45d0051d41b139c7004414d/e/ba40210f0cf61fe838ccdc8a

The code managing the oil cleaning working station is found at:

https://github.com/croningp/dropfactory/blob/master/software/working_station/clean_oil_parts.py

This file also manages the cleaning of the syringe.

### 2.1.2.9 Drying station

Link: https://github.com/croningp/dropfactory/blob/master/doc/working_stations/drying.md

Each drying station is a simple fan blowing air into a container that was previously cleaned with water (aqueous petri dish only) and acetone. We ensured that this additional air flow did not impact our droplet experiments via an airflow buffer. The drying stations are at the position 6, 7, and 8 of both the oil and aqueous wheels.

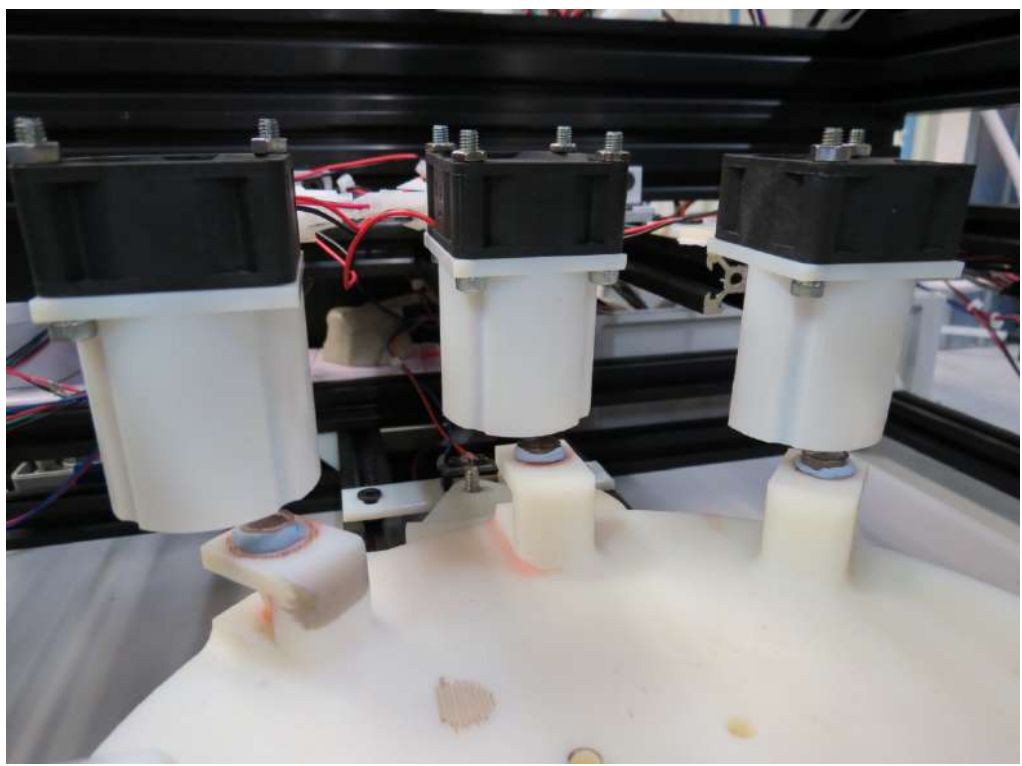

Supplementary Figure 61 - A photograph of the three oil drying stations.

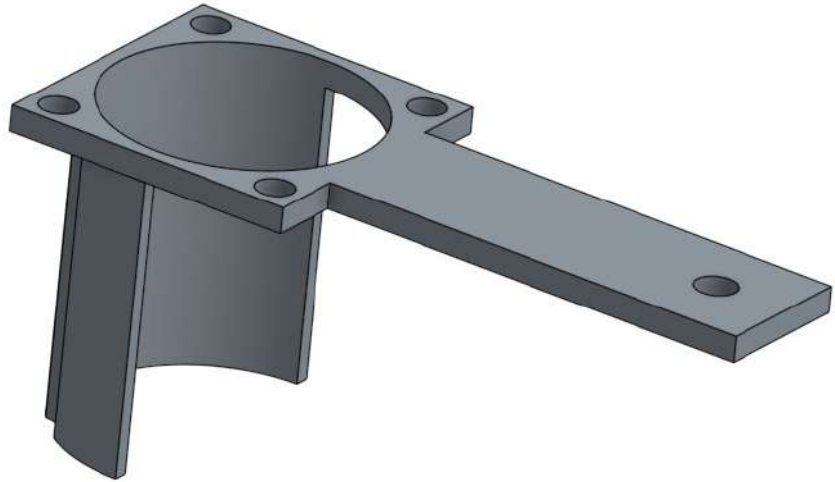

Supplementary Figure 62 - The Onshape CAD design file for the drying station airflow buffer. To the top of this is attached a fan.

The 3D files can be found at:

1. STL:
   https://github.com/croningp/dropfactory/blob/master/hardware/3d_parts/various/evaporator.stl
2. Onshape 3D model:
   https://cad.onshape.com/documents/62d832e8b2dc4f2c03b85d68/w/e45d0051d41b139c7004414d/e/ae1bb1b32d86c37772960515

### 2.1.2.10  Aqueous wheel

The aqueous wheel has 5 distinct working station types:

1. A **filling station** where the aqueous phase is poured into the dish:
   https://github.com/croningp/dropfactory/blob/master/doc/working_stations/dish_filling.md
2. A **droplet placement station** where the syringe places the droplets on the aqueous phase:
   https://github.com/croningp/dropfactory/blob/master/doc/working_stations/syringe.md
3. A **recording station** where droplets are being recorded via a webcam:
   https://github.com/croningp/dropfactory/blob/master/doc/working_stations/dish_recording.md
4. A **cleaning station** where the aqueous phase and droplets are sent to waste and containers are cleaned with acetone and water:
   https://github.com/croningp/dropfactory/blob/master/doc/working_stations/dish_cleaning.md

5. Three **drying stations** that blow air on the containers and dry them from the acetone remaining after cleaning:
   https://github.com/croningp/dropfactory/blob/master/doc/working_stations/drying.md

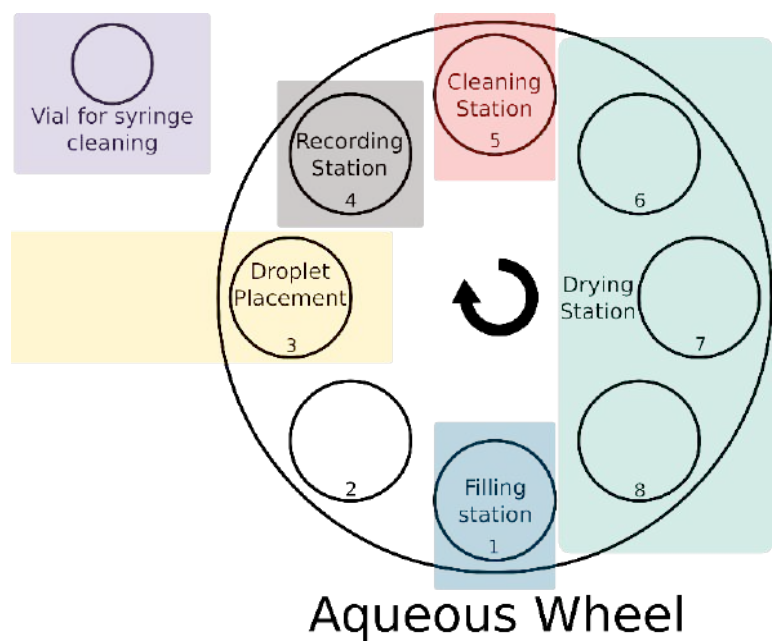

Supplementary Figure 63 - A schematic illustrating the stations located on the aqueous Geneva wheel.

### 2.1.2.11    Aqueous Filling station

Link:

https://github.com/croningp/dropfactory/blob/master/doc/working_stations/dish_filling.md

This station simply holds some tubes above the petri dish. In all of our experiments we used only one aqueous phase of TTAB at pH 13, but the holder has been designed to accommodate up to 9 tubes.

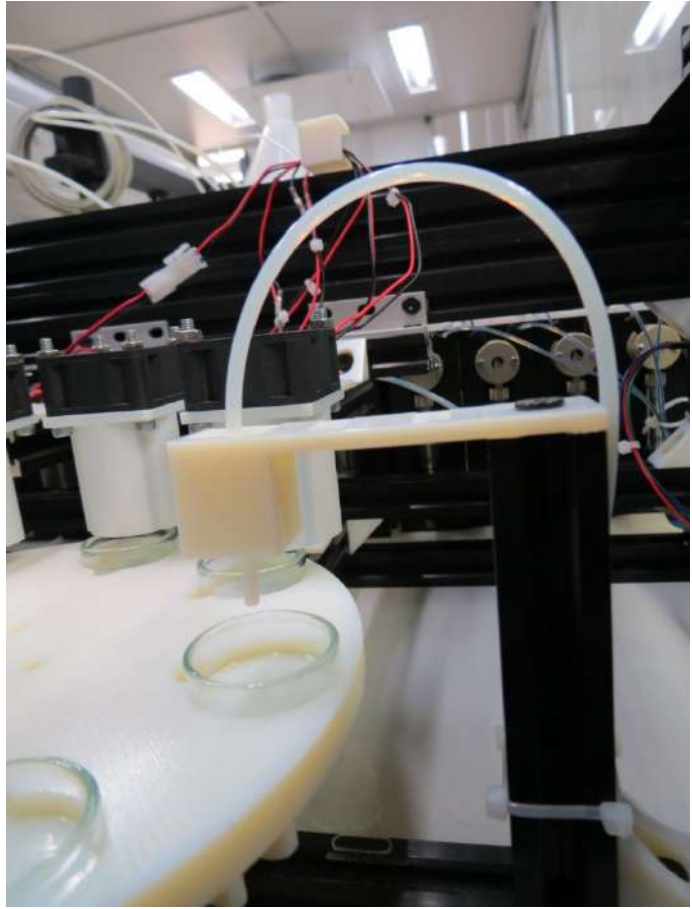

Supplementary Figure 64 - A photograph of the aqueous filling station.

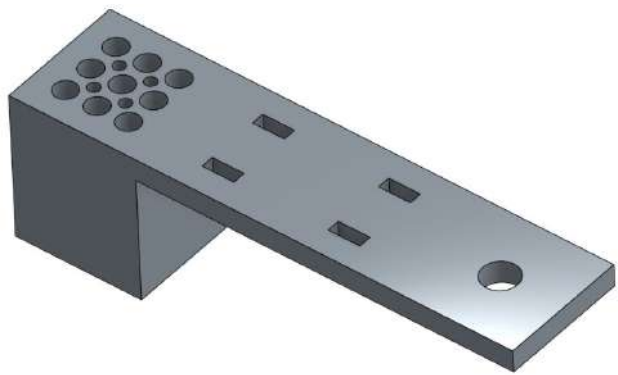

Supplementary Figure 65 - The Onshape CAD design file for the dish filling station tube guide.

The 3D design is available online:

1. STL:
https://github.com/croningp/dropfactory/blob/master/hardware/3d_parts/dish_filling/dish_filling.stl
2. Onshape 3D model:
https://cad.onshape.com/documents/62d832e8b2dc4f2c03b85d68/w/e45d0051d41b139c7004414d/e/02edab79fbbeda28022ade23

The code managing the dish filling working station is found at: https://github.com/croningp/dropfactory/blob/master/software/working_station/fill_petri_dish.py

This normalizes the quantity given in the experiment description file and fills the dish with the given ratios of each aqueous phase given their associated pump and for the defined volume.

### 2.1.2.12 Droplet placing station

Link: https://github.com/croningp/dropfactory/blob/master/doc/working_stations/syringe.md

The syringe is used for droplet placement, that is to move to collect the oil mixture, then move to the petri dish and place the droplets and finally cleans itself before moving to the next experiment. The experimental parameters are fully described in the experiment configuration file.

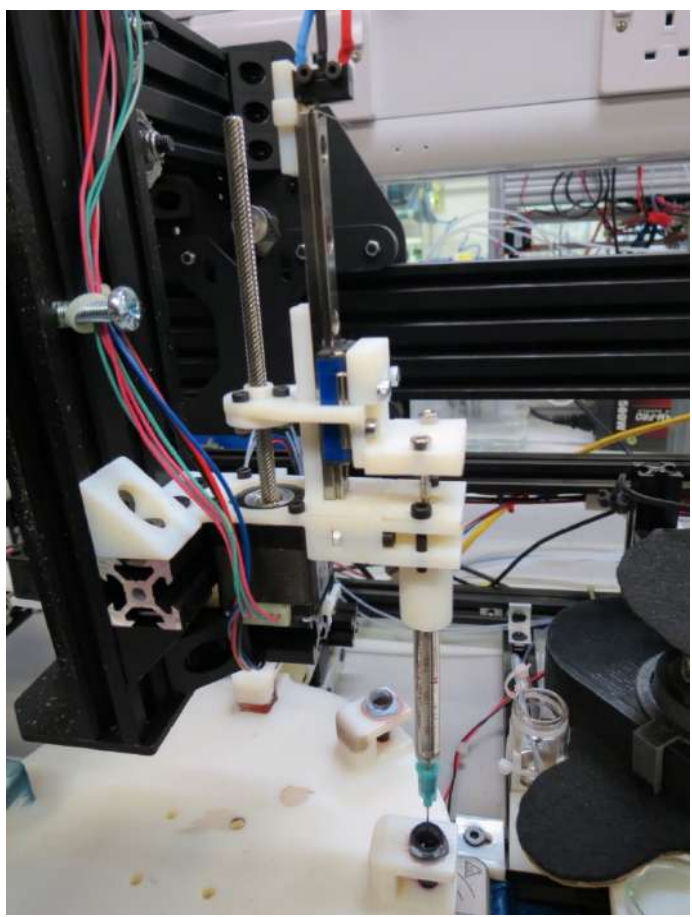

Supplementary Figure 66 - A photograph of the aqueous filling station.

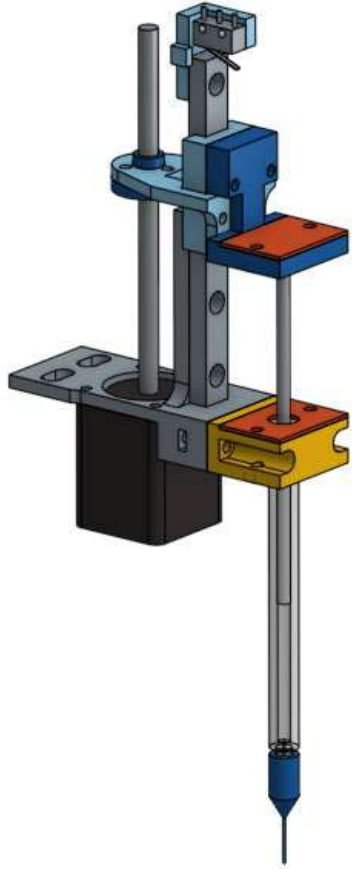

Supplementary Figure 67 - The Onshape CAD design file for the modular syringe driver.

The procedures for making droplets are as follows:

1. Syringe Oil Filling
    - move syringe on top of oil vial (position 3 of oil wheel)
    - move syringe down to dip needle into the oil
    - repeat 5x: pump and release 50μL of oil mixture, this is to ensure oils are mixed in addition to the action of the stirrer
    - pump 20uL more oil mixture than droplet making requires (total volume of requested droplets)
    - move syringe up
2. Droplet Making
    - move syringe on top of the centre of the petri dish (position 3 of aqueous wheel)
    - for each droplet defined in the config file:
    - move to given relative position, making sure it is not outside the petri dish
    - deliver given droplet volume, the droplet should not be released at this point but stick to the needle
    - move syringe down to touch the surface
    - move syringe up again above aqueous phase level
    - when all droplets are placed, move syringe up high above the dish
3. Syringe Cleaning

- o move syringe above of the dedicated syringe cleaning vial
- o empty syringe in the vial
- o wash the vial with 1.5 mL acetone
- o add 2.5mL acetone into vial
- o dip needle into acetone
- o repeat 3x: pump and deliver 100uL of acetone through the syringe, ensuring cleaning of remaining in oil mixture
- o move syringe up again above vial
- o empty vial of its contents
- o repeat 8x: pump and deliver 100uL of air through the syringe, ensuring drying of remaining acetone
- o wash the vial with 1 mL acetone

Our syringe driver is compatible with all µL scale Hamilton syringes: https://www.hamiltoncompany.com/products/syringes-and-needles/general-syringes/microliter-syringes/250-microL-Model-725-LT-SYR-NDL-Sold-Separately

We use dispensing needles from Weller reference KDS231P, see http://www.farnell.com/datasheets/514885.pdf

The 3D designs can be found at:

1. Modular Actuator: https://github.com/croningp/ModularSyringeDriver
2. STL Cleaning Vial Holder: https://github.com/croningp/dropfactory/blob/master/hardware/3d_parts/various/vial_holder.stl
3. Onshape Cleaning Vial Holder: https://cad.onshape.com/documents/62d832e8b2dc4f2c03b85d68/w/e45d0051d41b139c7004414d/e/640ac0deb1f80bf00c4bdb79

The code managing the syringes is split into two files:

1. *software/working_station/clean_oil_parts.py* for cleaning the syringes, this is because the pumps for acetone and waste are shared with the oil vial cleaning station. See https://github.com/croningp/dropfactory/blob/master/software/working_station/clean_oil_parts.py and https://github.com/croningp/dropfactory/blob/master/doc/working_stations/oil_cleaning.md
2. *software/working_station/make_droplets.py* that takes care of the sampling of oils and the droplet placement. See https://github.com/croningp/dropfactory/blob/master/software/working_station/make_droplets.py

## 2.1.2.13 Recording station

Link: https://github.com/croningp/dropfactory/blob/master/doc/working_stations/dish_recording.md

Once the droplets are placed on the aqueous phase, the dish is moved under a camera (position 4 of the aqueous wheel) for the droplet behaviours to be recorded and saved as a video file.

We use a simple webcam for recording and we need to ensure that the lighting conditions are very similar between each experiment. Especially, it is important to remove all light reflection on the water surface that hinders the performance of image analysis of the droplets.

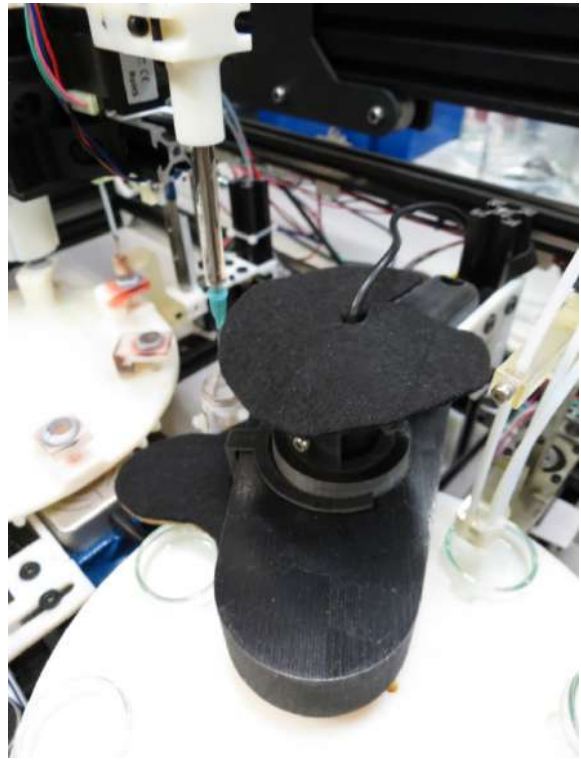

Supplementary Figure 68 - A photograph of the recording station

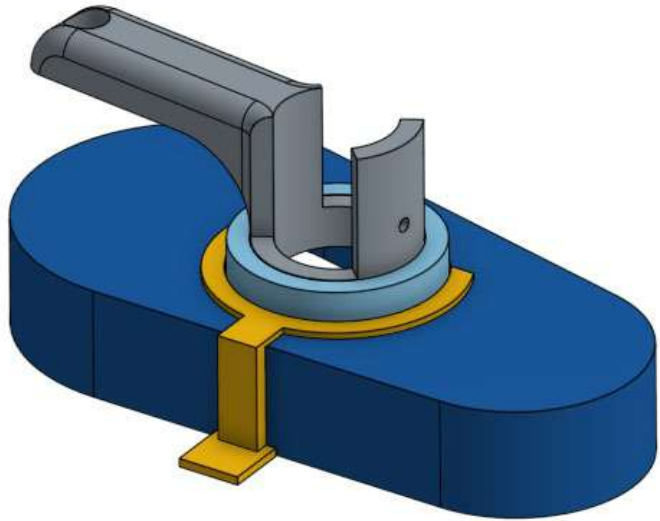

Supplementary Figure 69 - The 3D design of the recording station

The webcam we use is the Microsoft 6CH-00002: https://www.microsoft.com/accessories/en-gb/business/lifecam-cinema-for-business/6ch-00002

The 3D designs can be found at:

1. All STL files in the hardware/3d_parts/camera_holder folder: https://github.com/croningp/dropfactory/blob/master/hardware/3d_parts/camera_holder
2. Onshape 3D model: https://cad.onshape.com/documents/62d832e8b2dc4f2c03b85d68/w/e45d0051d41b139c7004414d/e/f1aad30ed184d979bb4387d0

The code managing the camera working station in *software/working_station/record_video.py*. This simply triggers the recording of a video of given duration into a specified file. See https://github.com/croningp/dropfactory/blob/master/software/working_station/record_video.py

It utilizes tools from our chemobot_tools library (https://github.com/croningp/chemobot_tools), and is interfaced and configured in the software/webcam folder, see

https://github.com/croningp/dropfactory/blob/master/software/webcam

### 2.1.2.14 Aqueous Cleaning Station

Link:

https://github.com/croningp/dropfactory/blob/master/doc/working_stations/dish_cleaning.md

This station handles three tubes, one to dispense water, one to dispense acetone, and one to empty the dish content to the waste. This later needs to be dipped into the dish, we use our modular linear actuator (https://github.com/croningp/dropfactory/blob/master/doc/modular_linear_actuator.md) for the up and down motion.

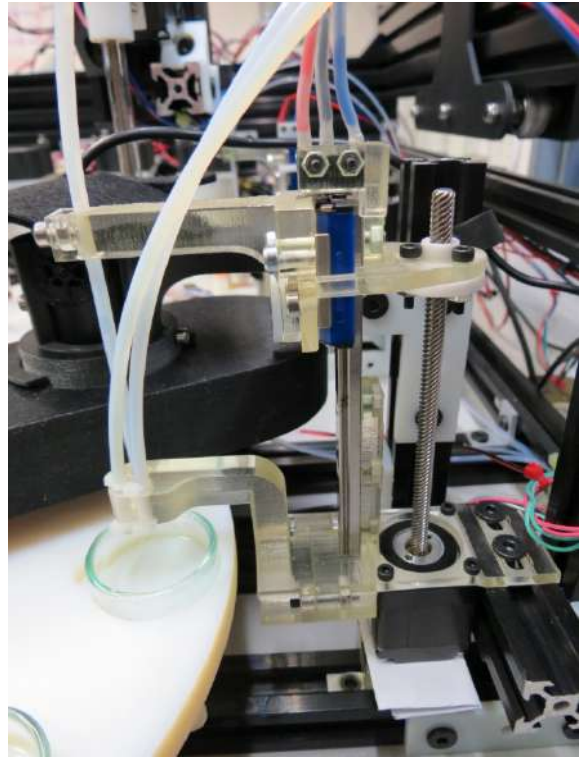

Supplementary Figure 70 - A photograph of the dish cleaning station. Two fixed tubes provide the water and acetone and one tube is mounted on an actuated z-axis for the waste removal of the dish content.

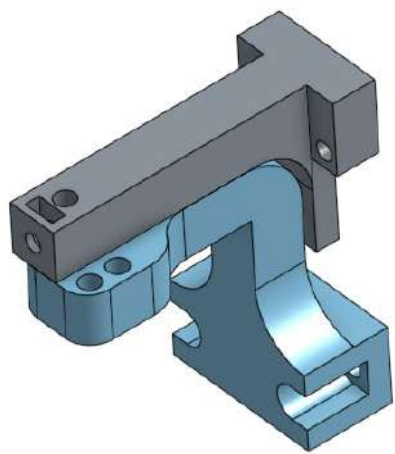

Supplementary Figure 71 - 3D design of the dish cleaning modular extensions to the modular syringe

The petri dish cleaning procedure is as follows:

1. empty petri dish of previous experiment's content

2. add 4.5mL of acetone into petri dish
3. empty petri dish
4. add 4.5mL of water into petri dish
5. empty petri dish
6. add 4.5mL of acetone into petri dish
7. empty petri dish
8. add 4.5mL of acetone into petri dish
9. empty petri dish

The 3D designs can be found here:

1. STL Mount: https://github.com/croningp/dropfactory/blob/master/hardware/3d_parts/dish_cleaning/dish_cleaning_mount.stl
2. STL Guide: https://github.com/croningp/dropfactory/blob/master/hardware/3d_parts/dish_cleaning/dish_cleaning_guide.stl
3. Modular Actuator: https://github.com/croningp/ModularSyringeDriver
4. Onshape 3D model: https://cad.onshape.com/documents/62d832e8b2dc4f2c03b85d68/w/e45d0051d41b139c7004414d/e/d76ad1c6bf725a9f379d21d0

The code managing the dish cleaning working station is here:

*software/working_station/clean_petri_dish.py*. See

https://github.com/croningp/dropfactory/blob/master/software/working_station/clean_petri_dish.py

### 2.1.2.15 Drying stations

Link: https://github.com/croningp/dropfactory/blob/master/doc/working_stations/drying.md

The drying stations for the aqueous wheel are the same as for the oil wheel.

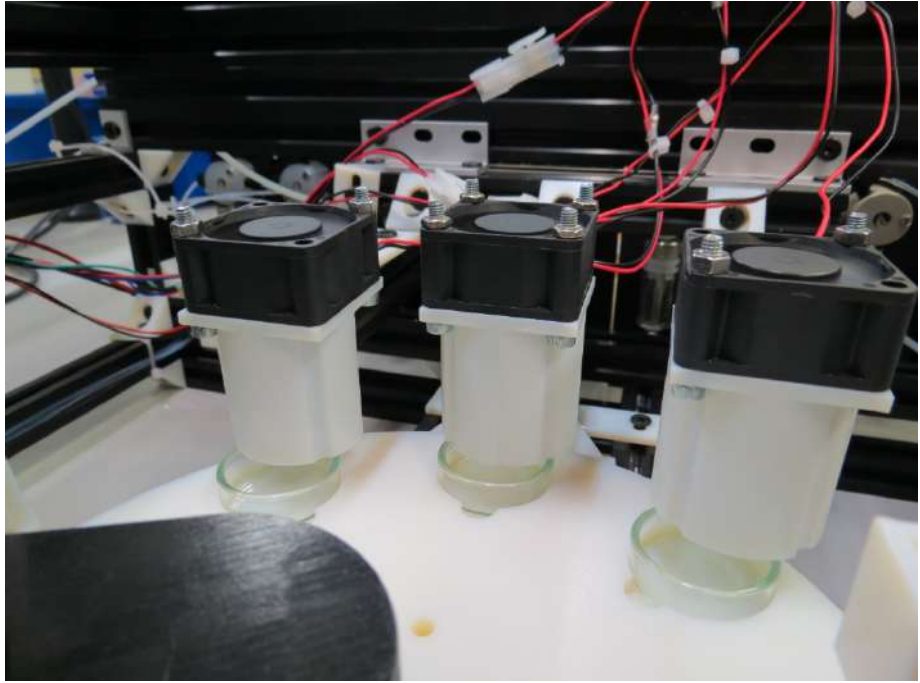

Supplementary Figure 72 - A photograph of the drying station. Each station is composed of a fan with a simple guide to canalize the flow of air to the dish.

### 2.1.2.16 Environment Monitoring

The temperature of the room was controlled via the air-conditioning unit fitted in the room, with the settings tailored to reduce temperature variation. The temperature and humidity were monitored using a SHT15 SparkFun Humidity and Temperature Sensor. It is available here: https://www.sparkfun.com/products/13683

The platform stores the temperature and humidity values in a *run_info.json* file in the experiment folder at the start of each experiment.

### 2.1.3 Software Control

The platform control, algorithms, and image analysis were all implemented in Python 2.7. Aside from the standard libraries, we are using the following libraries:

1. opencv: Image analysis with python binding. Version: cv2.version is '2.4.8'. http://opencv.org/
2. numpy: Scientific computing in Python. Version: numpy.version is '1.10.4'. http://www.numpy.org/
3. scipy: More scientific computing in Python. Version: scipy.version is '0.16.1'. http://www.scipy.org/scipylib/index.html
4. sklearn: Machine Learning in Python. Version: sklearn.version is '0.16.1'. http://scikit-learn.org/

5. seaborn: Statistical data visualization using matplotlib. Version: seaborn.version is '0.7.0'. https://github.com/mwaskom/seaborn
6. explauto: a library that implements the CA algorithm. The original repository is here: https://github.com/flowersteam/explauto and we modified it for the purpose of integrating it on Dropfactory: https://github.com/jgrizou/explauto
7. filetools: is python library to help handle files: https://github.com/jgrizou/filetools

### 2.1.3.1 Principles

A full experimental run of 1000 experiments require three steps:

1. **Perform experiments on the droplet system**. This task is given to the Dropfactory platform. Dropfactory has only one role, it looks for experiments (in the form of a *params.jon* file) to run in a specific folder and executes them. It then saves a video of the experiment as a *video.avi* file and an information file in the form of a *run_info.json* file with time, temperature and humidity information. All the code and information to run Dropfactory are available online: https://github.com/croningp/dropfactory
2. **Analyse the results of the experiments**. This task is given to a specific, standalone and independent software running in its own thread. This process looks for droplet videos (in the form of *video.avi* files) in a specific folder (typically generated by the platform once an experiment is finalized) and process the file in three steps. It first extract frame by frame the contour of each droplets, then attempt to stitch together the droplet in each frame as time sequences and finally computes a bunch of metrics about the droplet motion in the form of a *features.json* file. See section 2.1.5 title "Droplet Tracking" of this document. See also https://github.com/croningp/chemobot_tools and https://github.com/croningp/dropfactory_exploration
3. **Decide on what experiment to do next.** This task is given to a specific, standalone and independent software running in its own thread. This process looks for droplet features files (typically generated by the video droplet tracking in the form of a *features.json* file) and an experimental parameter files in a specific folder. It gathers all this information and, depending on the algorithm selected, output a new experimental file (*params.json* file) for the platform to execute. See https://github.com/croningp/dropfactory_exploration

The all process repeats until 1000 experiments are performed. In the following we details how each step of this process are implemented.

### 2.1.3.2 Running Experiments on Dropfactory

The code in the software folder of the dropfactory repository (https://github.com/croningp/dropfactory/tree/master/software) implements and orchestrates all the working stations (that are specialized modules implemented as threads and able to perform one simple task well) into a fully functional platform able to accept experimental files and execute them in a parallel fashion.

The *manager.py* file (https://github.com/croningp/dropfactory/blob/master/software/manager.py) is the entry point and only file to import for using Dropfactory. It contains an *add_XP(XP_dict)* function that adds an experimental configuration (*XP_dict*) to the manager. The *XP_dict* can be created as showed in *xp_maker.py* (https://github.com/croningp/dropfactory/blob/master/software/tools/xp_maker.py).

An experiment is fully described by a json file with the following fields. Note that there are helper tools to build such a file in *xp_maker.py*. Below is an example experimental file fully documented and explaining each field.

```
EXAMPLE_XP_DICT = {
    # Dropfactory outputs some information about the experimental conditions, such as the time of the day
    # it was run, the temperature, the humidity. The 'run_info' field tell the platform where to save
    # that information for this particular experiment. If the experiment video will be stored place in
    # the "xp_folder" folder, a good practice is to save it at the same place.
    # By convention we use RUN_INFO_FILENAME = 'run_info.json' (see software/tools/filenaming.py)
    'run_info': {
        'filename': os.path.join(xp_folder, RUN_INFO_FILENAME)
    },
    # 'min_waiting_time' is the minimum time a dish should stay at any station,
    # this is to ensure proper drying at the drying stations.
    'min_waiting_time': 60,  # in seconds
    # 'video_info' tells the platform how long the record an experiment for and where to save that video.
    # As with he 'run_info' field, it is a good practice is to save it at the same place.
    # By convention we use VIDEO_FILENAME = 'video.avi' (see software/tools/filenaming.py)
    'video_info': {
        'filename': os.path.join(xp_folder, VIDEO_FILENAME),
        'duration': 90  # in seconds
    },
    # 'arena_type' tell what type of dish the experiment should be using. Dish should be changed manually,
    # only one dish type can be present at the same time on the platform and the ARENA_TYPE field should
    # be changed accordingly in software/constants.py. This field is mostly a security/memory field,
    # we never used other dishes that a plain glass petri_dish.
    'arena_type': 'petri_dish',
    # 'oil_formulation' describe the composition of the oil droplets.
    # The number will be normalized to sum to 1.0.
    # The association between the compounds and the associated pumps is defined in software/constants.py.
```

```
        # Changes should be reported there accordingly.
        'oil_formulation': {
            'dep': 0.36,
            'octanol': 0.29,
            'octanoic': 0.0,
            'pentanol': 0.33
        },
        ## 'surfactant_volume' how much aqueous phase to pour in the dish
        'surfactant_volume': 3.5,  # in mL
        # 'surfactant_formulation' is similar 'oil_formulation' but for the aqueous phase,
        # which can be a mixture of multiple aqueous phases. The number will be normalized to sum to 1.0.
        # As for oils, the association between the compounds and the associated pumps is defined
        # in software/constants.py. Changes should be reported there accordingly.
        'surfactant_formulation': {
            'TTAB': 1.0
        },
        # 'droplets' is the placement information for droplet, it is a list where each elements
        # corresponds to one droplet. Each droplets is then described by its 'volume' (in uL) and
        # 'position' (in mm relative to the center of the dish). Here we have 4 droplets,
        # one at the center and three equally spread around on a circle of radius 5mm.
        # DEFAULT_DROPLET_VOLUME = 4 uL.
        'droplets': [
            {
                'volume': DEFAULT_DROPLET_VOLUME, # in uL
                'position': [0, 0] # relative position in mm from the dish center
            },
            {
                'volume': DEFAULT_DROPLET_VOLUME,
                'position': [-5, 0]
            },
            {
                'volume': DEFAULT_DROPLET_VOLUME,
                'position': [2.5, 4.33]
            },
            {
                'volume': DEFAULT_DROPLET_VOLUME,
                'position': [2.5, -4.33]
            }
        ]
}
```

Dropfactory is able to read these files and automatically execute the corresponding experiments. The code is segmented by functionalities as follows:

1. *software/arduino* (https://github.com/croningp/dropfactory/blob/master/software/arduino) holds the firmware for the two arduino boards that are used to control the entirety of the platform. It is based on our Arduino-CommandTools that allows to quickly and flexibly prototype Arduino based robots.

2. *software/pump*  (https://github.com/croningp/dropfactory/blob/master/software/pump) holds the pump configurations for the 10 Tricontinent C3000 pumps used to handle liquids for droplet experiments. That is oils and aqueous phases + waste management + cleaning liquids (acetone and water). It utilises our easy to use pycont python library.
3. *software/robot*  (https://github.com/croningp/dropfactory/blob/master/software/robot) contains all the utilities to actuate the platform, such as rotating the geneva wheels or precisely pumping and delivering liquids via our syringe systems. It is based on our commanduino tool-kit that allows to quickly and flexibly control Arduino based robots through Python.
4. *software/tools*  (https://github.com/croningp/dropfactory/blob/master/software/tools) holds various tools used to manage and organize dropfactory. The most important file is xp_manager.py that orchestrates the parallelized operation of the robot.
5. *software/webcam* (https://github.com/croningp/dropfactory/blob/master/software/webcam) contains the camera configuration for the MICROSOFT 6CH-00002 we use to video record the droplets. It is based on our *chemobot_tools* library used to detect and analyse droplets.
6. *software/working_station* (https://github.com/croningp/dropfactory/blob/master/software/working_station) contains all the individuals working station that fulfil a single task such as cleaning the oil containers, or placing droplet with the syringe. Those stations are implemented as threads and orchestrated by the *xp_manager.py* in the tools folder.

Finally, the remaining files are helpers used while developing the platform and testing all individual step of the processes.

## 2.1.4 Droplet Placement

Each experiment consisted of 4 droplets of 4μL placed in a Y pattern at the centre of a 32mm petri dish. The first droplet was placed at the centre of the dish, the three next droplets were placed on a circle of radius 5mm and at equally distributed along the circle by 120-degree angles. The relative coordinate from the centre of the petri dish, in mm and in the [x, y] coordinate of the Dropfactory were as follow: [0,0], [-5,0], [2.5, 4.33], [2.5, -4.33] as described in the example experiment json file described in the previous section 2.1.3.2.

Interestingly, we observed that the placement of the droplets had an influence on the droplet behaviour in the petri dish. During our first test of experiments (not reported and not used in this work), we used a 5 x 5 mm square pattern placement and noticed that the initial droplet placement was influencing the movement of the other three droplet later deposited on the aqueous phase. Supplementary Figure 72 shows both droplet placement patterns.

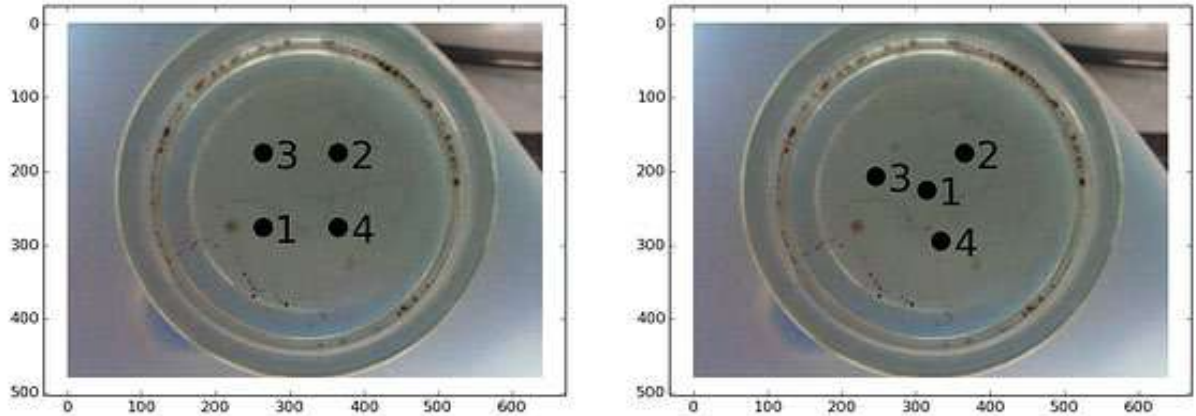

Supplementary Figure 73 - Visual representation of both droplet placement considered. On the left-hand side is the square pattern placement and right-hand side is the Y pattern placement. The black dots represent the position of the droplet and the associated number the ordering of the placement.

We analysed this effect by comparing a very large number of random experiments performed with the two patterns of droplet placement, respectively 2 runs of 1000 experiments for the square pattern placement, and 3 runs of 1000 experiments for the Y pattern placement used in this work.

For each droplet experiment, we extracted the position of the droplets through time and averaged the position of all droplets at each time step over each set of 1000 videos available for each placement condition. The result is 5 averaged trajectories of droplets, each representing 1000 experiments, 2 with a square pattern placement and 3 with a Y pattern placement. The logic behind this measure is that, because of the very large number of experiments, we should expect the average trajectories to cancelling one another and be stationary at the centre of the petri dish. If any bias is influencing the droplet motion, then the average trajectory would be biased too.

Supplementary Figure 73 shows these average trajectories. Each cross indicates the first frame, that is the start of a video, and the colours differentiate each experimental run. There is a clear trend with the square pattern droplet placement, the droplets are on average dragged towards the first droplet placed which introduces an important bias in the overall trajectory of the droplets. This discovery led us to use the Y pattern placement whose first droplet is placed in the centre of the dish which removes all bias on the droplet trajectories.

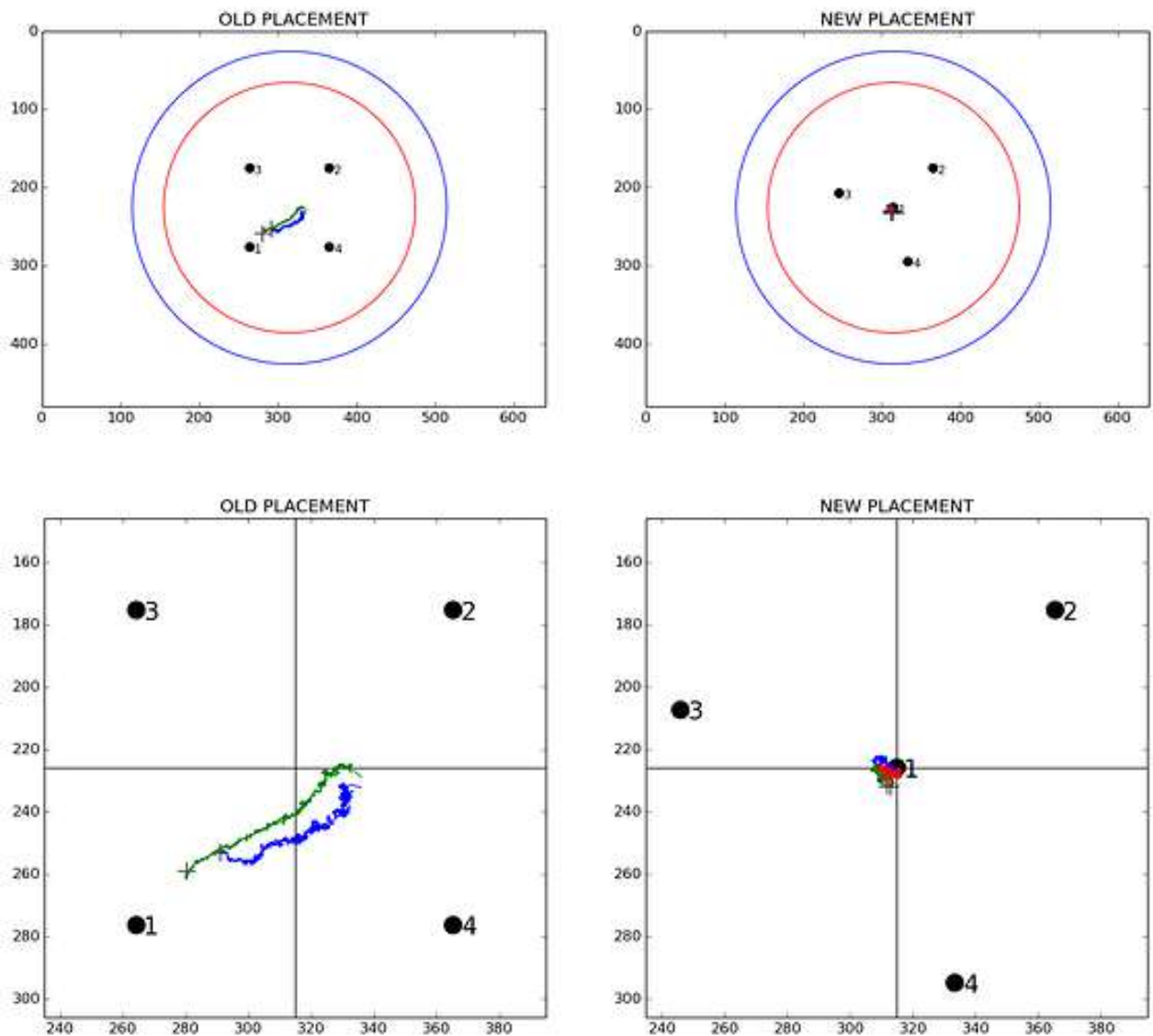

Supplementary Figure 74 - Average position of droplets in time for each run of 1000 experiments depending on droplet placement pattern. Left-hand side is the square pattern placement and right-hand side is the Y pattern placement. The black dot represents the position of the droplet and the associated number the ordering of the placement. The coloured trajectories are the averaged positions of droplets per frame over 1000 experiments. The + symbol is the starting position, i.e. the average position at the first frame of all videos. The placement of the first droplet does influence the initial position of the other droplets which seem to be attracted towards the first droplet placed, as seen on the left side of the figure where the + are not at the centre of the dish but closer to the droplet number 1. We remind that the placement is done a few seconds before the video recording starts. This discovery justified changing the droplet placement to the Y placement with the first droplet in the middle of the dish.

We hypothesise that the placement of the first droplet in a fresh aqueous phase results in the instantaneous release of oil on the surface of the aqueous phase which establishes a concentration gradient originating at the first droplets' initial position. This is supported by the observation that the first oil droplet often undergoes an initial high activity phase, lasting ca. 1 second, after it is placed. The first oil droplet placed can also usually be identified by viewing the recorded video, as it is often smaller and displays slightly different behavioural characteristics, which can also be observed in the 15-minute droplet displacement data. It has been shown that droplets are capable of chemotaxis due to surface tension imbalances [25,28] and we hypothesize that a similar phenomenon is responsible for the effect described above. Further analysis would be required to understand and validate the phenomenon and was not within the scope of this research.

The code used to generate this analysis can be found at: https://github.com/croningp/dropfactory_analysis/tree/master/analysis/drift_vs_droplet_placement

### 2.1.5 Droplet Tracking

The raw output of a droplet experiment is a video of black droplet moving in front of a white background. A video is of size 640x480 pixel and is recorded at 20fps. To extract information about the droplet behaviour we implemented a custom image tracking algorithm that comprises 4 main stages:

1. First, we **detect the location of the petri dish** in the video and define an area of interest for the tracking. This detection is done using the Hough Transform[41] and leverage information known about the petri dish used. Only droplets within that area of interest will be considered for analysis.
2. Second, we **extract the location and contour of each droplet** for each frame of the video and within the tracking area defined in the previous stage. This detection is done using a thresholding algorithm on the smoothed grey scale image of the droplet.
3. Third, given the position of each droplet through each frame of the video, we combine this information to **identify, tag, and extract droplet trajectories through time**. This is done using a proximity rule, droplets in subsequent frame that are close to one another are considered to be part of the same trajectory within some set constraints.
4. Fourth, given both the frame by frame information on droplet location and time sequences of individual droplets we **compute a set of averaged metrics** on the droplet behaviours, such as their speed, visible area, the number of droplets, covered distances, etc.

In the remaining of this subsection we describe each of the four analysis steps in more details. All the code associated with droplet tracking and analysis can be found at: https://github.com/croningp/chemobot_tools

#### 2.1.5.1 Detection of the dish and tracking area

To simplify the detection of droplets, we needed to narrow the area in which to focus our droplet detection algorithm. We know that the droplets are limited to the petri dish and observed that some experiments tend to generate droplets that stick to the walls of the petri dish – which are both hard to detect by image analysis and not relevant for this study. Hence, we defined a tracking area as within a circle centred at the petri dish centre and with a dimeter of 0.8 times the petri dish visible diameter.

The first task is to detect the petri dish in the image and a lot of prior information are available. We know that the size of the petri dish is constant with 32mm of outside diameter and that its position under the camera is fairly constant given the Geneva Wheel mechanism in use on the Dropfactory and the fixed position of the camera. However slight variations can be observed which we needed to detect.

The petri dish is the biggest circle like object observable in our video. The Hough Transform[41] is a feature extraction technique widely used in image analysis that can be applied to detecting circular objects in a 2D image and is implemented by default in OpenCV[39]. The *HoughCircles* function returns the circle in order of accumulators and our experimentation showed that the most likely circle was always the petri dish. The centre of the detected circle was used as the centre of the dish and the radius was set to 200 pixels, there were no need to detect the radius as the distance of the camera to the petri dish is fixed. To minimize chance of errors, we averaged the position of the dish in a given video by averaging the dish position detected every 100 frames (using the median value to minimize effect of possible outliers).

The code corresponding to the above detection can be found online at https://github.com/croningp/chemobot_tools/blob/master/chemobot_tools/droplet_tracking/tools.py as per the function named *find_petri_dish* and *get_median_dish_from_video*.

Finally, the tracking area is defined as a circle of the same centre but with a diameter of 0.8 times the one of the petri dish to exclude droplet struck on the walls of the petri dish from our analysis. Supplementary Figure 74 shows the detected petri dish circle in red and the corresponding tracking area in blue.

The code corresponding to the above detection can be found online at https://github.com/croningp/chemobot_tools/blob/master/chemobot_tools/droplet_tracking/tools.py as per the function named *create_dish_arena*.

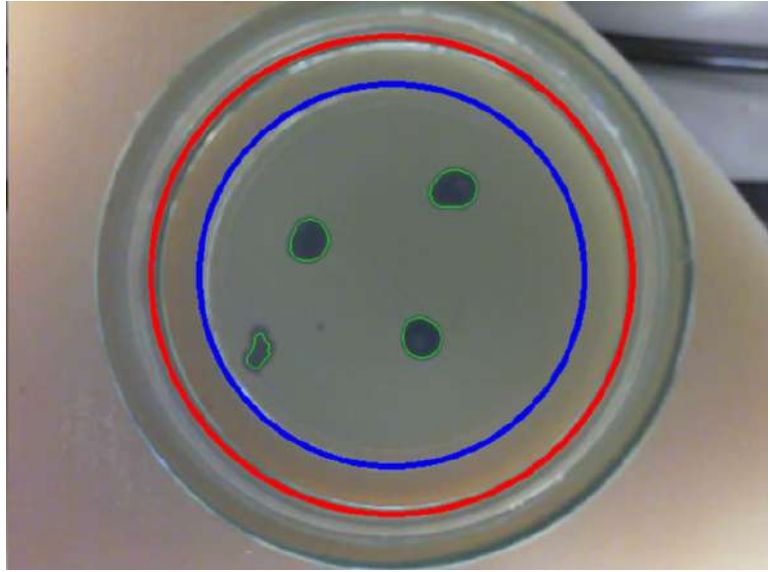

Supplementary Figure 75 – Red: The petri dish location and inside diameter detected using OpenCV and the Hough Transform algorithm. Blue: The tracking area defined as a circle of same centre and 0.8 times diameter as the petri dish. Only droplets in the blue area will be considered for further analysis. In green are the outer contours detected by the full droplet detection pipeline.

The information about the dish and tracking area are then stored into a *dish_info.json* file in the same folder as the video and is used for the subsequent step of our vision analysis pipeline.

### 2.1.5.2 Detection of droplet location and contour

Droplets are black and the background of the video / the bottom of the petri dish is white. To detect droplets, we transform the RGB image into a grey scale image that is then smoothed using a Gaussian blur (with a 5x5 filter and sigma computed as default in OpenCV) to ease processing. We finally apply a threshold that binarizes the image into non-droplet and droplet areas. If the grey channel of a pixel is above the threshold (i.e. whiter), the pixel is considered as a non-droplet, otherwise it is considered as being part of a droplet.

The threshold is not a constant value but computed for each video. To determine the threshold, we first compute the distribution of the pixel intensities within the tracking area average over all frames in the video. Because the distribution is averaged over the all video and the droplets are moving within the dish, the pixel intensity distribution is mostly representative of the background intensity. To extract the threshold value, we fit a Gaussian distribution on the pixel intensity distribution and define the threshold as the intensity above which 99.9% of the distribution is contained. This method enables to adapt the threshold to the variability of the light intensity in the room. The threshold value and decision curve are plotted and saved within each experimental folder.

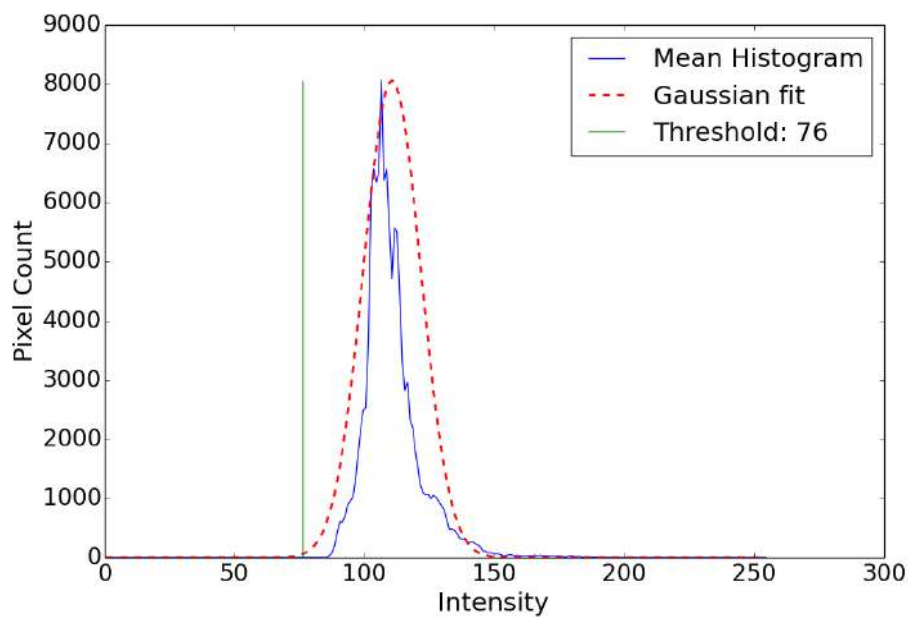

Supplementary Figure 76 – Illustration of the mechanism for defining the binarization threshold of a droplet video. The blue line shows the distribution of intensities of the average grey scale image of a full video. The red line shows the Gaussian distribution used to model the pixel intensity distribution. The green threshold is the value at which 99.9% of the probability distribution is comprised above that value, here 76. The threshold is then used in the binarization process.

The result is a new binary image of non-droplet and droplet areas. Supplementary Figure 76 illustrated these steps. Individual droplets, each being a group pixel isolated from each other's, can now be extracted and identified.

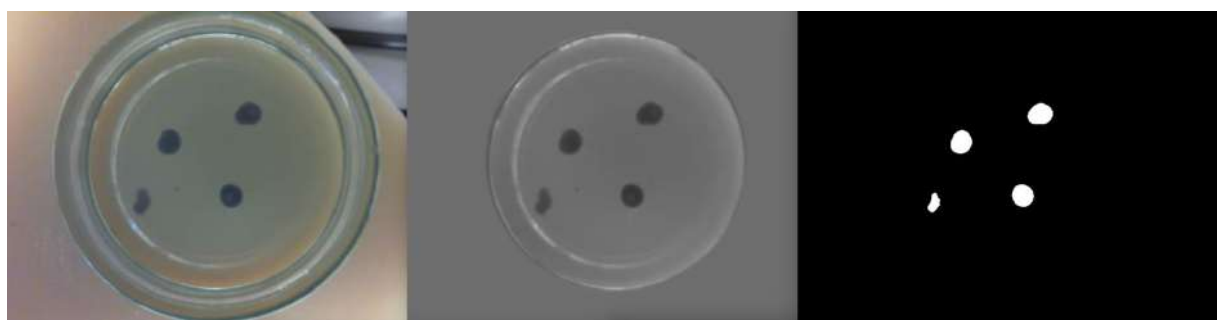

Supplementary Figure 77 – Visualization of the image processing pipeline for the detection of droplet. Left: The original RGB droplet video with petri dish of white background and 4 dark droplets at the surface of the aqueous phase. Middle: The grey scale image of the droplet highlighting the droplet in a single channel image. Right: The binarized image showing in white the pixel detected as being part of a droplet and in black the pixel that are not part of a droplet.

The built-in function *findContours* of OpenCV can extract contours of blobs of pixels from a binarized image using the method described in [42]. We set the mode to CV_RETR_EXTERNAL to retrieve only the extreme outer contours and the method to CV_CHAIN_APPROX_NONE to store all the contour points. The result is a list of the contours of each droplet blob, that is a list of pixel location that form the contours of each droplet, as seen on Supplementary Figure 74.

This process provides already valuable information about the droplet number, location and size, for each frame in a video. We now need to put that information together through time to reconstruct the motion of individual droplets.

The code associated to this section is available at:

- [https://github.com/croningp/chemobot_tools/blob/master/chemobot_tools/droplet_tracking/tools.py](https://github.com/croningp/chemobot_tools/blob/master/chemobot_tools/droplet_tracking/tools.py) for the 'lower level' functions:
    - *binarize_frame* for the pipeline of processing a RGB image into a binary image as described above.
    - *compute_frame_binarization_threshold* function for the mechanism of threshold detection
    - *compute_video_binarization_threshold* function for how we apply it to a full video, along with the *compute_avg_gray_frame* function
- [https://github.com/croningp/chemobot_tools/blob/master/chemobot_tools/droplet_tracking/simple_droplet_tracker.py](https://github.com/croningp/chemobot_tools/blob/master/chemobot_tools/droplet_tracking/simple_droplet_tracker.py) for the all pipeline of detecting droplets, see the *process_video* function.

### 2.1.5.3 Extracting individual droplet sequences

To extract information from individual droplet motion, we need to reconstruct individual trajectory of droplets. We implemented a proximity rule that assigns droplet ids based on their spatial proximity between frames. We added additional spatial, time and size constraints to ensure id assignment is limited to cases that physically make sense for our system:

1. **Space condition** - droplets need to be no more than 100 pixels apart from one frame to the next. If the distance is bigger than that the id assignment is probably wrong by linking two unrelated droplets.
2. **Time condition** - droplet sequences cannot be interrupted by more than 10 frames. That means that we allow for small tracking interruption in a time sequences of droplet position, those are either due to a miss detection by the image analysis or to a droplet moving out and back inside the tracking arena
3. **Size condition** - droplets cannot be smaller than 5 pixels in apparent radius. While developing the tracking we observed that blob detected of less than 5 pixels in radius could be due to false detection.

When a droplet cannot be linked with any previous droplets, a new droplet id is created and assigned to it. This can happen for various reasons: a droplet might split into two or more droplets, a droplet could leave the tracking area and reappear further away in space or time, two droplets might bump into one another which temporarily foul the tracking algorithm as tracking one droplet instead of two as during a fusion event or a contact between droplets.

The accuracy of the tracking of a single droplet all along an experiment is not critical in this work because only averaged high-level metrics are extracted from droplet videos. In addition, our aim is specifically to compare exploration algorithms on the same system rather than to measure droplet behaviours with the highest precision.

The practical implementation of the above grouping algorithm can be found in the https://github.com/croningp/chemobot_tools/blob/master/chemobot_tools/droplet_tracking/droplet_feature.py starting with the *aggregate_droplet_info* function and by following the program trail, especially the *track_droplets* and *group_stats_per_droplets_ids* functions.

### 2.1.5.4 Droplet Metrics Measured

Given the individual frame droplet location and the droplet sequences in time, we can compute a number of metrics that inform us about the dynamic of the droplets. We designed our metrics to be a single number informative about the all duration of an experiment. As a result, most metrics are averaged over the all duration of a video and longer droplet sequences have more weight in the averaging process. To compute each metrics in standard units (rather than pixels or pixel per frame), we know that the frame rate of the video is fixed and of 20 frames per seconds and that the petri dish internal diameter is of 28mm. We describe next the logical behind each metric and link to their implementation.

- **Ratio frame active** – The ratio of frames where at least 1 droplet is detected.
- **Average number of droplets** – The average number of droplets in the arena during the whole experiment (here 90s).
- **Average number of droplets final second** – Mean number of droplets in the arena during the final second
- **Average speed** – Mean droplet speed throughout the experiment, weighted by the duration of each droplet sequences. Video are recorded at 20 frames per seconds. The average speed is computed in $mms^{-1}$.
- **Maximum average single droplet speed** – The average speed of the fastest recorded uninterrupted droplet sequence. The average speed is computed in $mms^{-1}$.

- **Average droplet area.** The average droplet area is the visible area of droplets contours in an experiment as viewed from the top camera. The value is averaged for all droplets and weighted by the duration each droplet is alive. The average droplet area is given in mm$^2$.
- **Average circularity** – The circularity, or isoperimetric quotient, of a shape is defined as $f_{circ} = \frac{4\pi A}{P^2}$, where A is the droplet area (in mm$^2$) and P is the droplet perimeter (in mm). The average circularity is the circularity of each droplet in an experiment weighted by the duration each droplet sequence. The average circularity is a number between 0 and 1, with 1 being a perfect circle.
- **Median absolute circularity deviation -** the median absolute deviation of the circularity weighted by the duration of each droplet sequence. The logical behind this measure was to measure wobbling droplets that change their shape continuously while in motion.
- **Average spread** – the spread is the average distance of each droplet to the centre of mass of the combined droplet visible in the dish weighted by the visible area of each droplet. The logical behind this metric is to measure how close the droplets are moving together. The average spread is computed in mm.
- **Total droplet path length** – The total distance travelled by all droplets during the experiment. The path length is computed in mm.
- **Covered arena area** – The covered area is the ratio of the number of pixels in the tracking area that is visited at least once by a droplet during an experiment. The covered area is given as a ratio between 0 and 1.

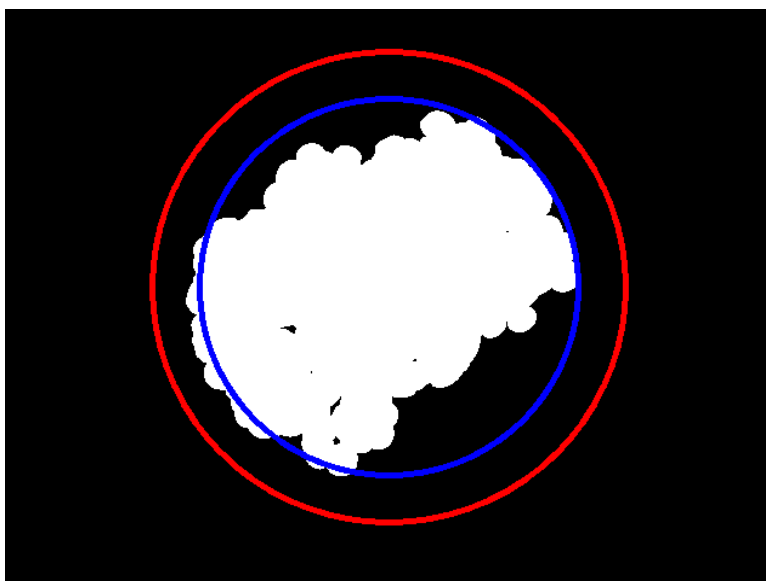

Supplementary Figure 78 - Visualization of the covered arena area metrics. In white are the superimposed droplet position for the full duration of a video. The covered area metric is the ratio of white pixel within the blue circle over the total number of pixel within the blue circle. It is a measure of how active the droplets have been.

The above metrics were computed for all experiments and their value are stored in a *droplet_features.json* file in the experiment folder.

The practical implementation of all the above metrics can be found in the https://github.com/croningp/chemobot_tools/blob/master/chemobot_tools/droplet_tracking/droplet_feature.py library starting line 436.

## 2.2 Algorithms Implementation

The exploration of an input to output mapping, for example exploring the droplet behaviours accessible from droplet recipes, is usually presented as the exploration an unknown function and more formally as follows.

A learning agent $A$ interacts with a system $S$ through the generation of experimental parameters $P$ and the observation of the experimental outcomes $O$. The goal of the agent is to learn the unknown function $f : P \to O$ defining the physical properties of the system. More specifically the agent might be interested in predicting the outcome of a given experiment through a forward model $\tilde{f} : P \to O$ or, more often, in inferring the experimental parameters that will lead to specific outcome using an inverse model $\tilde{f}^{-1} : O \to P$.

The goal for the agent is to estimate $\tilde{f}$ and $\tilde{f}^{-1}$ by collecting $(p, o)$ pairs through its interaction with the unknown system $f$, that is by producing a set of experimental parameters $p \in P$ and observing the outcome $o \in O$ where $o = f(p)$. This learning process can be difficult because:

1. $f$ is often non-linear making the learning of the forward function $\tilde{f}$ from examples not easy because the acquisition of $(p, o)$ pairs should be tailored and targeted at the non-linear portions of $f$ which are yet unknown.
2. $f$ is often redundant, meaning that many $p \in P$ will lead to the same $o \in O$. For example, many droplet recipes do not produce any movement of the droplets. In such cases learning the inverse function $\tilde{f}^{-1}$ becomes more difficult because most of the experiment performed will not generate new observations.
3. Sampling $f$ can be expensive, both in terms of time and budget, such that the collection of $(p, o)$ pairs can be a long and fastidious process. In our case, the experiments can only be made on the real system and in real time and make use of relatively expensive chemicals.

The method used to select the experiments to perform on the system can therefore make a significant difference in the quality of the estimation of both $\tilde{f}$ and $\tilde{f}^{-1}$. Data are precious and not equally useful; thus, their acquisition must be tailored to the specific system in study. This problem is one of exploration strategy and has been described extensively in the fields of robotics and machine learning[11].

Within the scope of this research, $P$ is a 4-dimensional space where each dimension represents the ratio of each oils in a droplet recipe and $p$ is a 4-dimensional vector representing one point in $P$, that is one specific experiment / oil recipe. $O$ is a 2-dimensional space where each dimension represents the average speed and the average number of droplet in an experiment and $o$ is a 2-dimensional vector representing the speed and division values from one specific droplet video analysed as described in the droplet tracking section 2.1.5. We note $(p, o)_i$ the pair of parameters and observations vector associated with the $i$th experiment.

In the following, we describe the two exploration algorithms used in this work, random parameter search (called 'random') and random goal exploration ('CA'), as well as the specific implementation, forward and inverse model estimators and the parameters values selected for this work. We finally detail the exploration metrics used to quantify and compare algorithms.

### 2.2.1 Random Parameter Search Algorithm

The random parameter search algorithm is a typical method used in high-throughput screening and does not use any information about the droplet or the observations made. For each experiment, the algorithm randomly samples 4 numbers between 0 and 1 from a uniform distribution and creates a vector $p$ that is passed to Dropfactory to be executed. The vector $p$ is normalized to sum to 1 such that each dimension represents the ratio of each oil. Each experiment is run on Dropfactory, the droplet video is saved and a vector of observation $o$ is outputted by the platform associated to each $p$.

### 2.2.2 Curious Algorithm: Random Goal Exploration

The curious algorithm used in this work is called random goal exploration[18] (see related algorithms in [43,44]). We recommend reading the associated literature and we summarize below the key concepts and implementation details of random goal exploration as well as the intuition as to why it is efficient for increased exploration.

The random goal exploration algorithm can be decomposed in 4 steps:

1. Randomly sample a target observation ($\hat{o}_i$) that we will try to observe from the system
2. Build an inverse model of the system ($\tilde{f}^{-1}$) using all previous observation $(p, o)_{1:i}$
3. Use the inverse model ($\tilde{f}^{-1}$) to infer the most probable experimental parameters that will lead to the target observation $p_i = \tilde{f}^{-1}(\hat{o}_i)$
4. Execute the experiment $p_i$ and observe the results $o_i$

Repeat for N iteration, in our case N =1000.

Each step requires to define some parameters and choose some models for the forward and inverse model. We used the implementation provided in the explauto Python library[45] which we adapted to interface with our setup. We describe our implementation next:

1. Randomly sample a target observation ($\hat{o}_i$) that we will to observe from the system:

Our observation space $P$ is 2-dimensional with a speed dimension given in mms$^{-1}$ and a division dimension given as the average number of droplet. $p = [s, d]$ with $s, d \in \mathbb{R}$. Given preliminary experiment and previous results reported on oil-in-water droplets [26,37,38], we decided to bound our observation space within [0, 20] for both dimensions. That was, for each dimension, more than twice the upper bound observed so far for this system. The algorithm then samples uniformly a value in [0, 20] as a target observation $\hat{o}_i$.

2. Build an inverse model of the system ($\tilde{f}^{-1}$) using all previous observation $(p, o)_{1:i}$:

The inverse model we use is one of the default one embedded in the explauto library[45], it requires to first build a forward model $\tilde{f}$ that is then inversed using an optimization algorithm that iterative sample $\tilde{f}$ to find the best $p_i$ that minimize the error on the observation ($\varepsilon = |\hat{o}_i - \tilde{o}_i|$), where $\hat{o}_i$ is the target observation and $\tilde{o}_i = \tilde{f}(p_i)$. The forward model $\tilde{f}$ is built using the locally weighted linear regressor[46] (with the parameter k set to 20, that is the number of neighbour point to take into account) and the inverse model is solved iteratively using the CMA-ES algorithm[47] (with sigma set to 0.01 and maxfevals set to 20).

If the number of parameters-observation pairs available is less than 20, the inverse model is not computed because the number of data available does allow for a viable computation of the model. For the 20 first experiments, the model is replaced by a uniform random sampling on experimental parameter.

3. Use the inverse model ($\tilde{f}^{-1}$) to infer the most probable experimental parameters that will lead to the target observation $p_i = \tilde{f}^{-1}(\hat{o}_i)$:

The output of step 2 after optimization of the CMAES algorithm is a new set of parameters $p_i$ that is the most likely to lead to the observation of the targeted observation $\hat{o}_i$ according to the current model of the system.

4. Execute the experiment $p_i$ and observe the results $o_i$:

We sent the experimental parameter to Dropfactory and add the pair $(p, o)_i$ to the list of all previous observation.

Finally, because of the limitation of our real-world setup, the results from one experiment is not available until it has been performed and analysed by Dropfactory. Because Dropfactory run several experiments in parallel using a queue system, the model used to predict the $i$ th experiment is using all the data up to the $i - 8$ th experiments.

Two powerful principles make the random goal exploration algorithm efficient as an exploration method:

- First, random goal exploration explicitly set targets in the output / observation space, which is the space of interest for most studies, rather than in the input / parameter space. This reduces inefficiency in exploration due to redundancies where many experimental conditions lead to the same state of the system (e.g. most of the possible droplet recipes lead to no droplet motion or most tested molecules show low affinity).

- Second, there is no need to understand the inner dynamics of the system to generate interesting observations. Random goal exploration only uses previous observations to construct a model of the system which limits possible biases introduced by human assumptions. This is important since by definition no model exists for unknown system, and, even when simplified models are available, they can only be as good as our current understanding which reduces discovery opportunities. Once exploration has been undertaken, a human scientist can probe interesting or useful systems for deeper understanding of further improvement, as we have done in this work for the temperature sensitivity of droplet behaviour.

The code associated to the algorithm can be found at: https://github.com/croningp/dropfactory_exploration/tree/master/explauto_tools

### 2.2.3 Exploration Measure

We quantified the amount of exploration achieved for each experimental run by computing the area of the alpha-shape containing all the experimental results at a given iteration for a given run on the 2D observation space of speed and division. An alpha-shape is a generalization of the convex hull of a set of points that is derived from the Delaunay triangulation, the method was first introduced in [48]. All alpha-shape were computed with alpha = 15.

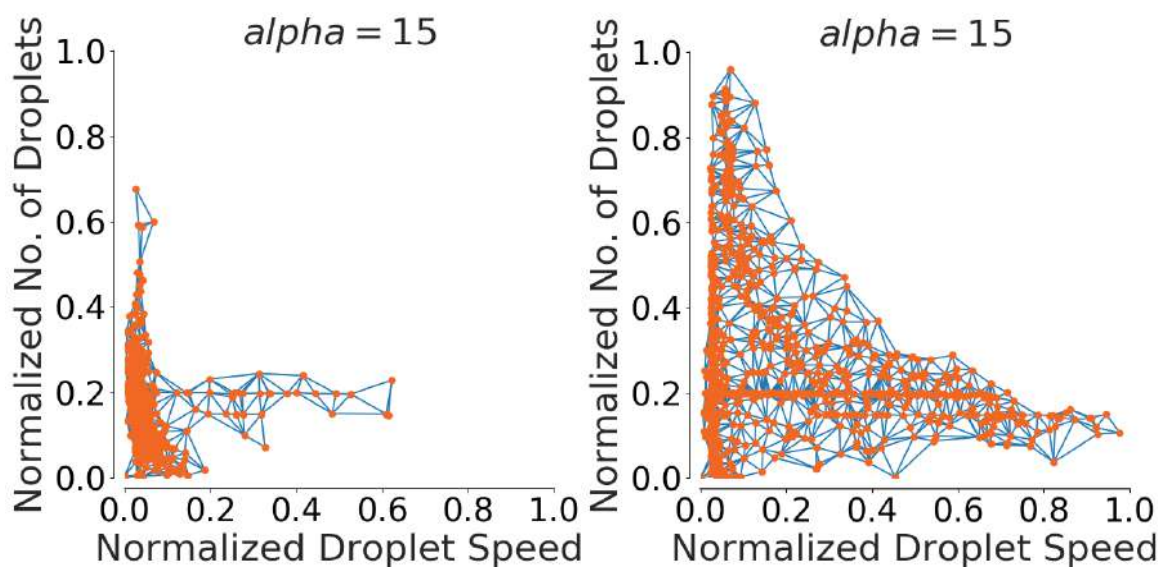

Supplementary Figure 79 - Visualization of the alpha-shape (with alpha = 15) used to quantify the explored space for each experiment for the random (left) and the CA (right) algorithm. Our measure of exploration is the area covered by the alpha-shape, that is the combined area of the blue triangles.

Our measure of exploration is the area covered by the alpha-shape, that is the combined area of the blue triangles on Supplementary Figure 78. The more the observations are spread the higher the exploration measure. The exploration value used in the main paper was normalized into percentages by dividing it by the area of the alpha-shape comprising all observations made using the platform, in our case the 6000 experiments (3 repeats of 2 methods). This normalization cannot be made absolute because we cannot know the theoretical minimal envelope comprising all possible observations of the droplet system given the experimental conditions.

This measure was judged more representative for our study than computing the area of the convex-hull encompassing the observation point or the τ-coverage method described in [44]. The convex-hull (alpha=0) is highly sensitive to outliers which can increase the convex-hull area significantly. The τ-coverage was adequate but judged more difficult to calibrate in an objective way due to the high sensitivity of the measure to value of the τ parameter. However, we note that all three exploration measures highlighted a strong and significant difference between random goal exploration and random parameter search methodology as reported in the paper.

The code to compute the coverage using the alpha-shape method can be found at: https://github.com/croningp/dropfactory_analysis/tree/master/figures/exploration_hull, it has been called concave hull in the code.

## 2.3 Experimental Procedures

### 2.3.1 Oil and Aqueous Phase Preparation

All surfactants and oils used in this work were purchased from Sigma-Aldrich Corporation and HoneyWell. Due to the need for maximum consistency throughout experiments, standard operating procedures were developed for oil and aqueous phase preparation which are shown here.

#### 2.3.1.1 Preparation of Oil Formulations – Standard Operating Procedure

1. Measure oil quantities into a glass bottle using a measuring cylinder (usually 500 mL)
2. Weigh out 0.5 gL$^{-1}$ Sudan Black B
3. Add the Sudan Black B to the mixture.
4. Ensure the dye is dissolved, use a magnetic stirrer.

Again, oils were allowed to equilibrate to experimental temperature for >12 hours.

#### 2.3.1.2 Preparation of Surfactant Solutions – Standard Operating Procedure

Note, a fixed mass of NaOH was used for consistency and the pH recorded, rather than adding base and aiming for a target pH. SOP:

1. Fill a plastic container with distilled water, be precise on the volume (usually 5 L)
2. Weight out the required amount of NaOH for pH adjustment (8 gL$^{-1}$)
3. Add NaOH to the distilled water. Shake to dissolve (care – heat generation on dissolution).
4. Wait overnight for the NaOH$_{(aq)}$ to stabilize in solution and cool down.
5. Weigh out the required amount of surfactant (20 mM TTAB, 6.728 gL$^{-1}$)
6. Add the surfactant to the solution, shake well.
7. Stir using a magnetic stirrer for multiple hours to dissolve, best to do overnight.

Solution is ready, but pH of solution is always recorded.

8. Ensure the pH meter is calibrated using buffers pH = 10, 13 before measuring and recording the pH.

Surfactant solutions were allowed to equilibrate to experimental temperature for >12 hours.

The average pH of the aqueous phase was measured at 13.12 with a standard deviation of 0.07. Below is a table listing all the recorded pH of the aqueous phase:

| Date | pH | Date | pH |
| --- | --- | --- | --- |
| 8/6/2016 | 13.042 | 9/3/2017 | 13.187 |
| 04/07/2016 | 13.190 | 10/03/2017 | 13.162 |
| 08/07/2016 | 13.087 | 17/03/2017 | 13.208 |
| 12/07/2016 | 13.091 | 30/03/2017 | 13.189 |
| 19/07/2016 | 13.102 | 10/04/2017 | 13.240 |
| 18/08/2016 | 13.088 | 10/04/2017 | 13.197 |
| 25/08/2016 | 13.103 | 18/04/2017 | 13.271 |
| 08/09/2016 | 13.087 | 18/04/2017 | 13.284 |
| 12/09/2016 | 13.104 | 01/05/2017 | 13.132 |
| 20/09/2016 | 13.056 | 23/05/2017 | 13.114 |
| 25/10/2016 | 13.068 | 05/06/2017 | 13.040 |
| 25/10/2016 | 13.087 | 26/07/2017 | 13.073 |
| 26/10/2016 | 13.088 | 26/07/2017 | 13.073 |
| 01/11/2016 | 13.091 | 22/08/2017 | 13.084 |
| 06/11/2016 | 13.115 | 22/08/2017 | 13.071 |
| 15/11/2016 | 13.027 | 01/12/2017 | 13.094 |

| | | | | |
|---|---|---|---|---|
| 24/11/2016 | 13.085 | | 05/12/2017 | 13.120 |
| 07/12/2016 | 13.297 | | 7/12/2017 | 13.143 |

### 2.3.2  $^1$H NMR Spectroscopy Experiments

The $^1$H NMR method used was based on that reported previously.[38] 3.5 mL of 20 mM TTAB (pH = c. 13) was placed in a glass vial. To this was added 4 × 4 µL droplets of the given oil formulation. After the given time, 1000 µL of the lower part of the aqueous phase was sampled, taking great care not to sample any oil phase droplets, of which 450 µL was used for analysis. All NMR measurements were performed using a two-channel Bruker Avance III HD 600 spectrometer equipped with a 5-mm BBFO probehead operating at 600.1 MHz for $^1$H. For acquisition of quantitative $^1$H solvent-suppressed experiments, standard presaturation sequence (zgpr from Bruker pulse program library) was used. Temperature was regulated at 298 K. Each spectrum was acquired in 4 scans. CW presaturation (1 mW) was applied on resonance during relaxation delay (2 s). Each sample included 5 mM maleic acid$_{(aq)}$ as an internal standard, present within a capillary inset within the NMR tube and processed on a Windows workstation using the TOPSPIN 3.2 software package.

For each recipe-temperature-timepoint combination at least 6 repeats were undertaken. Temperature when preparing the samples was regulated using the air-conditioning unit in the room. When regulated to a target of 22°C, temperature at the experimental location was 22.4 ±0.2°C. When regulated to a target of 28°C, temperature at the experimental location was 27.7 ±0.2°C. Following the integration with baseline correction method previously reported, these could then be converted into the concentrations of each oil dissolved in the aqueous phase and hence the remaining droplet compositions.

## 3  Supplementary Movies

In this section we provide a short description of each supplementary video as well as a link to watch them online. All the videos are available at the following link: https://www.youtube.com/playlist?list=PLBppiRCztuKo8gxq_kfcYM-5S_A-TlMU1

- **Supplementary Movie 1:** https://youtu.be/bY5OoRBJkf0

Operation of the parallelized droplet robot with details of each working station. The first ca. 45 seconds gives a general overview of the working platform, whilst the remainder of the video shows each working station in operation in detail.

- **Supplementary Movie 2:** https://youtu.be/E76t9LMbuts

The progression of the exploration for each run of each algorithm at an average temperature of 27°C. Each row represents an algorithm (top for CA and bottom for random) and each column represents an independent repeat. Each dot on the speed and division scatter plot represents a 90-second experiment. The CA consistently and rapidly identifies more varied droplet behaviours, and this differentiation only increases as more experiments are undertaken. Note how after only 50 experiments (approximately 1.5 hours of continuous experimentation) the CA is already exploring a significantly greater range of droplet behaviours than random.

- **Supplementary Movie 3:** https://youtu.be/6wPkWJDxN64

1st, 10th and 50th highest speed droplet recipes from one repeat of each method (top CA, bottom random). The 1st, 10th and 50th highest speed experiment observed with CA are significantly more active than their respectively ranked experiment from random.

- **Supplementary Movie 4:** https://youtu.be/zhTeDofB6mk

Effect of temperature on a droplet recipe during a 90s experiment. Recipes is composed of 1.9% octanoic acid, 47.9% DEP, 13.5% 1-octanol, and 36.7% 1-pentanol. From left to right, measured temperatures are 21.4°C, 25.1°C and 28.3°C. For this specific droplet recipe, the higher the temperature the earlier and the faster the droplets start moving.

- **Supplementary Movie 5:** https://youtu.be/80yAmBkzdmM

Effect of temperature on a droplet recipe during a 15-minute experiment. Recipes is composed of 1.9% octanoic acid, 47.9% DEP, 13.5% 1-octanol, and 36.7% 1-pentanol. From left to right, measured temperatures are 21.8°C, 25.8°C and 29.0°C. The measured droplet speed as quantified by our image processing is shown below each video. The video is sped up 5 times. The different phases of motion can clearly be observed as well as their temperature dependant onset timing.

- **Supplementary Movie 6:** https://youtu.be/zOURJEnbmV4

Effect of temperature on the methylene-blue dye released from a droplet recipe composed of 1.9% octanoic acid, 47.9% DEP, 13.5% 1-octanol, and 36.7% 1-pentanol during a 5min experiment. From left to right, average temperatures are 17.6°C and 28.6°C. The video shows two examples at each temperature, alongside the dye release over time as quantified by our image processing. The video is sped up 5 times. The dye is released immediately at the higher temperature whilst it is released only after 60s (12s in the speed up video) at the lower temperature.

# 4  Supplementary References